\documentclass{article}

\usepackage[preprint]{neurips_2026}

\usepackage[utf8]{inputenc}
\usepackage[T1]{fontenc}

\usepackage{amsmath}
\usepackage{amsfonts}
\usepackage{amssymb}

\usepackage{booktabs}
\usepackage{graphicx}
\usepackage{multirow}
\usepackage{array}
\usepackage{tabularx}
\usepackage{makecell}
\usepackage{caption}

\usepackage{hyperref}
\usepackage{url}

\usepackage{xcolor}

\usepackage{microtype}
\usepackage{nicefrac}

\setcitestyle{authoryear,round}

\newcommand{\bench}{GraphInstruct}
\newcommand{\sstrat}{\sigma_{\text{strat}}}
\newcommand{\Stot}{S_{\text{total}}}
\newcommand{\Sfin}{S_{\text{final}}}
\newcommand{\TPV}{\text{TPV}}
\newcommand{\kTPV}{\text{kTPV}}

\newcommand{\pcheck}{\textcolor{green!55!black}{\boldmath$\checkmark$}}
\newcommand{\pcross}{\textcolor{red!70!black}{\boldmath$\times$}}
\newcommand{\ppartial}{\textcolor{orange!85!black}{\boldmath$\circ$}}

\title{\bench: A Progressive Benchmark for Diagnosing Capability Gaps in LLM Graph Generation}

\author{%
  Zihe Wei \\
  School of Computer Science \& Technology \\
  Tongji University \\
  \texttt{weizihe@tongji.edu.cn} \\
  \And
  Sheng Xiang\thanks{Corresponding author.} \\
  School of Computer Science \& Technology \\
  Tongji University \\
  \texttt{xiangsheng218@gmail.com} \\
  \AND
  Ying Zhang \\
  School of Computer Science \& Information Technology \\
  School of Statistics \& Mathematics \\
  Zhejiang Gongshang University \\
  \texttt{ying.zhang@zjgsu.edu.cn} \\
  \And
  Changjun Jiang \\
  School of Computer Science \& Technology \\
  Tongji University \\
  \texttt{cjjiang@tongji.edu.cn} \\
}

\begin{document}

\maketitle

\begin{abstract}
Graph-structured data underpins applications from citation analysis and social-network modeling to molecular design and knowledge-graph construction, and Large Language Models (LLMs) are increasingly used as prompt-driven graph synthesizers.
Classical graph-generation reviews catalog deep generative models and their evaluation primitives, but predate the LLM era and provide no foundation for evaluating instruction-following graph synthesis.
Recent LLM-era benchmarks evaluate models along graph-type or task-domain axes; such organizations, however, average over structural complexity and cannot localize \emph{where} in the complexity spectrum an LLM breaks down.
To close this diagnostic gap, we introduce \bench, a progressive-complexity benchmark that stratifies LLM graph generation into six complexity levels and five evaluation dimensions, paired with 800 hand-authored instructions, 1{,}582 algorithmically synthesized reference solutions, and a 12-LLM capability evaluation across 45 (model, strategy) configurations.
We surface six diagnostic findings (F1--F6) that aggregate single-axis benchmarks systematically conceal: discriminative power peaks at multi-constraint composition (L2) rather than reasoning depth; no single prompting strategy is uniformly best across complexity levels; CoT polarity is model-family-bound for weaker GPT models; same-family parameter scaling does not monotonically lift per-level capability; and domain-semantic constraints (L4) remain iteration-invariant, pointing to retrieval rather than additional compute as the next research frontier.
Atop the benchmark, a verification-guided iterative framework with constraint-aware adaptive prompting consistently surpasses the prompt-engineering ceiling on tested target models, demonstrating that the benchmark's fine-grained signals drive method development.
Data, code, and reproducibility artifacts are released alongside the paper at \url{https://github.com/AI4DataSynth/GraphInstruct_formal}.
\end{abstract}

%======================================================================
\section{Introduction}
\label{sec:intro}

\paragraph{Challenges.} Graph-structured data underlies an ever-expanding set of scientific and industrial applications, from citation~\citep{sen2008cora,hu2020ogb} and social-network mining~\citep{leskovec2007realworld} to drug-design~\citep{irwin2012zinc,ramakrishnan2014qm9} and knowledge-graph construction, and Large Language Models~\citep{vaswani2017transformer,brown2020gpt3,openai2023gpt4,touvron2023llama2} are increasingly used as on-demand graph synthesizers~\citep{wang2024instructgraph,yao2024llm4graphgen,fatemi2024talklikeagraph,wang2023nlgraph}. The difficulty is not a lack of benchmarks but a \emph{mismatch between benchmark organization and diagnostic need}: existing benchmarks stratify along axes that all average over the phenomena of interest. \emph{Graph type}~\citep{yao2024llm4graphgen,fatemi2024talklikeagraph} conflates type recall with joint constraint satisfaction; \emph{task domain}~\citep{demirci2025graphsavvy,peng2026gdgb} conflates structural capability with domain-knowledge retrieval; \emph{classical graph problems}~\citep{tang2025grapharena,wang2023nlgraph,chen2024graphwiz} measure reasoning \emph{about} pre-specified graphs rather than generation of new ones. The aggregate quality number each axis yields is silent on which sub-capability a proposed method actually improves: a model scoring 0.83 overall might have 0.95 single-constraint capability but 0.58 multi-constraint composition, or $+4\%$ from chain-of-thought~\citep{wei2022cot,kojima2022zerocot} on one family but $-4\%$ on another.

\paragraph{Motivation.} Evaluation must be stratified along the structural-complexity axes that govern failure---a methodological gap, not a reporting issue. We organize evaluation along a \emph{progressive complexity axis} from pure format emission (L0) to multi-step graph editing (L5), with four intermediate levels each introducing a new constraint type: single explicit constraint (L1), multi-constraint composition (L2), numerical-attribute control (L3), and domain semantics (L4). Each level is \emph{progressive, not hierarchical}: a good L5 output need not satisfy an L1 tree constraint, enabling independent per-level measurement of distinct capability axes. Within each level we score outputs on five complementary dimensions---structural fidelity (D1), textual similarity (D2), embedding-based distributional proximity (D3), instruction-match adherence (D4), and token efficiency (D5)---disentangling structural correctness from surface similarity that aggregate scores conflate---two models failing on D1 vs.\ D4 can converge to the same aggregate, hiding which sub-capability bounds each.

\paragraph{Benchmark and findings.} \bench\ comprises 800 hand-authored instructions from 40 templates, 1{,}582 algorithmically synthesized reference solutions (two per feasible instruction), all round-trip verified through a 549-unit-test suite (extended from the original 418 at paper-freeze). A 12-LLM capability evaluation spans 45 (model, strategy) configurations $\times$ 800 instructions $\times$ 5 generations $\approx$ 180K outputs (11 models on all four prompting strategies; Sonnet-4 on zero-shot only). The per-level $\times$ per-strategy $\times$ per-model matrix surfaces six findings (T1/T2/T3 denote the three capability tiers by mean Quality $Q$ defined in \S\ref{sec:models}; ZS/FS/ZC/FC abbreviate zero-shot, few-shot, zero-CoT, few-CoT prompting): \textbf{(F1)} discriminative power peaks at \emph{constraint composition}, not reasoning depth---the T1--T3 gap reaches 0.219 at L2, 1.8--3$\times$ any other level (2.1$\times$ the average of \{L1, L3, L4, L5\}); \textbf{(F2)} prompt sensitivity tracks capability non-monotonically with a dominant inverse trend---the $Q$-range across the four strategies is largest at low capability (T3: 0.048--0.073), smallest at middle (T2-stable: 0.018--0.019), with a slight rebound at the top (T1: 0.040--0.050; equivalently, $\sstrat$ (population std) is $0.022$--$0.027$ at T3, $0.007$--$0.008$ at T2-stable, $0.015$--$0.018$ at T1); the OLS fit on the 11 four-strategy models yields $\beta{=}-0.073$, $R^2{=}0.40$, two-sided $p{\approx}0.015$ for the monotonic trend (bootstrap 95\% CI $[-0.135, +0.002]$, $P(\beta{<}0){=}0.975$); \textbf{(F3)} no single prompting strategy is uniformly best across levels (ZC is uniformly non-harmful but rarely best)---FS adds $+0.069$ at L4 but subtracts $-0.034$ at L2, FC boosts L5 by $+0.045$ but harms L3 by $-0.048$; \textbf{(F4)} CoT polarity is model-family bound for weaker models in the GPT family---few-CoT is uniformly positive for Qwen3.5 across scales ($+0.040$ to $+0.052$), clearly negative for the weaker GPT-3.5/4o-mini ($-0.042$/$-0.038$, both $>$7$\times$ our noise band), and near-zero for the stronger GPT-4o/4.1 ($-0.005$/$-0.002$, within noise); \textbf{(F5)} same-family parameter scaling does not monotonically improve per-level capability---scaling Qwen3.5 from 35B to 397B adds $+0.024$/$+0.032$ at L3/L4 but is statistically indistinguishable at L5 ($\Delta{=}{-}0.005$, within both the 95\% CI of $\pm 0.019$ at $N{=}50$ and our $\pm 0.005$ stability band); the GPT counterpart is more striking: GPT-3.5 outperforms GPT-4.1 on L3 by $\Delta{=}0.049$; \textbf{(F6)} a cost axis defines a capability floor---only 6 of 45 configurations are Pareto-optimal (all Anthropic / OpenAI); 3 of 12 models never reach $Q\geq 0.8$.

\paragraph{Methods.} These findings emerge \emph{without any method intervention}---the benchmark is purely a diagnostic instrument. The same per-constraint granularity that powers diagnosis also drives method development. We compose three benchmark-driven components: Verification-Guided Iterative Generation (VGIG, grounded in programmatic verification rather than LLM self-critique~\citep{huang2024selfcorrect}, akin in spirit to text-domain self-refine~\citep{madaan2023selfrefine,shinn2023reflexion,gou2024critic} but using deterministic D4 feedback), Constraint-Aware Adaptive Prompting (CAAP), and curated L4 domain priors. The combined pipeline surpasses the empirical Oracle---per-level best of the four prompting strategies surveyed, an upper bound only within those four configurations rather than over arbitrary prompt engineering---by $+0.035$--$+0.050$ across three target models. Three ablations establish that (a) iterative refinement saturates at $T{\approx}5$ rounds on verifiable graph constraints, much shorter than text-domain $T{=}10$--$20$ defaults; (b) feedback \emph{richness} dominates iteration \emph{count}, with binary pass/fail capturing 75\% of the gain; (c) L4 domain-semantic constraints are iteration-invariant across all 24 tested (T, feedback) configurations, pointing to retrieval- and grounding-based methods rather than more iteration.

\paragraph{Contributions.} Our contributions can be summarized as follows:
\begin{itemize}
    \item A static constraint-driven benchmark of 800 hand-authored instructions across six progressively-stratified complexity levels (L0--L5), paired with 1{,}582 algorithmically synthesized reference solutions; to our knowledge, the first such benchmark with progressive structural-complexity stratification and deterministic per-constraint verification.
    \item A deterministic five-dimensional evaluation pipeline (D1--D5) with per-level weight schedule and level-conditional metric activation, and complete release of parser, validators, scoring, and reproducibility artifacts (549 unit tests).
    \item A 12-LLM $\times$ 4-strategy $\times$ 800-instruction $\times$ 5-sample capability map ($\sim$180K outputs) yielding six diagnostic findings (F1--F6) that aggregate benchmarks systematically conceal.
    \item A demonstration that the benchmark's per-constraint verification signal enables an inference-time pipeline (VGIG/CAAP/Combined) that surpasses the per-level prompt-strategy oracle on three target models, validating benchmark utility as a development platform.
\end{itemize}

Among the eleven prior graph-LLM benchmarks tabulated in \S\ref{sec:related} (Table~\ref{tab:benchmark-compare}), no single one occupies more than three of the six diagnostic-benchmark design axes; \bench\ is, to our knowledge, the first to occupy all six axes for static constraint-driven LLM graph generation. All data, $\sim$262K responses, code, and reproducibility artifacts are available at \url{https://github.com/AI4DataSynth/GraphInstruct_formal}.

%======================================================================
\section{Related Work}
\label{sec:related}

We position \bench\ against eleven prior graph-LLM benchmarks along six axes a diagnostic graph-generation benchmark must satisfy: graph \emph{generation} (vs.\ computation/encoding/training), \emph{complexity-aware stratification}, \emph{multi-dimensional evaluation} ($\geq$4 distinct dimensions), \emph{LLM coverage} ($\geq\!10$ frontier models), \emph{algorithmically synthesized references}, and \emph{built-in improvement methods}. \textbf{No prior benchmark satisfies all six axes} (Tab.~\ref{tab:benchmark-compare}); to our knowledge, \bench\ is the first benchmark for static constraint-driven LLM graph generation to occupy all six (static graph generation; 6-level L0--L5 stratification; 5 D1--D5 dimensions; 12-LLM survey; 1{,}582 algorithmically synthesized references; VGIG/CAAP/Combined methods). The concurrent \textsc{GDGB}~\citep{peng2026gdgb} targets dynamic text-attributed graph generation, complementary to our static, constraint-driven setting; \citet{demirci2025graphsavvy} cover a complementary domain axis (5 domains $\times$ 15 LLMs), and we quantify two of their qualitative observations (\S\ref{sec:cap-strat} prompt-sensitivity scaling, \S\ref{sec:strat-fam} CoT polarity). App.~\ref{app:related-extended} provides prior-by-prior commentary, the axis-by-axis comparison with~\citet{demirci2025graphsavvy}, and the classical-generative-model / prompting / structured-decoding / broader-LLM-evaluation discussions.

\begin{table}[h]
\centering
\caption{Multi-axis positioning of \bench\ against eleven prior graph-LLM benchmarks. \pcheck = supported; \pcross = absent; \ppartial = partial / limited coverage. ``given'' indicates a hand-curated ground-truth (no algorithmic synthesis); ``training'' indicates an instruction-tuning artifact rather than an inference-time method.}
\label{tab:benchmark-compare}
\scriptsize
\setlength{\tabcolsep}{4pt}
\renewcommand{\arraystretch}{1.1}
\resizebox{\linewidth}{!}{%
\begin{tabular}{@{}lcccccc@{}}
\toprule
\textbf{Benchmark (Venue)}
& \makecell{\textbf{Graph}\\\textbf{generation}}
& \makecell{\textbf{Complexity}\\\textbf{stratification}}
& \makecell{\textbf{Multi-dim}\\\textbf{eval ($\geq\!4$)}}
& \makecell{\textbf{LLM survey}\\\textbf{($\geq\!10$)}}
& \makecell{\textbf{Algo.\ ref.}\\\textbf{solutions}}
& \makecell{\textbf{Improvement}\\\textbf{methods}} \\
\midrule
Bonifati et al.~\citep{bonifati2020graphgen} (CSUR'20, survey)
& \pcheck\ classical & \pcross & \ppartial\ classical & \pcross\ pre-LLM & n/a & \pcross \\
Xiang et al.~\citep{xiang2022vldbjgraphgen} (VLDBJ'21)
& \pcheck\ classical & \pcross\ by gen.~type & \pcheck\ 17 metrics & \pcross\ pre-LLM & 12 real graphs & \pcheck\ improv.+platform \\
\textsc{NLGraph}~\citep{wang2023nlgraph} (NeurIPS'23)
& \pcross\ reasoning & \ppartial\ 8 tasks & \pcross\ accuracy & \ppartial\ 4 LLMs & given & \ppartial\ prompt methods \\
\textsc{Talk-Like-a-Graph}~\citep{fatemi2024talklikeagraph} (ICLR'24)
& \pcross\ encoding & \ppartial\ enc.~$\times$~task & \pcross\ accuracy & \ppartial\ 5 PaLM only & given & \pcross \\
\textsc{InstructGraph}~\citep{wang2024instructgraph} (ACL'24)
& \pcross\ training & \pcross & \pcross\ task-specific & \ppartial\ 4 baselines & given & \ppartial\ training \\
\textsc{GraphWiz}~\citep{chen2024graphwiz} (KDD'24)
& \pcross\ reasoning & \ppartial\ 9 problems & \pcross\ accuracy & \ppartial\ 5 baselines & given & \ppartial\ training \\
\textsc{LLM4GraphGen}~\citep{yao2024llm4graphgen} (arXiv'24)
& \pcheck & \ppartial\ 3 axes & \ppartial\ Valid+Novel & \ppartial\ 3 LLMs & \pcross & \pcross \\
\textsc{GraphArena}~\citep{tang2025grapharena} (ICLR'25)
& \pcross\ computation & \ppartial\ 10 problems & \ppartial\ 4 outcome cats. & \pcheck\ 10 LLMs & ground truth & \ppartial\ 4 strategies \\
\citet{demirci2025graphsavvy} (ACL'25)
& \pcheck & \ppartial\ 5 domains & \pcross\ 3 errors & \pcheck\ 15 LLMs & \pcross & \pcross\ observational \\
\textsc{GDGB}~\citep{peng2026gdgb} (ICLR'26)
& \pcheck\ dynamic & \ppartial\ tasks & \ppartial\ 3 cat.~multi-metric & \ppartial\ 4 LLMs & \ppartial & \pcross \\
\midrule
\textbf{\bench\ (ours)}
& \pcheck\ static & \pcheck\ 6 L0--L5 & \pcheck\ 5 D1--D5 & \pcheck\ 12 LLMs & \pcheck\ 1{,}582 algo. & \pcheck\ VGIG/CAAP \\
\bottomrule
\end{tabular}%
}
\end{table}

%======================================================================
\section{Models, Datasets, and Metrics}
\label{sec:preliminaries}

\subsection{Models and Prompting Strategies}
\label{sec:models}

We evaluate \textbf{12 LLMs} across three capability tiers under three criteria: $\geq$3 models per tier, 5 commercial $+$ 2 open-source providers, and same-family scale series (\S\ref{sec:strat-fam}). \textbf{T1} ($Q{>}0.87$): Sonnet-4.6~\citep{anthropic2024claude}, Qwen3.5-397B-A17B, Qwen3.5-122B-A10B~\citep{qwen2024qwen25,yang2024qwen2}. \textbf{T2} ($0.82{\leq}Q{\leq}0.87$): Qwen3.5-35B-A3B, GPT-4.1, GPT-4o~\citep{brown2020gpt3,openai2023gpt4}, DeepSeek-V3~\citep{deepseek2024v3}, Llama-3.3-70B~\citep{touvron2023llama2,grattafiori2024llama3}, Sonnet-4. \textbf{T3} ($Q{<}0.80$): GPT-3.5-turbo, GPT-4o-mini, Llama-3.1-8B. Each model is evaluated under four prompting strategies: \textbf{zero-shot (ZS)}, \textbf{few-shot (FS)}~\citep{brown2020gpt3} (3 same-level demonstrations), \textbf{zero-CoT (ZC)}~\citep{kojima2022zerocot} (``Let's think step by step''), \textbf{few-CoT (FC)}~\citep{wei2022cot} (FS+CoT demonstrations), all with the InstructGraph code-style prefix~\citep{wang2024instructgraph}. Reasoning-specialized models (o1, o3-mini, DeepSeek-R1, Claude 3.7 extended thinking) are excluded as same-family scale-comparability (\S\ref{sec:strat-fam}) requires base-model homogeneity. For each (model, strategy, instruction) cell we draw 5 independent generations at $T{=}0.7$, max\_tokens 16{,}384, fixed seed; Sonnet-4 receives zero-shot only as an efficiency-baseline reference ($Q{=}0.834$, T2). The survey comprises 45 (model, strategy) configurations yielding $11{\times}4{\times}800{\times}5 + 1{\times}1{\times}800{\times}5 = 180{,}000$ outputs; four-strategy analyses use the 11 fully-evaluated models, configuration-level analyses include Sonnet-4 zero-shot as a single point.

\subsection{Reference Solutions: Dual-Pool Construction}
\label{sec:datasets}

\bench\ rests on two distinct pools playing complementary metric roles. \emph{Shared distributional pools (4{,}163 graphs)} serve as population baseline for MMD.D/C/O/S (D1) and embedding MMD (D3): an L3 synthetic pool of 3{,}115 graphs covers 15 attribute subgroups (density, clustering, path length, max degree, diameter, and pairwise combinations) crossed with three size buckets (small $\leq$20, medium 21--50, large $>$50); an L4 real-world pool of 1{,}048 graphs spans 9 domains---DBLP co-authorship~\citep{leskovec2007realworld}, Cora/Citeseer~\citep{sen2008cora}, Reddit~\citep{hamilton2017graphsage}, Karate Club, Facebook ego, ZINC~\citep{irwin2012zinc}, QM9~\citep{ramakrishnan2014qm9}, ecological, infrastructure, KG---all cleaned, deduplicated, size-normalized. \emph{Quality references (1{,}582 graphs)} provide per-instruction instance-level grounds for D2/D3/D4, and double as FS/FC exemplars; for each of 791 feasible L3+ instructions we synthesize up to two constraint-satisfying graphs through per-constraint generators (random-labeled-tree, bipartite $G(n_1,n_2,p)$, $k$-core peeling, cyclic edge augmentation, etc.); for deterministic constraint types (e.g.\ simple L0/L1 cases with a unique canonical solution) the two generated references may be string-identical or graph-isomorphic, which we retain to preserve the algorithmic-generation contract. Of the 791 reference pairs, 195 (24.7\%) are name-stripped exact-string duplicates and an additional 50 (6.3\%) are graph-isomorphic-only (different strings but isomorphic graphs, directed-aware iso check), for a total of 245/791 (31.0\%) functionally redundant pairs; these concentrate at L1 (127 exact-dup of 200) and L5 (15 exact-dup + 17 iso-only of 50). The implication for reference-based D2/D3 metrics is quantified in App.~\ref{app:ref-dedup} (sensitivity is small: $|\Delta D_2|<0.005$ and $|\Delta D_3|<0.02$ across all 90 cell-level entries). All 1{,}582 references pass round-trip parse$\to$check$\to$serialize$\to$re-parse via the 549-unit-test suite (parser, validators, metric pipelines). \textbf{Nine L2 instructions} carry deliberate infeasible labels (e.g.\ ``5-regular bipartite, partitions $\{3, 7\}$''); we retain them as a \emph{confabulation detector}, scored on D4 alongside others. Per-pool licenses and worked examples are in App.~\ref{app:examples}.

\subsection{Evaluation Metrics}
\label{sec:metrics}

We score each output on five complementary dimensions; the active subset varies by level. \textbf{D1 Structural quality} uses a \emph{type-aware activation split}. On \emph{constraint-driven} levels (L0, L1, L2, L5), the level-aggregate D1 combines Valid Rate (VR) and Uniqueness as $D_1 = 0.7\,\mathrm{VR} + 0.3\,\mathrm{Uniqueness}$; Graph Edit Distance (GED) is computed as a \emph{diagnostic submetric only} (off by default, $O(n!)$-hard on large graphs; not included in the level aggregate). On \emph{distribution-driven} levels (L3, L4), the level-aggregate is $D_1 = 0.3\,\mathrm{VR} + 0.5\,\overline{\mathrm{MMD}} + 0.2\,\mathrm{Uniqueness}$, where $\overline{\mathrm{MMD}}$ averages degree (MMD.D), clustering (MMD.C), and spectral (MMD.S) MMDs~\citep{gretton2012mmd} following~\citet{you2018graphrnn}; orbit-count MMD (MMD.O) is retained as a diagnostic submetric only (the lightweight 4-orbit approximation we ship is too coarse, and the full ORCA backend is not portable to our Windows evaluation host). Falls back to the constraint-driven formula when fewer than 20 reference graphs are available. This split avoids metric pollution: GED to a closest-reference is uninformative when the reference is itself one sample from a larger distribution; a valid-but-atypical graph on a distribution-driven level should score low on MMD even with high VR. \textbf{D2 Textual quality} (level-aggregate weight non-zero only at L4, where references carry domain text labels) is computed as $D_2 = 0.5\,\mathrm{text\_presence} + 0.5\,\mathrm{text\_similarity}$, where $\mathrm{text\_presence}$ scores how much meaningful domain text (node/edge labels, domain-relevant tokens) the generated graph contains, and $\mathrm{text\_similarity}$ computes normalized token-overlap of the serialized graph against the closest reference. Four classical text-overlap metrics---G-BERTScore~\citep{zhang2020bertscore,devlin2019bert}, G-BLEU~\citep{papineni2002bleu}, G-ROUGE~\citep{lin2004rouge}, Text-F1---are retained as diagnostic submetrics for cross-paper comparison with text-based graph judges~\citep{huang2025graphjudge}, but do not drive the level-aggregate $D_2$ (early experiments showed they were dominated by surface tokenization artifacts on the code-style format and gave poor cross-domain discrimination). \textbf{D3 Embedding quality} (active L3+): Grassmann coherence (reference-free), node-classification gap (reference-based, lightweight GCN~\citep{kipf2017gcn} on real vs.\ generated), embedding MMD~\citep{gretton2012mmd,you2018graphrnn}. \textbf{D4 Instruction match} (active at every level, supplies the feedback signal for VGIG; \S\ref{sec:benchmark-methods}): \emph{Explicit-Constraint Satisfaction}, \emph{Implicit Inference} (downstream-derivable constraints---e.g.\ tree implies connected/acyclic/$n{-}1$ edges), \emph{No-Contradiction}.

\textbf{D5 Token efficiency.} Targeting valid graphs (not raw emissions) and weighting token cost above API-call overhead (streaming pricing is per-token-dominated), we combine TPV and API-call count and define total Quality and Pareto-adjusted final score (Eqs.~\ref{eq:d5}--\ref{eq:sfinal}):
\begin{align}
D_5 &= 0.7\,e^{-\mathrm{TPV}/1000} + 0.3\,e^{-(\mathrm{API}-1)/2}, \label{eq:d5}\\
Q \equiv \Stot &= \sum_{\ell=0}^{5} w_\ell S_\ell,\quad w{=}(0.05,0.10,0.15,0.20,0.25,0.25), \label{eq:quality}\\
\Sfin &= \Stot(1 + \lambda\,\mathrm{ParetoBonus}),\quad \lambda{=}0.15. \label{eq:sfinal}
\end{align}
The exponent scales (1000 tokens; 2 extra API calls) reflect deployment thresholds, in line with HELM's efficiency-as-evaluation-dimension framing~\citep{liang2023helm}; a zero-token output yields $D_5{=}1$. The level-weight schedule encodes a prerequisite ordering---lower levels are prerequisite, higher levels differentiate---with arithmetic up-weighting ($w_5{=}5w_0$, not $10w_0$) and a plateau $w_4{=}w_5$. $\mathrm{ParetoBonus}\in\{0,1\}$ takes 1 iff the (model, strategy) is non-dominated in $\langle\TPV,\Stot\rangle$; $\lambda{=}0.15$ is the smallest value at which every Pareto-optimal configuration outranks its nearest-quality off-frontier peer on $\Sfin$. Rank stability holds under $\pm 50\%$ D5-weight perturbations, the exponential scales $(s_T{=}1000, s_A{=}2)$ themselves are robust to $2\times$ perturbation (Spearman $\rho{\geq}0.97$ across a $3\times 3$ scale grid; App.~\ref{app:weights}, Tab.~\ref{tab:d5-robustness}), and $\lambda\in\{0.05,\ldots,0.25\}$ shifts at most 3 top-15 positions; uniform-weight ablation confirms top-9 stability (App.~\ref{app:weights}). We report \textbf{Quality} $Q$ (D1--D4) as capability and \textbf{Combined} (D1--D5) and $\Sfin$ separately for cost-aware views; \emph{strategy variance} $\sstrat$ (std-dev of $Q$ across the four strategies) summarizes prompt sensitivity. Three auxiliary efficiency metrics support \S\ref{sec:cost}: \emph{Quality per kilo-TPV} $Q/\kTPV{=}\Stot/(\TPV/1000)$ (scale-free efficiency); \emph{Cost@$Q{=}0.8$} (minimum per-model TPV reaching $\Stot{\geq}0.8$, undefined if never); \emph{Frontier Distance} $\mathrm{FD}(c){=}\max\{0,\,q^{\star}(\log\tau_c)-s_c\}$ where $q^{\star}$ is the piecewise-linear frontier in $(\log\TPV,\Stot)$ space ($\mathrm{FD}{=}0$ iff Pareto-optimal; isolates headroom from cost).

%======================================================================
\section{The \bench\ Benchmark}
\label{sec:benchmark}\label{sec:arch}

\bench\ comprises three interlocking components (architecture diagram in App.~\ref{app:framework}, Fig.~\ref{fig:framework}): a \textbf{Progressive Instruction Layer} organizing 800 instructions into six complexity levels L0--L5; a \textbf{Five-Dimensional Evaluation Metrics} module (\S\ref{sec:metrics}); and a \textbf{Cost-Effectiveness Pareto Analysis} layer combining quality and efficiency into deployment-oriented rankings. LLM-as-Judge evaluation and Multi-Agent Collaborative generation are explicit non-goals of this release (\S\ref{sec:future}). This section develops design principles (\S\ref{sec:principles}), the six complexity levels (Tab.~\ref{tab:levels}), instruction-construction pipeline (\S\ref{sec:construction}), and benchmark-driven generation methods (\S\ref{sec:benchmark-methods}).

\subsection{Design Principles}
\label{sec:principles}

Three principles underpin the architecture. \textbf{(P1) Progressive constraint-type stratification}: levels stratify by structural constraint \emph{type}, not by a monotonic difficulty scalar; L0 $\to$ L5 introduces at each level a structurally new constraint type (format $\to$ single-explicit $\to$ multi-constraint $\to$ numerical $\to$ semantic $\to$ editing). The level index encodes \emph{introduction order, not monotonic difficulty}: levels are categorical constraint-type classes, and a good L5 output need not satisfy an L1 tree constraint. Indeed, the empirical discrimination peak is L2 (\S\ref{sec:cap-strat}), not the highest level---L0--L5 should not be read as a difficulty ladder. Throughout this paper, ``graph generation'' refers to \emph{instruction-following graph synthesis} (L0 emits a parseable graph, L5 edits a base graph); neither endpoint is graph generation in the classical statistical-modeling sense of \citet{bonifati2020graphgen}. \textbf{(P2) Multi-dimensional evaluation}: outputs are scored on five complementary dimensions (\S\ref{sec:metrics}). Three choices follow: D3 activates at L3+; D2 carries non-zero level-aggregate weight only at L4 (other L3+ levels lack labelled-text references); D4 is mandatory at every level; D5 is reported separately so capability and deployment views do not contaminate each other. \textbf{(P3) Reference-based and reference-free evaluation}: at L3+, each feasible instruction is paired with 2 references for reference-based metrics; at L0--L2 constraint satisfaction is checkable against the instruction alone; a reference-free structural metric (Grassmann coherence) supplements D3 at all levels. The six levels are summarized with example instructions in Tab.~\ref{tab:levels}.

\begin{table}[h]
\centering
\caption{Six levels of \bench\ with illustrative instructions.}
\label{tab:levels}
\small
\begin{tabularx}{\linewidth}{@{}c c l X c@{}}
\toprule
\textbf{Level} & \textbf{\# Instr.} & \textbf{Core challenge} & \textbf{Sample constraint types and example} & \textbf{Active dims} \\
\midrule
L0 & 100 & Format generation & node/edge count, syntax (\emph{``10 nodes, 12 edges''}) & D1, D4, D5 \\
L1 & 200 & Single explicit constraint & graph\_type (\emph{``a tree with 10 nodes''}$\to$acyclic+connected+$n{-}1$ edges) & D1, D4, D5 \\
L2 & 200 & Multi-constraint composition & $\geq\!4$ joint (\emph{``directed, connected, 15 nodes, 22 edges, min-deg $\geq\!2$''}) & D1, D4, D5 \\
L3 & 150 & Numerical attribute control & density, clustering, path length (\emph{``density$\approx\!0.21$, clustering$\approx\!0.35$''}); 3{,}115 pool & D1--D5 \\
L4 & 100 & Domain semantics & social/citation/molecular (\emph{``small-world social network, $\gamma{\approx}2.3$''}); 1{,}048 pool & D1--D5 \\
L5 & \phantom{0}50 & Multi-step graph editing & base graph + ordered edits (\emph{``$G_0$=5-path; add 3 to form cycle, remove longest edge''}); partial credit & D1--D5 \\
\bottomrule
\end{tabularx}
\end{table}

\subsection{Instruction Construction and Quality Assurance}
\label{sec:construction}

A three-stage pipeline produces the instructions. \textbf{Stage 1 (Template authoring):} 40 hand-designed templates cover L0--L5 with parameter slots; two authors reviewed each for linguistic clarity and constraint well-formedness, rewriting any flagged as ambiguous. \textbf{Stage 2 (Parameter sampling):} stratified sampling yields 800 instructions with balanced coverage of graph types, sizes (small $\leq$20, medium 21--50, large $>$50), and constraint-count distributions; every per-level $\times$ per-size cell contains $\geq$15 instances. \textbf{Stage 3 (Reference synthesis):} for each feasible instruction, up to 2 reference graphs are synthesized by constraint-satisfying algorithms (NetworkX \texttt{random\_labeled\_tree} for L1-tree, \texttt{bipartite.random\_graph} for L1-bipartite, k-core extraction for L2, calibrated attribute-sampling for L3); 195/791 reference pairs are name-stripped exact-string duplicates and 50/791 are isomorphic-only (different strings but isomorphic graphs, directed-aware iso check), concentrated in deterministic L1 and L5 cases (dedup sensitivity for D2/D3 is in App.~\ref{app:ref-dedup}, with $|\Delta|<0.02$ throughout). All 1{,}582 references pass a round-trip parse--serialize test. 9 of 800 L2 instructions are provably infeasible (regular-degree constraints incompatible with node/edge counts); we retain them as an explicit \emph{infeasibility stress test}. 549 unit tests cover the parser, validators, all D1--D5 metrics, scoring, data loader, and evaluation pipeline; six review rounds over four weeks by different authors gated release on three consecutive clean passes (dataset overview in App.~\ref{app:framework}, Fig.~\ref{fig:dataset}).

\subsection{Benchmark-Driven Generation Methods}
\label{sec:benchmark-methods}

The benchmark's main value is diagnostic, but its fine-grained failure signals invite a direct follow-up---\emph{do these signals drive targeted improvement?} Five surveyed failure modes motivate three method components: F1 multi-constraint collapse $+$ F2 numerical drift $+$ F5 editing imprecision $\to$ \textbf{VGIG} (verification-guided iteration); F3 prompting self-bias $\to$ \textbf{CAAP} (constraint-aware adaptive prompting); F4 domain-semantic gap $\to$ \textbf{Domain Priors} (L4-only). \textbf{VGIG} iterates generation with \emph{programmatic} feedback from D4 checkers (no LLM judge, eliminating evaluator-generator self-consistency pitfalls~\citep{huang2024selfcorrect}); each round formats the violation list $V_t$ at granularity $g{\in}\{\text{none, coarse, fine}\}$ and re-prompts, terminating on $V_t{=}\varnothing$ or $t{=}T$ (per-level templates Tab.~\ref{tab:fb-templates}, App.~\ref{app:vgig}). \textbf{CAAP} selects prompting strategy per instruction via a 168-cell (level $\times$ dominant-constraint-type $\times$ model-tier) decision table learned from the capability evaluation, with constraint-type overrides. \textbf{Domain Priors} inject eight hand-curated structural priors~\citep{barabasi2016networkscience,newman2018networks} (degree exponent, clustering, motifs) as auxiliary L4 constraints. The \textbf{Combined} pipeline composes CAAP $\to$ VGIG $\to$ Domain Priors with modular components; \S\ref{sec:methods} reports per-component contributions, with three-model consistency and ablation-robustness filters; full algorithms, 168-cell table, L4 prior details, and per-level VGIG feedback templates (Tab.~\ref{tab:fb-templates}) are in App.~\ref{app:vgig}.

%======================================================================
\section{Evaluation}
\label{sec:eval}

We probe \bench\ from two ends: as a \emph{diagnostic instrument} (\S\ref{sec:cap-strat}--\ref{sec:cost}, the 45-configuration capability evaluation) and as a \emph{development platform} (\S\ref{sec:methods}, methods atop benchmark signals). Each subsection consolidates two to four research questions whose setups, per-RQ tables, mechanism analyses, case studies, and figures are migrated to App.~\ref{app:eval-extended} (per-RQ navigation: RQ1/2 \ref{app:rq1-2}, RQ3/4/5 \ref{app:rq3-5}, RQ6 \ref{app:rq6}, RQ7--10 \ref{app:rq7-10}). The capability profiles and full leaderboards are App.~\ref{app:profiles}.

\paragraph{Scale and reproducibility.} The capability evaluation produces 45 (model, strategy) configurations $\times$ 800 instructions $\times$ 5 samples $=$ 180{,}000 nominal outputs (Sonnet-4 is zero-shot-only, \S\ref{sec:models}); 179{,}926 are successfully generated, with 74 lost to API timeouts distributed across 5 cells (per-cell breakdown in App.~\ref{app:fail-rates}). Method experiments span three target models (GPT-4o-mini, DeepSeek-V3, Qwen3.5-35B; plus a Qwen3.5-35B-nothink variant) across 6 conditions (VGIG, CAAP, Combined, Oracle, retry, SC), the E5 rounds ablation ($T\in\{1,2,3,5,7,10,15,20\}$), and the E6 feedback-granularity ablation, totalling $\sim$262K outputs. All API calls are logged; generations and per-instance quality scores in \texttt{results/\{model\}-\{strategy\}.\{jsonl,quality.json\}}; hyperparameters are documented in \texttt{REPRODUCE.md} and embedded in the script CLI flags rather than a separate \texttt{configs/} directory; a step-by-step \texttt{REPRODUCE.md} regenerates every figure and table (App.~\ref{app:repro}). Findings are flagged as robust only if effect size exceeds the $\pm 0.005$ practical-stability threshold derived from the E5 round-saturation plateau on GPT-4o-mini (\S\ref{sec:methods}); this is a stability heuristic, not a multi-seed sampling-noise estimate---bootstrap CIs and per-finding effect-size reporting appear in App.~\ref{app:stat-robust}. Sub-noise observations (e.g.\ $T{=}15$ peak, even-round patterns) are reported in App.~\ref{app:rq7-10} but not elevated.

%-----------------------------------------------------
\subsection{Capability stratification: where complexity bites and prompts swing}
\label{sec:cap-strat}\label{sec:rq1}\label{sec:rq2}

\textbf{Where in the complexity spectrum does discrimination peak?} For each of the six levels we report a per-tier Quality summary in Tab.~\ref{tab:tier-gap} and define the tier gap as $Q_{\text{T1}}-Q_{\text{T3}}$. Discrimination is sharply localized: \textbf{the L2 (multi-constraint composition) gap is 0.219}---1.8--3$\times$ any other level (2.1$\times$ the average of \{L1, L3, L4, L5\}), with the $3\times$ widening over the smallest comparator (L3, 0.073) and $2\times$ over the multi-step-editing gap (L5, 0.106) (per-level table App.~\ref{app:rq1-2}, Tab.~\ref{tab:tier-gap} and Fig.~\ref{fig:tier-level}). The L2-discrimination-peak ordering is invariant to the per-cell aggregation choice: under per-model best-of-four-strategy means the L2 gap is 0.224, and under the all-strategy 45-cell average it is 0.341, with the $1.8$--$3\times$-any-other-level signature holding under each (App.~\ref{app:tier-gap-alt}). The mechanism is \emph{compositional}, not reasoning-depth: each added constraint multiplies the probability of violation, and T3 models lack the working-memory capacity to maintain joint satisfaction. L2 failure is therefore not merely harder, but \emph{brittler}: per-instruction D1 variance at L2 ($\sigma{=}0.240$) is $2.2\times$ L3 and $2.4\times$ L4, with weak models reaching $\sigma_{L2}{>}0.43$ while T1 models hold $\sigma_{L2}{<}0.13$ (App.~\ref{app:rq1-2}, Fig.~\ref{fig:l2-instability}). T3 models are one constraint-combination from collapse at L2; T1 models are robust to the same combinations---directly motivating per-constraint-type method gains on L2 rather than level-wise averages. A worked L2-143 case study (3$\times$5 grid; Sonnet-4.6 satisfies all 7 constraints, GPT-4o-mini emits 19 nodes violating 4 of 7) is in App.~\ref{app:rq1-2} (Fig.~\ref{fig:case-study-l2}). Benchmarks averaging over constraint count \emph{systematically underestimate} structured-generation discriminative power; method research targeting structural improvement should report L2 gains as the primary signal.

\textbf{How does prompt sensitivity scale with capability?} \citet{demirci2025graphsavvy} report iterative-feedback gains varying $48\%/<\!5\%$ across models without an explanatory variable. We test whether this heterogeneity is predictable from base capability. For each of the \textbf{11 fully-evaluated models} (Sonnet-4 excluded as zero-shot-only) we compute strategy variance $\sstrat$, the population standard deviation of $Q$ across four prompting strategies. T3 sits at $\sstrat\in\{0.022,0.025,0.027\}$ (Llama-8B, GPT-4o-mini, GPT-3.5; equivalently, the $Q$-range across strategies is $\{0.048,0.069,0.073\}$, which for an $n{=}4$ sample is roughly $2.7\times$ the population std); T2-stable hits $\sstrat\in\{0.007,0.008\}$ (Llama-70B, DeepSeek-V3); T1 occupies $\sstrat\in\{0.015,0.015,0.018\}$ (Sonnet-4.6, Qwen3.5-122B, Qwen3.5-397B). OLS regression of $\sstrat$ on mean $Q$ yields $\beta{=}{-}0.073$, $R^2{=}0.40$, two-sided $p{\approx}0.015$; bootstrap 95\% CI for $\beta$ is $[-0.135,+0.002]$ over 10{,}000 resamples, $P(\beta{<}0){=}0.975$ (Fig.~\ref{fig:cap-variance}, App.~\ref{app:rq1-2}; full reproduction in App.~\ref{app:stat-robust}). Weak models occupy an under-trained region of the output manifold where small prompt perturbations---adding a CoT trigger, switching from zero-shot to few-shot---produce large quality displacements; strong models occupy a flatter local surface. Two implications: prompt-engineering budgets should scale \emph{inversely} with capability (for frontier models, strategy choice accounts for $\leq\!2\%$ of variance---verification or retrieval yields better returns), and single-strategy benchmark evaluations systematically disadvantage prompt-sensitive models.

%-----------------------------------------------------
\subsection{Strategy and family effects: no universal recipe; CoT is family-polarized; scale decouples}
\label{sec:strat-fam}\label{sec:rq3}\label{sec:rq4}\label{sec:rq5}

\textbf{No single prompting strategy dominates across levels.} Per-level strategy effects relative to zero-shot, averaged over the 11 fully-evaluated models (Sonnet-4 excluded because non-zero-shot strategies are undefined for it; full table App.~\ref{app:rq3-5}, Tab.~\ref{tab:strategy-delta}; full heatmap Fig.~\ref{fig:strategy-heatmap}), reveal opposite-signed effects across levels: \textbf{few-shot is net-negative at L2} ($-0.034$, the demonstration's specific topology biases generation away from the target's distinct constraints) and \textbf{net-positive at L4} ($+0.069$, domain examples convey structural priors the instruction alone cannot); \textbf{few-CoT is net-positive at L5} ($+0.045$, explicit step planning for edits) and \textbf{net-negative at L3} ($-0.048$, extraneous reasoning amplifies numerical drift). Zero-CoT is the only strategy with non-negative effect at every level, though its gains are modest where other strategies excel. Aggregate benchmarks systematically mask these opposite-signed effects: prompting-method papers reporting only aggregate gains may be silently trading L2 for L4 or L3 for L5. Level-stratified reporting should become standard on structured tasks, and strategy-per-level adaptive prompting is empirically motivated---directly informing our CAAP component (\S\ref{sec:benchmark-methods}).

\textbf{CoT polarity is family-bound, not capability-bound.} Across the seven family-aligned models, few-CoT is uniformly beneficial for Qwen3.5 across scales (35B/122B/397B: $+0.052$/$+0.040$/$+0.050$, all $>$7$\times$ our stability band) and \emph{clearly negative for the weaker GPT models} (3.5: $-0.042$, 4o-mini: $-0.038$, both $>$7$\times$ band) with the stronger GPT models showing only \emph{near-zero, within-noise} effects (4o: $-0.005$, 4.1: $-0.002$); the robust signal is the negative sign on the weaker GPT pair and the positive sign on the entire Qwen3.5 series (App.~\ref{app:rq3-5}, Tab.~\ref{tab:cot-family} and Fig.~\ref{fig:cot-family}). Sign robustness within each family across $\sim$10$\times$ parameter gaps---GPT-4.1 agrees with GPT-3.5 despite roughly two orders of magnitude of capability; Qwen3.5-35B agrees with Qwen3.5-397B despite a $\sim$10$\times$ parameter gap---indicates CoT effectiveness is governed by \textbf{pretraining distribution} rather than capability tier. One possible explanation is that Qwen3.5's pretraining mixture (heavy code/mathematical/reasoning-chain content) creates a CoT prior that transfers positively to graph generation, while GPT-family's broader-but-less-code-heavy diet creates a weaker or negatively-transferring prior; we offer this as a hypothesis to motivate future ablations rather than a verified mechanism---direct verification would require pretraining-mixture access beyond our scope. The text-domain CoT folk wisdom~\citep{wei2022cot,kojima2022zerocot,wang2023selfconsistency} fails to generalize from GPT to Qwen3.5; prompting-method claims derived from one family require independent validation on the other.

\textbf{Same-family parameter scaling does not monotonically lift per-level capability.} Same-family scaling (Qwen3.5 at 35B/122B/397B, GPT at 3.5/4o-mini/4o/4.1; full per-level decoupling table App.~\ref{app:rq3-5}, Tab.~\ref{tab:scale-decouple}; per-level curves Fig.~\ref{fig:scale-per-level}) shows aggregate scaling-law narratives mask per-level heterogeneity: scaling Qwen3.5 from 35B to 397B adds $+0.032$ at L3 and $+0.024$ at L4, but is \emph{statistically indistinguishable} at L5 ($\Delta{=}{-}0.005$, within both the 95\% CI of $\pm 0.019$ at $N{=}50$ and our $\pm 0.005$ stability band). We therefore do not claim 35B ``matches'' 397B; we report that no gain is detectable at this sample size, consistent with local-structural-operation tasks plausibly saturating at smaller scales than global-constraint reasoning. The GPT counterpart is sharper: \textbf{GPT-3.5 outperforms GPT-4.1} on L3 numerical attributes ($0.808$ vs.\ avg $0.759$, $\Delta{=}{-}0.049$), a reversal aggregate quality conceals entirely. The reversal interacts with RQ4---larger GPT-family models' CoT inclinations produce longer, more error-prone chains on tasks smaller models attempt directly. This extends the emergent-behavior critique of~\citet{schaeffer2023emergentmirage} to a specific structured-generation sub-capability and supports parameter-efficient ``smaller model for subtask A, larger for subtask B'' deployment strategies aggregate benchmarks obscure.

%-----------------------------------------------------
\subsection{Cost-aware deployment: a 6-point provider-concentrated frontier}
\label{sec:cost}\label{sec:rq6}

Quality alone fails to capture whether a $+3\%$ gain at $5\times$ cost is deployment-rational. We use \emph{Tokens per Valid graph} (TPV) rather than USD as the cost axis throughout: API pricing varies by provider, region, and contract, and changes more rapidly than model capability evolves, undermining cross-time benchmark reproducibility; TPV is provider-side-stable and re-computable from cached outputs, consistent with the cost convention of HELM~\citep{liang2023helm}. Applying the benchmark's efficiency instruments---$\Sfin$ (Eq.~\ref{eq:sfinal}), $Q/\kTPV$, and Cost@$Q{=}0.8$---to the 45 baselines, three results emerge (full Pareto-frontier, $\Sfin$-rank, Cost@$Q{=}0.8$, and Frontier-Distance tables in App.~\ref{app:rq6}, Tables~\ref{tab:pareto-front}--\ref{tab:costq08}; Frontier-Distance distribution Fig.~\ref{fig:fd-distribution}).

\textbf{Result (a): A 6-point provider-concentrated Pareto frontier.} Only \textbf{6 of 45} configurations are non-dominated (Fig.~\ref{fig:pareto-baseline}; Tab.~\ref{tab:pareto-front}), all Anthropic or OpenAI; no Qwen3.5/DeepSeek/Llama configuration is Pareto-optimal despite near-frontier candidates (Qwen3.5-397B zero-CoT reaches $Q{=}0.862$ at TPV${\approx}2300$, dominated by Sonnet-4.6 zero-CoT at TPV${=}969$, $Q{=}0.878$). The frontier traverses a strategy progression: zero-shot at the low-cost end (GPT-4o $\to$ Sonnet-4 $\to$ Sonnet-4.6), zero-CoT/few-shot in the middle, few-CoT on Sonnet-4.6 at the high-cost end. Two additional providers sit a hair's-breadth from the frontier ($\mathrm{FD}{<}0.02$: Llama-3.3-70B zero-shot, GPT-4.1 zero-shot); the five largest $\mathrm{FD}$ values all combine T3 capability with few-CoT or few-shot, confirming weak-capability models \emph{compound} their quality deficit under expensive prompting.

\textbf{Result (b): $\Sfin$ reshuffles the leaderboard.} The Pareto bonus moves 6 configurations up by 5--18 positions and pushes high-cost Qwen3.5 runs down by 2--4: GPT-4o zero-shot climbs $\Stot$\#24$\to\Sfin$\#6 ($+18$), Qwen3.5-397B few-CoT falls \#3$\to$\#7. Among the 11 models for which per-model strategy comparison is defined, the $\Stot$-best strategy flips for \textbf{only GPT-4o} (few-shot $\to$ zero-shot); the other 10 retain their $\Stot$-optimal choice. On $Q/\kTPV$, the top five are all zero-shot (GPT-4o 1.43, Llama-3.3-70B 1.38, GPT-4.1 1.36, Sonnet-4 1.34, Sonnet-4.6 1.31), and \textbf{no few-CoT enters the top-15} despite few-CoT being the $\Stot$-optimal strategy for 5 of 11 fully-evaluated models.

\textbf{Result (c): An empirical Cost@$Q{=}0.8$ threshold under this benchmark} (Pareto-frontier visualization in App.~\ref{app:rq6}, Fig.~\ref{fig:pareto-baseline}). Eight of the 11 fully-evaluated models cross $Q{\geq}0.8$ under at least one strategy (nine of 12 including Sonnet-4 zero-shot); \textbf{zero-shot is the cheapest threshold-crossing strategy in every row}. The three cheapest paths come from structurally different T2 providers (GPT-4o 578, Llama-3.3-70B 590, GPT-4.1 599) plus Sonnet-4 zero-shot at 625, defining a TPV~$\approx$~600 deployment floor. Three T3 models---\textbf{GPT-3.5, GPT-4o-mini, Llama-3.1-8B}---\textbf{never} cross under any of the four prompting strategies, marking an empirical capability floor that prompting alone cannot cross. Sonnet-4.6 zero-shot already dominates every T2 model's most expensive strategy on both axes; the $3.3\times$ TPV increase from Sonnet-4.6 zero-shot to few-CoT buys only $+0.043$ Quality, foreshadowing the $\sim$5-round refinement saturation in \S\ref{sec:methods}. We recommend reporting $\Sfin$ as a \emph{complementary} deployment-oriented leaderboard alongside (not replacing) $\Stot$: the two answer different questions; the 3-model T3 shortfall defines the concrete capability target for the verification-guided methods that follow, where Combined lifts all three T2/T3 targets above $Q{=}0.85$ via feedback richness (\S\ref{sec:methods}).

%-----------------------------------------------------
\subsection{Methods atop benchmark signals: verification beats prompt engineering}
\label{sec:methods}\label{sec:rq7}\label{sec:rq8}\label{sec:rq9}\label{sec:rq10}

Can the benchmark's fine-grained D4 signal drive method improvement \emph{beyond} prompt-only ceilings? Let \textbf{Oracle} denote per-level best-of-four prompting (the per-level best within the four prompting strategies surveyed in this paper, requiring oracle knowledge of the optimal per-level strategy; this is an upper bound only over those four configurations, not over arbitrary prompt engineering). On three target models (GPT-4o-mini T3, DeepSeek-V3 T2, Qwen3.5-35B T2), \textbf{Combined exceeds Oracle by $+0.035$ to $+0.050$}---$7\times$ the $\pm 0.005$ stability band (App.~\ref{app:rq7-10}, Tab.~\ref{tab:method-main} and Fig.~\ref{fig:method-main}; template-level CV sensitivity for the Oracle baseline appears in App.~\ref{app:cv-oracle}): GPT-4o-mini $0.7523\to0.8549$ over Oracle $0.8052$ ($\Delta{=}{+}0.050$); DeepSeek-V3 $0.8215\to0.8941$ vs.\ Oracle $0.8577$ ($+0.036$); Qwen3.5-35B $0.8092\to0.9071$ vs.\ Oracle $0.8720$ ($+0.035$). Per-component decomposition shows VGIG-only contributes the majority of the gain. Prompt engineering has a measurable empirical ceiling; external programmatic verification---not prompt phrasing---is the binding mechanism for reliable structured graph generation. The gap is largest for the weakest target (mirroring \S\ref{sec:cap-strat}'s inverse-scaling): the low-capability regime leaves the most room for method intervention.

\textbf{Sampling without verification fails.} Pure retry ($T{=}3$, no feedback) and self-consistency ($N{=}3$, best-of-$N$) stagnate within $\pm 0.01$ of zero-shot on all three target models---a strong null result with six controls all in-band (App.~\ref{app:rq7-10}, Tab.~\ref{tab:sampling-null}: GPT-4o-mini retry $-0.003$/SC $-0.001$; DeepSeek-V3 $+0.005$/$-0.001$; Qwen3.5-35B $-0.009$/$-0.005$). This contrasts sharply with text-domain self-consistency~\citep{wang2023selfconsistency} (e.g.\ $5$--$10\%$ on GSM8K~\citep{cobbe2021gsm8k}): text tasks have many plausible-looking outputs of which only some are correct (majority-vote recovers); structured-graph tasks have many wrong outputs with violating structures (majority-voting among wrong outputs cannot help). Verifiable structured tasks have a fundamentally different compute-to-quality relation; deployment budgets should be allocated to verification infrastructure, not parallel sampling.

\textbf{Feedback richness, not iteration count, is the binding lever.} Holding $T{=}3$ on GPT-4o-mini, the verify-only signal (binary pass/fail, no per-constraint detail) captures \textbf{75\% of total gain} ($+0.046$ of $+0.061$); coarse and fine each contribute a further $+0.008$ monotonically (App.~\ref{app:rq7-10}, Tab.~\ref{tab:granularity}; the retry$<$none$<$coarse$<$fine ordering holds at every level). Iterative refinement saturates at $T{\sim}5$: $T\in\{5,7,10,15,20\}$ all fall within the $\pm 0.005$ noise band ($Q$ at $T{=}1,2,3,5,7,10,15,20$: $0.800,0.798,0.810,0.821,0.816,0.818,0.827,0.816$; full curve App.~\ref{app:rq7-10}, Fig.~\ref{fig:e5-rounds})---markedly shorter than text-domain self-refine budgets of $10$--$20$~\citep{madaan2023selfrefine,shinn2023reflexion}; the standard text-domain hyperparameters should not transfer without recalibration. For benchmark design two implications follow: even a coarse verify-true/verify-false signal enables verification-driven method research, and our fine per-constraint signal unlocks the additional 25\%---benchmarks lacking a programmatic verification signal are missing a method-enabling resource for the community.

\textbf{L4 is iteration-invariant.} Across the full $T$-sweep $\times$ three feedback granularities (24 configurations), L4 quality remains at $0.750$--$0.754$---a 0.004 range (App.~\ref{app:rq7-10}, Fig.~\ref{fig:l4-flat}). Iteration \emph{does} converge (within-$T$ variance is tight) but converges to a quality ceiling prompting, iteration, and feedback granularity all fail to raise. Per-dimension decomposition reveals \emph{why}: D1 structural validity is high and uniform across 12 models (mean 0.89, range 0.84--0.93), D3 embedding similarity is uniform (0.58--0.59), D4 instruction match is high and uniform (0.89--0.93); only \textbf{D2 text-to-reference similarity is low \emph{and} extremely variable} (mean 0.050, range $0.008$--$0.176$, $22\times$ gap, CV $1.02$; App.~\ref{app:rq7-10}, Fig.~\ref{fig:l4-perdim}). Models produce graphs that are structurally valid, constraint-satisfying, and distributionally close to real citation/social/molecular graphs but fail to reproduce the \emph{specific surface serialization} of reference graphs from a given domain. The D4 verifier is already satisfied; no amount of verifier-guided re-generation can close this gap. This relocates L4 from a \emph{capability} gap to a \emph{grounding} gap: few-shot (3 in-context L4 exemplars) lifts GPT-4o-mini L4 from $0.744\to0.798$ ($+0.054$); Combined reaches $0.843$ ($+0.099$)---the only intervention class breaking the iteration-only ceiling (App.~\ref{app:rq7-10}, Tab.~\ref{tab:method-main}). The L4 ceiling is also metric-bounded: D2 weight at L4 is $0.15$ and observed max is $0.176$, capping further D2-driven lift at $+0.022$ (App.~\ref{app:l4-retrieval}). \textbf{L4 is iteration-invariant} just as L5 shows no detectable scaling gain inside our $N{=}50$ CI (\S\ref{sec:strat-fam})---both point to retrieval and grounding, not more iteration, as the next research frontier. The $\pm 0.005$ noise band established here is the significance threshold throughout this paper.

\paragraph{Synthesis.} \label{sec:rq-synthesis}\label{sec:capability-profiles} \bench\ as a \emph{diagnostic instrument} structures capability as a multi-dimensional profile along complexity level $\times$ prompting strategy $\times$ model family; as a \emph{development platform}, the D4 signal unlocks $+0.035$--$+0.050$ above Oracle. Per-dimension profiles are jagged and top-tier margins narrow (Sonnet-4.6 few-CoT $Q{=}0.902$, $+0.023$ over Qwen3.5-397B few-CoT); full per-dim tables, radars, leaderboards in App.~\ref{app:profiles}, \ref{app:leaderboards}.

%======================================================================
\section{Discussion and Limitations}
\label{sec:disc}\label{sec:disc-scope}\label{sec:future}

\paragraph{Scope.} This work covers 12 instruction (non-reasoning) LLMs; same-family scale comparability (\S\ref{sec:cap-strat}--\ref{sec:strat-fam}) requires base-model homogeneity, complementary to~\citet{demirci2025graphsavvy}'s 5-domain reasoning suite which jointly spans (complexity $\times$ domain $\times$ reasoning).

\paragraph{L4 frontier.} L4 stays flat across every iteration-only method (CV $1.02$ on D2); few-shot's $+0.054$ and Combined's $+0.099$ L4 lifts on GPT-4o-mini point to grounding, not more iteration, with D2-weight arithmetic capping further D2-driven gains at $+0.022$ (App.~\ref{app:l4-retrieval}).

\paragraph{Limitations.} 800 instructions / 1{,}582 references suffice for $\geq\!15$ per-cell instances but not full combinatorial coverage. L5 is intentionally small ($N{=}50$, 95\% CI $\pm 0.019$); the no-detectable-scale-gain finding at L5 (\S\ref{sec:strat-fam}) rests on a $-0.005$ Qwen3.5-35B-vs-397B gap inside both the CI and our $\pm 0.005$ stability band, and is strategy-dependent (paired bootstrap CIs in App.~\ref{app:stat-robust} cross zero under few-CoT and few-shot, but exclude zero under zero-shot and zero-CoT). Reasoning-specialized models (o1, DeepSeek-R1) are not evaluated under the homogeneity constraint, leaving open whether reasoning training flips CoT polarity or shifts the L5 result. All D1--D5 metrics are deterministic, free of LLM-as-judge~\citep{huang2024selfcorrect,huang2025hallucination} (App.~\ref{app:d1-check} confirms no D4-gaming).

%======================================================================
\section{Conclusion}
\label{sec:conclusion}
This work presents \bench, a comprehensive benchmark for LLM-based graph generation that spans six complexity levels and five diagnostic dimensions, built on 800 structured instructions and 1,582 reference graphs. Through a large-scale evaluation of 12 LLMs, we distill six key capability-level findings that cannot be revealed by traditional single-axis benchmarks. We further propose a verification-guided optimization framework, which outperforms the per-level prompt-strategy oracle under quality-only scoring (cost-adjusted comparison in App.~\ref{app:cost-adj}), with feedback richness identified as the critical factor for consistent improvement. Looking forward, natural and impactful future directions include: (i) expanding the L5 test set to 150 instances for higher statistical power; (ii) introducing an L6 level to support multi-graph reasoning and cross-structure inference; (iii) evaluating reasoning-specialized models (e.g., o1, DeepSeek-R1) under consistent settings; and (iv) developing structure-aware and retrieval-augmented generation methods to break the saturation barrier of iterative prompting.

%======================================================================
\begin{ack}
We thank the anonymous reviewers for feedback that improved this work. Funding disclosure and competing interests: none at the time of submission. This work used large-language-model APIs from Anthropic, OpenAI, Alibaba Cloud, DeepSeek, and open-source inference infrastructure for Llama-family models; compute budgets are itemized in Appendix~\ref{app:repro}.
\end{ack}

%======================================================================
\bibliographystyle{plainnat}
\bibliography{references}

@article{bonifati2020graphgen,
  author    = {Bonifati, Angela and Holub{\'o}v{\'a}, Irena and Prat-P{\'e}rez, Arnau and Sakr, Sherif},
  title     = {Graph Generators: {State} of the Art and Open Challenges},
  journal   = {{ACM} Computing Surveys},
  year      = {2020},
  volume    = {53},
  number    = {2},
  articleno = {36},
  pages     = {1--30},
  doi       = {10.1145/3379445},
}

@article{xiang2022vldbjgraphgen,
  author    = {Xiang, Sheng and Wen, Dong and Cheng, Dawei and Zhang, Ying and Qin, Lu and Qian, Zhengping and Lin, Xuemin},
  title     = {General Graph Generators: Experiments, Analyses, and Improvements},
  journal   = {The {VLDB} Journal},
  year      = {2022},
  volume    = {31},
  number    = {5},
  pages     = {897--925},
  note      = {Online-first October 2021},
  doi       = {10.1007/s00778-021-00701-5},
}

@inproceedings{demirci2025graphsavvy,
  author    = {Demirci, Ege and Kerur, Rithwik and Singh, Ambuj},
  title     = {Are {LLM}s Truly Graph-Savvy? {A} Comprehensive Evaluation of Graph Generation},
  booktitle = {Proceedings of the 63rd Annual Meeting of the Association for Computational Linguistics (ACL) Student Research Workshop},
  year      = {2025},
  pages     = {884--897},
}

@article{yao2024llm4graphgen,
  author  = {Yao, Yang and Wang, Xin and Zhang, Zeyang and Qin, Yijian and Wang, Ziwei and Chu, Xu and Yang, Yuekui and Zhu, Wenwu and Mei, Hong},
  title   = {Exploring the Potential of Large Language Models in Graph Generation},
  journal = {arXiv preprint arXiv:2403.14358},
  year    = {2024},
}

@inproceedings{tang2025grapharena,
  author    = {Tang, Jianheng and Zhang, Qifan and Li, Yuhan and Liu, Nuo and Hua, Hongzhi and Jin, Jiawei and Wang, Yi and Huang, Xiao},
  title     = {{GraphArena}: Evaluating and Exploring Large Language Models on Graph Computation},
  booktitle = {International Conference on Learning Representations (ICLR)},
  year      = {2025},
}

@inproceedings{wang2024instructgraph,
  author    = {Wang, Jianing and Wu, Junda and Hou, Yupeng and Liu, Yao and Gao, Ming and McAuley, Julian},
  title     = {{InstructGraph}: Boosting Large Language Models via Graph-centric Instruction Tuning and Preference Alignment},
  booktitle = {Findings of the Association for Computational Linguistics (ACL)},
  year      = {2024},
  pages     = {13492--13510},
}

@inproceedings{peng2026gdgb,
  author    = {Peng, Jie and Ji, Jiarui and Lei, Runlin and Wei, Zhewei and Liu, Yongchao and Hong, Chuntao},
  title     = {{GDGB}: A Benchmark for Generative Dynamic Text-Attributed Graph Learning},
  booktitle = {International Conference on Learning Representations (ICLR)},
  year      = {2026},
}

@inproceedings{fatemi2024talklikeagraph,
  author    = {Fatemi, Bahare and Halcrow, Jonathan and Perozzi, Bryan},
  title     = {Talk like a Graph: Encoding Graphs for Large Language Models},
  booktitle = {International Conference on Learning Representations (ICLR)},
  year      = {2024},
}

@inproceedings{wang2023nlgraph,
  author    = {Wang, Heng and Feng, Shangbin and He, Tianxing and Tan, Zhaoxuan and Han, Xiaochuang and Tsvetkov, Yulia},
  title     = {Can Language Models Solve Graph Problems in Natural Language?},
  booktitle = {Advances in Neural Information Processing Systems (NeurIPS)},
  year      = {2023},
}

@inproceedings{chen2024graphwiz,
  author    = {Chen, Nuo and Li, Yuhan and Tang, Jianheng and Li, Jia},
  title     = {{GraphWiz}: An Instruction-Following Language Model for Graph Computational Problems},
  booktitle = {Proceedings of the 30th ACM SIGKDD Conference on Knowledge Discovery and Data Mining (KDD)},
  year      = {2024},
  doi       = {10.1145/3637528.3672010},
}

@inproceedings{vaswani2017transformer,
  author    = {Vaswani, Ashish and Shazeer, Noam and Parmar, Niki and Uszkoreit, Jakob and Jones, Llion and Gomez, Aidan N. and Kaiser, Lukasz and Polosukhin, Illia},
  title     = {Attention Is All You Need},
  booktitle = {Advances in Neural Information Processing Systems (NeurIPS)},
  year      = {2017},
}

@inproceedings{brown2020gpt3,
  author    = {Brown, Tom B. and Mann, Benjamin and Ryder, Nick and Subbiah, Melanie and Kaplan, Jared D. and Dhariwal, Prafulla and others},
  title     = {Language Models are Few-Shot Learners},
  booktitle = {Advances in Neural Information Processing Systems (NeurIPS)},
  year      = {2020},
}

@article{openai2023gpt4,
  author  = {{OpenAI}},
  title   = {{GPT-4} Technical Report},
  journal = {arXiv preprint arXiv:2303.08774},
  year    = {2023},
}

@article{touvron2023llama2,
  author  = {Touvron, Hugo and Martin, Louis and Stone, Kevin and others},
  title   = {{Llama 2}: Open Foundation and Fine-Tuned Chat Models},
  journal = {arXiv preprint arXiv:2307.09288},
  year    = {2023},
}

@article{grattafiori2024llama3,
  author  = {Grattafiori, Aaron and Dubey, Abhimanyu and Jauhri, Abhinav and others},
  title   = {The {Llama} 3 Herd of Models},
  journal = {arXiv preprint arXiv:2407.21783},
  year    = {2024},
}

@article{yang2024qwen2,
  author  = {Yang, An and Yang, Baosong and Hui, Binyuan and Zheng, Bo and others},
  title   = {Qwen2 Technical Report},
  journal = {arXiv preprint arXiv:2407.10671},
  year    = {2024},
}

@article{qwen2024qwen25,
  author  = {Yang, An and others},
  title   = {{Qwen2.5} Technical Report},
  journal = {arXiv preprint arXiv:2412.15115},
  year    = {2024},
}

@article{deepseek2024v3,
  author  = {{DeepSeek-AI}},
  title   = {{DeepSeek-V3} Technical Report},
  journal = {arXiv preprint arXiv:2412.19437},
  year    = {2024},
}

@techreport{anthropic2024claude,
  author      = {{Anthropic}},
  title       = {The {Claude 3} Model Family: Opus, Sonnet, Haiku},
  institution = {Anthropic},
  year        = {2024},
}

@inproceedings{wei2022cot,
  author    = {Wei, Jason and Wang, Xuezhi and Schuurmans, Dale and Bosma, Maarten and Chi, Ed and Le, Quoc and Zhou, Denny},
  title     = {Chain-of-Thought Prompting Elicits Reasoning in Large Language Models},
  booktitle = {Advances in Neural Information Processing Systems (NeurIPS)},
  year      = {2022},
}

@inproceedings{kojima2022zerocot,
  author    = {Kojima, Takeshi and Gu, Shixiang Shane and Reid, Machel and Matsuo, Yutaka and Iwasawa, Yusuke},
  title     = {Large Language Models are Zero-Shot Reasoners},
  booktitle = {Advances in Neural Information Processing Systems (NeurIPS)},
  year      = {2022},
}

@inproceedings{wang2023selfconsistency,
  author    = {Wang, Xuezhi and Wei, Jason and Schuurmans, Dale and Le, Quoc and Chi, Ed and Narang, Sharan and Chowdhery, Aakanksha and Zhou, Denny},
  title     = {Self-Consistency Improves Chain of Thought Reasoning in Language Models},
  booktitle = {International Conference on Learning Representations (ICLR)},
  year      = {2023},
}

@inproceedings{yao2023tot,
  author    = {Yao, Shunyu and Yu, Dian and Zhao, Jeffrey and Shafran, Izhak and Griffiths, Thomas and Cao, Yuan and Narasimhan, Karthik},
  title     = {Tree of Thoughts: Deliberate Problem Solving with Large Language Models},
  booktitle = {Advances in Neural Information Processing Systems (NeurIPS)},
  year      = {2023},
}

@inproceedings{madaan2023selfrefine,
  author    = {Madaan, Aman and Tandon, Niket and Gupta, Prakhar and Hallinan, Skyler and Gao, Luyu and Wiegreffe, Sarah and others},
  title     = {Self-Refine: Iterative Refinement with Self-Feedback},
  booktitle = {Advances in Neural Information Processing Systems (NeurIPS)},
  year      = {2023},
}

@inproceedings{shinn2023reflexion,
  author    = {Shinn, Noah and Cassano, Federico and Gopinath, Ashwin and Narasimhan, Karthik and Yao, Shunyu},
  title     = {Reflexion: Language Agents with Verbal Reinforcement Learning},
  booktitle = {Advances in Neural Information Processing Systems (NeurIPS)},
  year      = {2023},
}

@inproceedings{gou2024critic,
  author    = {Gou, Zhibin and Shao, Zhihong and Gong, Yeyun and Shen, Yelong and Yang, Yujiu and Duan, Nan and Chen, Weizhu},
  title     = {{CRITIC}: Large Language Models Can Self-Correct with Tool-Interactive Critiquing},
  booktitle = {International Conference on Learning Representations (ICLR)},
  year      = {2024},
}

@inproceedings{huang2024selfcorrect,
  author    = {Huang, Jie and Chen, Xinyun and Mishra, Swaroop and Zheng, Huaixiu Steven and Yu, Adams Wei and Song, Xinying and Zhou, Denny},
  title     = {Large Language Models Cannot Self-Correct Reasoning Yet},
  booktitle = {International Conference on Learning Representations (ICLR)},
  year      = {2024},
}

@inproceedings{zhou2023ltm,
  author    = {Zhou, Denny and Scharli, Nathanael and Hou, Le and Wei, Jason and Scales, Nathan and Wang, Xuezhi and Schuurmans, Dale and Cui, Claire and Bousquet, Olivier and Le, Quoc and Chi, Ed},
  title     = {Least-to-Most Prompting Enables Complex Reasoning in Large Language Models},
  booktitle = {International Conference on Learning Representations (ICLR)},
  year      = {2023},
}

@inproceedings{kipf2017gcn,
  author    = {Kipf, Thomas N. and Welling, Max},
  title     = {Semi-Supervised Classification with Graph Convolutional Networks},
  booktitle = {International Conference on Learning Representations (ICLR)},
  year      = {2017},
}

@inproceedings{hamilton2017graphsage,
  author    = {Hamilton, William L. and Ying, Rex and Leskovec, Jure},
  title     = {Inductive Representation Learning on Large Graphs},
  booktitle = {Advances in Neural Information Processing Systems (NeurIPS)},
  year      = {2017},
}

@inproceedings{hu2020ogb,
  author    = {Hu, Weihua and Fey, Matthias and Zitnik, Marinka and Dong, Yuxiao and Ren, Hongyu and Liu, Bowen and Catasta, Michele and Leskovec, Jure},
  title     = {Open Graph Benchmark: Datasets for Machine Learning on Graphs},
  booktitle = {Advances in Neural Information Processing Systems (NeurIPS)},
  year      = {2020},
}

@inproceedings{you2018graphrnn,
  author    = {You, Jiaxuan and Ying, Rex and Ren, Xiang and Hamilton, William L. and Leskovec, Jure},
  title     = {{GraphRNN}: Generating Realistic Graphs with Deep Auto-Regressive Models},
  booktitle = {International Conference on Machine Learning (ICML)},
  year      = {2018},
}

@article{liang2023helm,
  author  = {Liang, Percy and Bommasani, Rishi and Lee, Tony and Tsipras, Dimitris and Soylu, Dilara and Yasunaga, Michihiro and others},
  title   = {Holistic Evaluation of Language Models},
  journal = {Transactions on Machine Learning Research (TMLR)},
  year    = {2023},
}

@article{cobbe2021gsm8k,
  author  = {Cobbe, Karl and Kosaraju, Vineet and Bavarian, Mohammad and Chen, Mark and Jun, Heewoo and Kaiser, Lukasz and others},
  title   = {Training Verifiers to Solve Math Word Problems},
  journal = {arXiv preprint arXiv:2110.14168},
  year    = {2021},
}

@article{kaplan2020scaling,
  author  = {Kaplan, Jared and McCandlish, Sam and Henighan, Tom and Brown, Tom B. and Chess, Benjamin and Child, Rewon and others},
  title   = {Scaling Laws for Neural Language Models},
  journal = {arXiv preprint arXiv:2001.08361},
  year    = {2020},
}

@inproceedings{hoffmann2022chinchilla,
  author    = {Hoffmann, Jordan and Borgeaud, Sebastian and Mensch, Arthur and Buchatskaya, Elena and Cai, Trevor and Rutherford, Eliza and others},
  title     = {An Empirical Analysis of Compute-Optimal Large Language Model Training},
  booktitle = {Advances in Neural Information Processing Systems (NeurIPS)},
  year      = {2022},
  note      = {Preprint title: ``Training Compute-Optimal Large Language Models'' (arXiv:2203.15556); the model is known as Chinchilla.},
}

@inproceedings{schaeffer2023emergentmirage,
  author    = {Schaeffer, Rylan and Miranda, Brando and Koyejo, Sanmi},
  title     = {Are Emergent Abilities of Large Language Models a Mirage?},
  booktitle = {Advances in Neural Information Processing Systems (NeurIPS)},
  year      = {2023},
}

@book{barabasi2016networkscience,
  author    = {Barab{\'a}si, Albert-L{\'a}szl{\'o} and P{\'o}sfai, M{\'a}rton},
  title     = {Network Science},
  publisher = {Cambridge University Press},
  year      = {2016},
}

@book{newman2018networks,
  author    = {Newman, Mark},
  title     = {Networks},
  edition   = {2nd},
  publisher = {Oxford University Press},
  year      = {2018},
}

@inproceedings{papineni2002bleu,
  author    = {Papineni, Kishore and Roukos, Salim and Ward, Todd and Zhu, Wei-Jing},
  title     = {{BLEU}: A Method for Automatic Evaluation of Machine Translation},
  booktitle = {Proceedings of the 40th Annual Meeting of the Association for Computational Linguistics (ACL)},
  year      = {2002},
}

@inproceedings{lin2004rouge,
  author    = {Lin, Chin-Yew},
  title     = {{ROUGE}: A Package for Automatic Evaluation of Summaries},
  booktitle = {Text Summarization Branches Out: Proceedings of the ACL Workshop},
  year      = {2004},
}

@inproceedings{zhang2020bertscore,
  author    = {Zhang, Tianyi and Kishore, Varsha and Wu, Felix and Weinberger, Kilian Q. and Artzi, Yoav},
  title     = {{BERTScore}: Evaluating Text Generation with {BERT}},
  booktitle = {International Conference on Learning Representations (ICLR)},
  year      = {2020},
}

@article{gretton2012mmd,
  author  = {Gretton, Arthur and Borgwardt, Karsten M. and Rasch, Malte J. and Sch{\"o}lkopf, Bernhard and Smola, Alexander},
  title   = {A Kernel Two-Sample Test},
  journal = {Journal of Machine Learning Research (JMLR)},
  volume  = {13},
  pages   = {723--773},
  year    = {2012},
}

@inproceedings{devlin2019bert,
  author    = {Devlin, Jacob and Chang, Ming-Wei and Lee, Kenton and Toutanova, Kristina},
  title     = {{BERT}: Pre-training of Deep Bidirectional Transformers for Language Understanding},
  booktitle = {Proceedings of the 2019 Conference of the North American Chapter of the Association for Computational Linguistics (NAACL)},
  year      = {2019},
}

@article{sen2008cora,
  author  = {Sen, Prithviraj and Namata, Galileo and Bilgic, Mustafa and Getoor, Lise and Galligher, Brian and Eliassi-Rad, Tina},
  title   = {Collective Classification in Network Data},
  journal = {AI Magazine},
  volume  = {29},
  number  = {3},
  pages   = {93--106},
  year    = {2008},
  doi     = {10.1609/aimag.v29i3.2157},
}

@article{irwin2012zinc,
  author  = {Irwin, John J. and Sterling, Teague and Mysinger, Michael M. and Bolstad, Erin S. and Coleman, Ryan G.},
  title   = {{ZINC}: A Free Tool to Discover Chemistry for Biology},
  journal = {Journal of Chemical Information and Modeling},
  volume  = {52},
  number  = {7},
  pages   = {1757--1768},
  year    = {2012},
}

@article{ramakrishnan2014qm9,
  author  = {Ramakrishnan, Raghunathan and Dral, Pavlo O. and Rupp, Matthias and von Lilienfeld, O. Anatole},
  title   = {Quantum Chemistry Structures and Properties of 134 Kilo Molecules},
  journal = {Scientific Data},
  volume  = {1},
  pages   = {140022},
  year    = {2014},
}

@article{leskovec2007realworld,
  author  = {Leskovec, Jure and Kleinberg, Jon and Faloutsos, Christos},
  title   = {Graph Evolution: Densification and Shrinking Diameters},
  journal = {ACM Transactions on Knowledge Discovery from Data (TKDD)},
  volume  = {1},
  number  = {1},
  year    = {2007},
  doi     = {10.1145/1217299.1217301},
}

@article{huang2025hallucination,
  author    = {Huang, Lei and Yu, Weijiang and Ma, Weitao and Zhong, Weihong and Feng, Zhangyin and Wang, Haotian and others},
  title     = {A Survey on Hallucination in Large Language Models: Principles, Taxonomy, Challenges, and Open Questions},
  journal   = {ACM Transactions on Information Systems (TOIS)},
  volume    = {43},
  number    = {2},
  articleno = {42},
  year      = {2025},
  doi       = {10.1145/3703155},
}

@article{datasheets2021gebru,
  author  = {Gebru, Timnit and Morgenstern, Jamie and Vecchione, Briana and Vaughan, Jennifer Wortman and Wallach, Hanna and Iii, Hal Daum{\'e} and Crawford, Kate},
  title   = {Datasheets for Datasets},
  journal = {Communications of the {ACM}},
  volume  = {64},
  number  = {12},
  pages   = {86--92},
  year    = {2021},
  doi     = {10.1145/3458723},
}

@inproceedings{huang2025graphjudge,
  author    = {Huang, Haoyu and Chen, Chong and Sheng, Zeang and Li, Yang and Zhang, Wentao},
  title     = {Can {LLM}s be Good Graph Judge for Knowledge Graph Construction?},
  booktitle = {Proceedings of the 2025 Conference on Empirical Methods in Natural Language Processing (EMNLP)},
  year      = {2025},
}

%======================================================================
\newpage
\appendix

\section{Framework Diagram and Dataset Overview}
\label{app:framework}

\begin{figure}[h]
\centering
\includegraphics[width=0.9\linewidth]{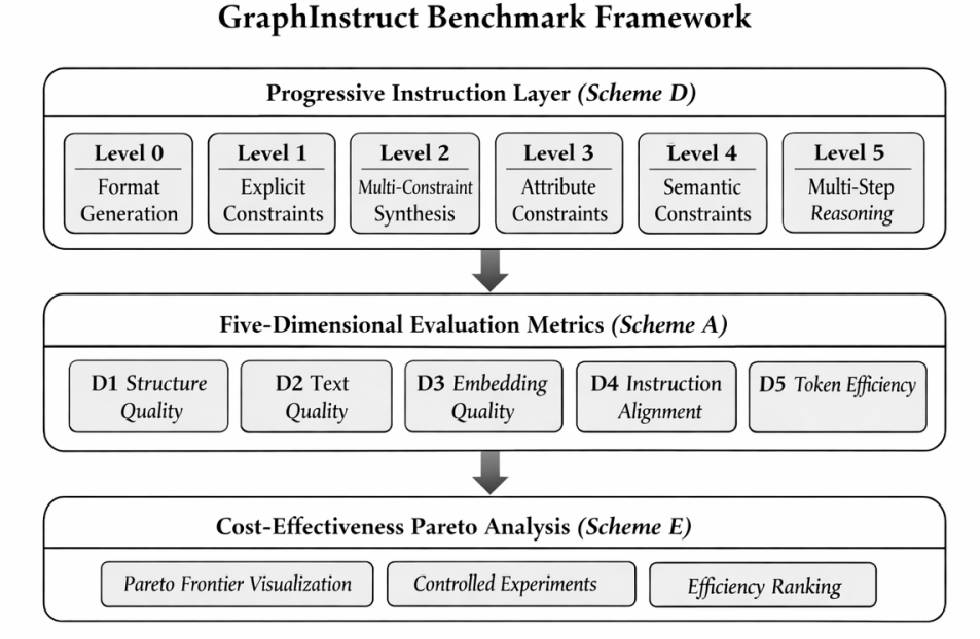}
\caption{The \bench\ benchmark framework. The Progressive Instruction Layer (L0--L5) feeds into the Five-Dimensional Evaluation Metrics pipeline (D1--D5); the Pareto analysis layer integrates Quality (D1--D4) and Efficiency (D5) into deployment-oriented rankings. LLM-as-Judge and Multi-Agent Collaboration are optional future modules.}
\label{fig:framework}
\end{figure}

\begin{figure}[h]
\centering
\includegraphics[width=0.9\linewidth]{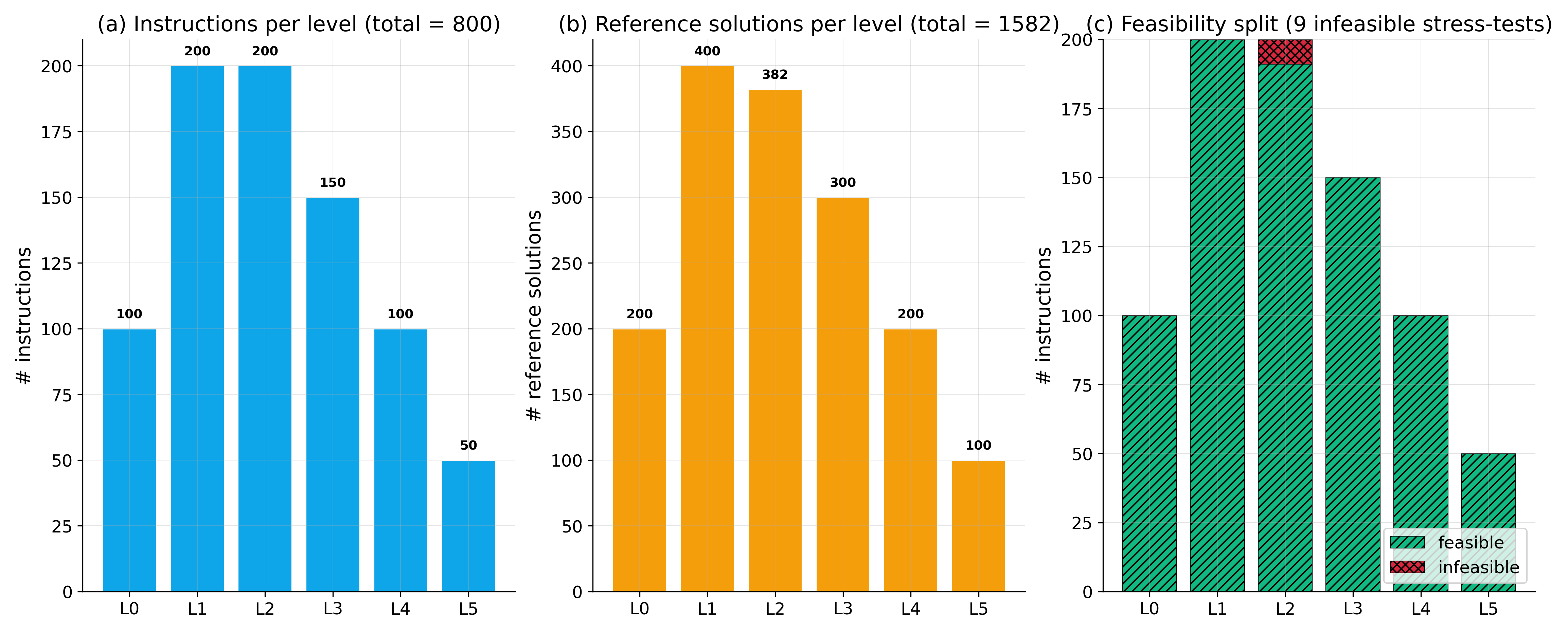}
\caption{\bench\ dataset overview. Left: per-level instruction count. Center: graph-size distribution. Right: constraint-count distribution, showing progressive increase from L0--L1 to L2 (4+ constraints) and compound semantic/editing constraints at L4--L5.}
\label{fig:dataset}
\end{figure}

%======================================================================
\section{Related Work: Extended Discussion}
\label{app:related-extended}

This appendix supplements \S\ref{sec:related} with prior-by-prior commentary, the axis-by-axis comparison with \citet{demirci2025graphsavvy}, and the classical-generative-model / prompting / structured-decoding / broader-LLM-evaluation discussions migrated here for length. The full positioning table is Tab.~\ref{tab:benchmark-compare} in \S\ref{sec:related}.

\paragraph{Closest concurrent: GDGB and Demirci et al.} \textsc{GDGB}~\citep{peng2026gdgb} (ICLR'26) targets dynamic text-attributed graph generation with three task categories (continuation, retrieval, prediction) and four LLMs; it is complementary to our static, constraint-driven setting. \citet{demirci2025graphsavvy} (ACL'25) cover a complementary domain axis (5 application domains $\times$ 15 LLMs); they observe iterative-feedback gains varying $48\%/<\!5\%$ across models without identifying an explanatory variable, and qualitatively note that prompting strategies differ in effectiveness across models. Our \S\ref{sec:cap-strat} and \S\ref{sec:strat-fam} quantify two of their qualitative observations: prompt sensitivity tracks capability with a dominant inverse trend ($\beta{=}-0.073$, $R^2{=}0.40$, 95\% bootstrap CI for $\beta$ $[-0.135, +0.002]$; non-monotonic at the top tier) and CoT polarity is family-specific. Axis-by-axis: their benchmark stratifies by application domain, ours by structural complexity; their generation method is prompt-only, ours adds a verification-guided pipeline; they evaluate 15 LLMs on 5 domains, we evaluate 12 LLMs on 6 levels with same-family scale series enabling RQ4/RQ5. The two benchmarks are complementary, jointly covering (complexity $\times$ domain) of LLM graph generation.

\paragraph{Earlier graph-generation benchmarks.} \textsc{LLM4GraphGen}~\citep{yao2024llm4graphgen} stratifies along three property axes (graph type, scale, hardness) but evaluates only 3 LLMs and reports two metrics (Validity, Novelty); no algorithmically synthesized references. \textsc{InstructGraph}~\citep{wang2024instructgraph} (ACL'24) introduces the code-style format we adopt and a hallucination taxonomy, but as a training/instruction-tuning artifact rather than an inference-time diagnostic. \textsc{GraphArena}~\citep{tang2025grapharena} (ICLR'25) evaluates 10 LLMs on 10 classical computation problems with four prompting strategies, surfacing a coverage notion of capability that aggregates over algorithmic problem types but does not stratify generation complexity. Earlier surveys~\citep{bonifati2020graphgen,xiang2022vldbjgraphgen} predate the LLM era and frame graph generation as a classical statistical-modeling task; our work treats LLMs as on-demand graph synthesizers and asks where their failures cluster.

\paragraph{Reasoning-about-graphs benchmarks.} \textsc{NLGraph}~\citep{wang2023nlgraph} (NeurIPS'23) evaluates LLM reasoning on 8 classical graph problems (BFS/DFS/MST/...); \textsc{Talk-Like-a-Graph}~\citep{fatemi2024talklikeagraph} (ICLR'24) evaluates encoding $\times$ task combinations for PaLM-only; \textsc{GraphWiz}~\citep{chen2024graphwiz} (KDD'24) trains a problem-solver model on 9 classical graph problems. All three measure reasoning \emph{about} pre-specified graphs rather than \emph{generation of new graphs}; our axis is distinct.

\paragraph{Prompting and structured decoding.} Chain-of-thought prompting~\citep{wei2022cot,kojima2022zerocot} and self-consistency~\citep{wang2023selfconsistency} are text-domain primitives; we test their transfer to structured graph generation in \S\ref{sec:strat-fam} (CoT family polarity) and \S\ref{sec:methods} (self-consistency null result). Self-refine~\citep{madaan2023selfrefine,shinn2023reflexion,gou2024critic} and tree-of-thought~\citep{yao2023tot,zhou2023ltm} establish a $T{\sim}10$--$20$ refinement budget; we re-calibrate this to $T{\sim}5$ on verifiable graph constraints. LLM self-critique limitations~\citep{huang2024selfcorrect,huang2025hallucination} motivate our use of programmatic verification rather than LLM-as-judge.

\paragraph{Broader LLM evaluation.} Scaling-law studies~\citep{kaplan2020scaling,hoffmann2022chinchilla} and emergent-behavior critiques~\citep{schaeffer2023emergentmirage} contextualize our \S\ref{sec:strat-fam} scale-decoupling finding. The MMD evaluation primitives we adopt for D1 and D3 trace to~\citet{gretton2012mmd,you2018graphrnn}; G-BERTScore/G-BLEU/G-ROUGE adapt~\citet{zhang2020bertscore,papineni2002bleu,lin2004rouge,devlin2019bert} to graph-as-string output, in the spirit of LLM-based graph-judge frameworks~\citep{huang2025graphjudge}. Grassmann coherence is implemented per its standard subspace-distance formulation.

%======================================================================
\section{Worked Example per Level}
\label{app:examples}

We give one representative instruction per level (L0--L5), in the order they appear in \texttt{data/instructions/level\_X.json}. Each entry shows the natural-language instruction, the explicit and implicit constraint specifications, the size bucket, the feasibility flag, and one of the two algorithmically synthesized reference solutions in our InstructGraph code-style format. L4 entries additionally carry a \texttt{domain} field; L5 entries additionally carry a \texttt{base\_graph} field denoting the graph the LLM is asked to edit.

\paragraph{L0 -- Format generation (\texttt{L0-001}).}
\begin{itemize}\setlength\itemsep{0pt}
\item \textbf{instruction}: ``Create a graph with 3 nodes.''
\item \textbf{explicit\_constraints}: \texttt{["num\_nodes=3"]}
\item \textbf{implicit\_constraints}: \texttt{["directed=false"]}
\item \textbf{graph\_sizes}: \texttt{["small"]}; \textbf{feasible}: \texttt{true}
\end{itemize}
Reference solution (1 of 2):
\begin{verbatim}
Graph[name='L0-001-ref1', nodes=3] {
    node_list = ['0', '1', '2'];
    edge_list = [('0','1'), ('0','2'), ('1','2')];
}
\end{verbatim}

\paragraph{L1 -- Single explicit constraint (\texttt{L1-001}).}
\begin{itemize}\setlength\itemsep{0pt}
\item \textbf{instruction}: ``Generate a tree with 5 nodes.''
\item \textbf{explicit\_constraints}: \texttt{["graph\_type=tree", "num\_nodes=5"]}
\item \textbf{implicit\_constraints}: \texttt{["num\_edges=4", "acyclic=true", "connected=true"]}
\item \textbf{graph\_sizes}: \texttt{["small"]}; \textbf{feasible}: \texttt{true}
\end{itemize}
Reference solution (1 of 2):
\begin{verbatim}
Graph[name='L1-001-ref1', nodes=5] {
    node_list = ['0', '1', '2', '3', '4'];
    edge_list = [('0','3'), ('0','2'), ('1','2'), ('1','4')];
}
\end{verbatim}

\paragraph{L2 -- Multi-constraint composition (\texttt{L2-001}).}
\begin{itemize}\setlength\itemsep{0pt}
\item \textbf{instruction}: ``Generate a connected 3-regular graph with 8 nodes.''
\item \textbf{explicit\_constraints}: \texttt{["degree=3", "num\_nodes=8", "connected=true", "directed=false"]}
\item \textbf{implicit\_constraints}: \texttt{["num\_edges=12"]} (from $|E|{=}\sum d_v / 2$)
\item \textbf{graph\_sizes}: \texttt{["small"]}; \textbf{feasible}: \texttt{true}
\end{itemize}
Reference solution (1 of 2):
\begin{verbatim}
Graph[name='L2-001-ref1', nodes=8] {
    node_list = ['0', '1', '2', '3', '4', '5', '6', '7'];
    edge_list = [
        ('0','1'), ('0','3'), ('0','6'), ('1','5'), ('1','3'),
        ('2','7'), ('2','3'), ('2','6'), ('4','6'), ('4','5'),
        ('4','7'), ('5','7')];
}
\end{verbatim}

\paragraph{L3 -- Numerical attribute control (\texttt{L3-001}).}
\begin{itemize}\setlength\itemsep{0pt}
\item \textbf{instruction}: ``Generate a community-structured graph with 12 nodes divided into 2 communities using a stochastic block model. The graph should be connected, be undirected, have density at most 0.451, have modularity at least 0.25, have clustering coefficient at least 0.519, and have average path length at most 2.0213.''
\item \textbf{explicit\_constraints}: \texttt{["num\_nodes=12", "connected=true", "directed=false", "density<=0.451", "modularity>=0.25", "clustering\_coefficient>=0.519", "average\_path\_length<=2.0213"]}
\item \textbf{implicit\_constraints}: \texttt{[]} (numerical attributes are themselves the contract)
\item \textbf{graph\_sizes}: \texttt{["small"]}; \textbf{feasible}: \texttt{true}
\end{itemize}
Reference solution (1 of 2; note per-node \texttt{block} attributes and graph-level metadata):
\begin{verbatim}
Graph[name='L3-001-ref1', nodes=12] {
    node_list = ['0','1','10','11','2','3','4','5','6','7','8','9'];
    edge_list = [
        ('0','1'), ('0','2'), ('0','3'), ('0','5'), ('1','3'),
        ('1','4'), ('1','5'), ('1','8'), ('2','3'), ('2','5'),
        ('2','7'), ('3','5'), ('3','6'), ('4','6'), ('4','8'),
        ('5','6'), ('6','7'), ('6','8'), ('6','10'), ('6','11'),
        ('7','8'), ('7','9'), ('7','10'), ('7','11'), ('8','11'),
        ('9','11'), ('10','11')];

    0.block=0; 1.block=0; 2.block=0; 3.block=0; 4.block=0; 5.block=0;
    6.block=1; 7.block=1; 8.block=1; 9.block=1; 10.block=1; 11.block=1;

    k = 2; model = 'sbm'; subgroup = 'SBM-S1';
    partition = [{0, 1, 2, 3, 4, 5}, {6, 7, 8, 9, 10, 11}];
}
\end{verbatim}

\paragraph{L4 -- Domain semantics (\texttt{L4-001}).}
\begin{itemize}\setlength\itemsep{0pt}
\item \textbf{instruction}: ``Generate a social network with 10 users in an online community. Members form connections based on shared interests and interactions. The network should have density at most 0.7595, have minimum degree 1, be connected, and be undirected.''
\item \textbf{explicit\_constraints}: \texttt{["num\_nodes=10", "density<=0.7595", "min\_degree=1", "connected=true", "directed=false"]}
\item \textbf{implicit\_constraints}: \texttt{[]}; \textbf{domain}: \texttt{"social"}
\item \textbf{graph\_sizes}: \texttt{["small"]}; \textbf{feasible}: \texttt{true}
\end{itemize}
Reference solution (1 of 2; note per-node string \texttt{label} attributes and graph-level \texttt{domain} / \texttt{edge\_type}):
\begin{verbatim}
Graph[name='L4-001-ref1', nodes=10, domain='social'] {
    node_list = ['0', '1', '2', '3', '4', '5', '6', '7', '8', '9'];
    edge_list = [
        ('0','3'), ('0','5'), ('0','6'), ('0','7'), ('0','8'),
        ('1','3'), ('1','4'), ('1','5'), ('1','6'), ('1','7'), ('1','9'),
        ('2','3'), ('2','4'), ('2','5'), ('2','7'), ('3','4'),
        ('3','5'), ('3','6'), ('3','7'), ('3','8'), ('3','9'),
        ('4','5'), ('4','6'), ('4','7'), ('5','6'), ('5','7'),
        ('5','8'), ('5','9'), ('6','7'), ('6','8'), ('7','8')];

    0.label='Alice'; 1.label='Cora'; 2.label='Hugo'; 3.label='Grace';
    4.label='Yuki'; 5.label='Wendy'; 6.label='Mona'; 7.label='Amber';
    8.label='Hope';  9.label='Jill';

    domain = 'social'; edge_type = 'colleagues';
}
\end{verbatim}

\paragraph{L5 -- Multi-step graph editing (\texttt{L5-024}).}
\begin{itemize}\setlength\itemsep{0pt}
\item \textbf{instruction}: ``Given a 4-cycle (0-1-2-3) with one diagonal edge (0,2). Nodes 0 and 2 have degree 3, while nodes 1 and 3 have degree 2. Add the minimum number of edges to make the graph 3-regular (all vertices degree 3).''
\item \textbf{explicit\_constraints}: \texttt{["num\_nodes=4", "connected=true", "degree=3", "directed=false", "task\_type=make\_regular"]}
\item \textbf{implicit\_constraints}: \texttt{["num\_edges=6", "directed=false"]}
\item \textbf{graph\_sizes}: \texttt{["small"]}; \textbf{feasible}: \texttt{true}
\end{itemize}
Base graph (the LLM is asked to transform this):
\begin{verbatim}
Graph[name='L5-024-base', nodes=4] {
    node_list = ['0', '1', '2', '3'];
    edge_list = [('0','1'), ('0','3'), ('0','2'), ('1','2'), ('2','3')];
}
\end{verbatim}
Reference solution (1 of 2; the unique additional edge is $(1,3)$):
\begin{verbatim}
Graph[name='L5-024-ref1', nodes=4] {
    node_list = ['0', '1', '2', '3'];
    edge_list = [
        ('0','1'), ('0','2'), ('0','3'), ('1','2'),
        ('1','3'), ('2','3')];
}
\end{verbatim}

\paragraph{L4 reference-pool provenance and licensing.}
The L4 reference pool comprises 1{,}048 graphs drawn from nine public corpora. Each subset retains its upstream license; we redistribute under terms compatible with research use. The full per-source attribution is in \texttt{DATA\_LICENSE.md} accompanying the release; a summary appears in Tab.~\ref{tab:l4-provenance}.

\begin{table}[h]
\centering
\caption{L4 real-world reference-pool provenance summary (full per-source citations in \texttt{DATA\_LICENSE.md}).}
\label{tab:l4-provenance}
\small
\begin{tabular}{lll}
\toprule
\textbf{L4 sub-pool} & \textbf{Upstream source(s)} & \textbf{License / terms} \\
\midrule
\texttt{citation.pkl} & DBLP, Cora & SNAP terms; open-access \\
\texttt{social.pkl} & Reddit hyperlinks, Karate Club, Facebook ego & MIT / public-domain / SNAP \\
\texttt{biological.pkl} & ZINC, QM9, MUTAG (TUDataset) & ZINC research / CC0 / CC-BY \\
\texttt{infrastructure.pkl} & SNAP \texttt{roadNet-PA} (subset) & SNAP terms \\
\texttt{communication.pkl} & SNAP \texttt{email-Eu-core} & SNAP terms \\
\texttt{ecological.pkl} & Public food-web databases & Public-domain compilations \\
\texttt{general.pkl} & Wikidata subgraph snapshots & CC0 \\
\texttt{ba-\{s,m,l\}.pkl} & Synthetic BA random graphs (ours) & CC-BY-4.0 \\
\bottomrule
\end{tabular}
\end{table}

All L4 graphs are size-normalized via BFS / random-walk subgraph sampling, deduplicated by Weisfeiler-Lehman hash, and stripped of node / edge attributes other than what D2 (token-level reference similarity) and D3 (embedding-MMD / node-classification gap) require.

%======================================================================
\section{Scoring Formulas and Weight Ablation}
\label{app:weights}

This appendix gives the explicit per-level dimension weights used by all experiments, recaps the score formulas from \S\ref{sec:metrics}, and reports a perturbation analysis of the weights and the D5 hyperparameters.

\paragraph{Per-level dimension weights.} The level score $S_\ell = \sum_d w_{\ell,d}\, S_{d,\ell}$ uses the per-level dimension weights in Tab.~\ref{tab:dim-weights}. D2 (token-level reference similarity) is active only at L4, where reference graphs carry domain-specific surface text; D3 (embedding similarity) is active only at L3--L5, where reference \emph{distributions} are well-defined. D4 (instruction match) is the largest weight at every level, reflecting the design intent that joint constraint satisfaction is the primary diagnostic signal.

\begin{table}[h]
\centering
\caption{Per-level dimension weights (default scoring; rows sum to 1 within each level). ``D2/D3 active'' lists the discriminative metrics at each level; for the others the corresponding weight is 0.}
\label{tab:dim-weights}
\small
\begin{tabular}{ccccccc}
\toprule
\textbf{Level} & \textbf{D1 (struct)} & \textbf{D2 (text)} & \textbf{D3 (embed)} & \textbf{D4 (instr)} & \textbf{D5 (eff)} & \textbf{D2/D3 active} \\
\midrule
L0 & 0.10 & 0.00 & 0.00 & 0.60 & 0.30 & none \\
L1 & 0.15 & 0.00 & 0.00 & 0.70 & 0.15 & none \\
L2 & 0.15 & 0.00 & 0.00 & 0.70 & 0.15 & none \\
L3 & 0.15 & 0.00 & 0.15 & 0.50 & 0.20 & D3 \\
L4 & 0.10 & 0.15 & 0.05 & 0.55 & 0.15 & D2, D3 \\
L5 & 0.15 & 0.00 & 0.15 & 0.50 & 0.20 & D3 \\
\bottomrule
\end{tabular}
\end{table}

\paragraph{Score formulas (recap).} Total Quality (Eq.~\ref{eq:quality}), D5 efficiency (Eq.~\ref{eq:d5}), and Pareto-adjusted final score (Eq.~\ref{eq:sfinal}) are defined in \S\ref{sec:metrics}. The combined score reported throughout the paper is the level-weighted Quality, $Q \equiv \Stot = \sum_{\ell=0}^{5} w_\ell\, S_\ell$ with $w = (0.05, 0.10, 0.15, 0.20, 0.25, 0.25)$. Two evaluation modes are supported by \texttt{graphinstruct.scoring}:
\begin{itemize}\setlength\itemsep{0pt}
\item \textbf{Default} keeps D5 in the per-level weights as shown in Tab.~\ref{tab:dim-weights} and feeds the cost-aware $\Sfin$ (Eq.~\ref{eq:sfinal}; $\lambda{=}0.15$ unless noted).
\item \textbf{Quality-only} (the mode used for every $\Stot$ leaderboard in \S\ref{sec:eval} and App.~\ref{app:eval-extended}--\ref{app:leaderboards}) zeroes the D5 weight and renormalises D1--D4 to sum to 1, isolating quality from cost. The published $\Stot$ values are produced in this mode.
\end{itemize}

\paragraph{Weight perturbation analysis.} We probe how sensitive the 45-cell quality-only $\Stot$ ranking is to alternative weight schemes by recomputing $\Stot$ on the same per-instruction dimension scores under several perturbations (\texttt{scripts/weight\_ablation.py}). Tab.~\ref{tab:weight-ablation} reports the overlap with the default top-9, top-15, and top-20 sets, the top-15 Jaccard, and the largest position shift in the default top-15. Small perturbations (D4 weight $+0.05$ at every level) leave the ranking essentially unchanged. Aggressive perturbations restructure the ranking, with D4 emerging as the dominant ranking signal: a D4-only scheme retains 8 of 9 default top-9 entries, while a D1-only scheme retains only 3 of 9. Uniform weighting (D1$=$D2$=$D3$=$D4$=$0.25 at every level) sits between these extremes and shifts as many as 12 positions inside the top-15, primarily because L0--L2 lack meaningful D2/D3 signal so the uniform scheme injects noise from inactive dimensions at low levels.

\begin{table}[h]
\centering
\caption{Weight perturbation analysis on the 45-cell quality-only ranking. ``Default'' is the per-level scheme of Tab.~\ref{tab:dim-weights} with D5 zeroed. All schemes are quality-only (D5$=$0, D1--D4 renormalised to 1).}
\label{tab:weight-ablation}
\small
\begin{tabular}{lcccc}
\toprule
\textbf{Scheme} & \textbf{Top-9 retained} & \textbf{Top-15 retained} & \textbf{Top-15 Jaccard} & \textbf{Max shift, top-15} \\
\midrule
Default                          & 9 / 9 & 15 / 15 & 1.000 & 0  \\
D4 weight $+0.05$ at every level & 9 / 9 & 15 / 15 & 1.000 & 1  \\
Uniform D1$=$D2$=$D3$=$D4$=$0.25 & 5 / 9 & 11 / 15 & 0.579 & 12 \\
D4-only                          & 8 / 9 & 12 / 15 & 0.667 & 8  \\
D1-only                          & 3 / 9 &  9 / 15 & 0.429 & 20 \\
\bottomrule
\end{tabular}
\end{table}

The pattern matches the design rationale: D4 (instruction match) carries the most diagnostic information because parser-passing graphs already get near-1 D1, and D5 captures cost rather than capability. Quality-only zeroing of D5 separates capability from cost; the cost-aware view is reported separately as $\Sfin$ in App.~\ref{app:rq6}.

\paragraph{D5 exponential-scale robustness.} The default D5 form $0.7\,e^{-\mathrm{TPV}/1000} + 0.3\,e^{-(\mathrm{API}-1)/2}$ uses two free hyperparameters: the TPV scale ($s_T{=}1000$ tokens) and the API-call scale ($s_A{=}2$ extra calls). To verify rank stability under reasonable alternative choices we recompute total Quality $Q$ for all 45 (model, strategy) baseline configurations under a $3\times 3$ grid $s_T\in\{500,1000,2000\}$, $s_A\in\{1,2,4\}$ (each scale halved, doubled, and at default), holding all other weights fixed (\texttt{scripts/d5\_robustness.py}). Spearman $\rho$ and Kendall $\tau$ against the default ranking are uniformly high (Tab.~\ref{tab:d5-robustness}): $\rho\in[0.966, 1.000]$ across all 9 cells, with the top-5 set retaining at least 4 of 5 members in every cell. The most aggressive perturbation ($s_T{=}500$, $s_A{=}4$, halving the TPV scale and doubling the API scale) yields the lowest $\rho{=}0.966$; the default and adjacent scales yield $\rho{>}0.999$. The exponential form is robust to scale specification within a $2\times$ envelope; the choice $(1000, 2)$ encodes typical streaming-API per-token-dominated pricing rather than a load-bearing modeling choice.

\begin{table}[h]
\centering
\caption{D5 exponential-scale robustness: Spearman $\rho$ and Kendall $\tau$ of the 45-cell Quality ranking against the default $(s_T,s_A){=}(1000,2)$. Top-5 Jaccard is the overlap of the top-5 set with the default top-5.}
\label{tab:d5-robustness}
\small
\begin{tabular}{cc cccc}
\toprule
$s_T$ & $s_A$ & Spearman $\rho$ & Kendall $\tau$ & Top-1 stable & Top-5 Jaccard \\
\midrule
500  & 1 & 0.971 & 0.875 & no  & 0.667 \\
500  & 2 & 0.970 & 0.877 & no  & 0.667 \\
500  & 4 & 0.966 & 0.869 & no  & 0.667 \\
1000 & 1 & 1.000 & 0.996 & yes & 1.000 \\
\textbf{1000} & \textbf{2} & \textbf{1.000} & \textbf{1.000} & \textbf{---}   & \textbf{1.000} \\
1000 & 4 & 1.000 & 0.994 & yes & 1.000 \\
2000 & 1 & 0.988 & 0.925 & no  & 1.000 \\
2000 & 2 & 0.991 & 0.937 & no  & 1.000 \\
2000 & 4 & 0.990 & 0.933 & no  & 1.000 \\
\bottomrule
\end{tabular}
\end{table}

%======================================================================
\section{VGIG Pseudocode, CAAP Decision Tree, and Feedback Templates}
\label{app:vgig}\label{app:caap}

This appendix gives the implementation-level details of the two improvement methods (VGIG and CAAP) evaluated in App.~\ref{app:rq7-10}. The reference implementation lives at \texttt{graphinstruct/improvements/} in the released code (\texttt{runners.py}, \texttt{caap.py}, \texttt{feedback.py}, \texttt{domain\_priors.py}).

\paragraph{Algorithm 1: VGIG iterative refinement.} For each instruction, $K$ independent refinement chains are run (each chain at a slightly jittered temperature for diversity). Each chain does at most $T{+}1$ generation calls: 1 round-0 baseline plus up to $T$ feedback-driven refinement steps. A chain terminates early when the satisfaction rate reaches 1, or when two consecutive parse failures occur.

\begin{table}[h]
\centering
\caption{VGIG (\texttt{runners.VGIGRunner}). Inputs: instruction $I$, model $M$, max rounds $T$, num chains $K$, feedback level $\ell\in\{\text{none}, \text{coarse}, \text{fine}\}$, base temperature $\tau$. Output: $K$ refined samples. In all reported method experiments (VGIG, CAAP, Combined, retry, SC) we use $K{=}3$ chains; this contrasts with the $k{=}5$ independent generations of the baseline capability survey (\S\ref{sec:models}). The chain-count is fixed across method conditions for apples-to-apples cost comparison; sample-size impact on confidence intervals is addressed in App.~\ref{app:stat-robust}.}
\label{alg:vgig}
\small
\begin{tabular}{@{}rl@{}}
\toprule
1  & \textbf{for} chain $k = 1, \dots, K$ \textbf{do} \\
2  & \quad $\tau_k \gets \mathrm{clip}(\tau + \delta_k, [0.1, 1.5])$ \quad \textit{// jittered round-0 temperature} \\
3  & \quad $\hat{\tau}_k \gets \max(0.2, 0.6\,\tau_k)$ \quad \textit{// lower temperature for refinement} \\
4  & \quad $x \gets M(\text{prompt}_0(I);\ \tau_k)$ \\
5  & \quad $s \gets \text{satisfaction\_rate}(x, I)$ \\
6  & \quad \textbf{if} $I.\text{feasible}$ is false \textbf{then} emit $x$; \textbf{continue} \\
7  & \quad $f \gets 0$ \quad \textit{// consecutive parse-failure counter} \\
8  & \quad \textbf{for} $t = 1, \dots, T$ \textbf{do} \\
9  & \qquad \textbf{if} $s \geq 1.0$ \textbf{then break} \\
10 & \qquad \textbf{if} $\text{parse}(x)$ failed \textbf{then} \\
11 & \qquad \quad $f \gets f + 1$; \textbf{if} $f \geq 2$ \textbf{then break} \\
12 & \qquad \quad $x' \gets M(\text{prompt}_0(I);\ \tau_k)$; update $x, s$ if better; \textbf{continue} \\
13 & \qquad $f \gets 0$ \\
14 & \qquad $\text{fb} \gets \text{generate\_feedback}(x, I, \ell)$ \\
15 & \qquad \textbf{if} $\text{fb}$ is empty \textbf{then break} \\
16 & \qquad $x' \gets M(\text{refine\_prompt}(I, x, \text{fb}, t);\ \hat{\tau}_k)$ \\
17 & \qquad $s' \gets \text{satisfaction\_rate}(x', I)$ \\
18 & \qquad \textbf{if} $s' > s$ \textbf{then} $x \gets x'$;\ $s \gets s'$ \\
19 & \quad \textbf{end for} \\
20 & \quad emit $x$ as chain-$k$ output \\
21 & \textbf{end for} \\
\bottomrule
\end{tabular}
\end{table}

The chain temperatures use offsets $\delta_k \in \{-0.10, 0.00, +0.10, -0.05, +0.05\}$ cycling over chains (\texttt{runners.py:\_CHAIN\_TEMP\_OFFSETS}), giving deterministic per-chain temperature schedules at any base $\tau$. \texttt{generate\_feedback} (\texttt{feedback.py}) materialises per-violation feedback at the requested granularity (Tab.~\ref{tab:fb-templates}). \texttt{refine\_prompt} embeds the previous serialised graph, the feedback list, and the round counter $t$ into a structured refinement instruction.

\paragraph{Algorithm 2: CAAP per-instruction strategy selection.} CAAP selects a (strategy, prompt-augmentation) pair conditional on the instruction's level and the target model's tier. The decision rules are derived from the per-level signed strategy effects in Tab.~\ref{tab:strategy-delta} and the per-family CoT polarity in Tab.~\ref{tab:cot-family}.

\begin{table}[h]
\centering
\caption{CAAP (\texttt{caap.select\_strategy}). Inputs: instruction $I$ (with level, explicit constraints, optional domain), model name $m$. Output: \texttt{CAAPDecision} = (strategy, extras dict, rationale).}
\label{alg:caap}
\small
\begin{tabular}{@{}rl@{}}
\toprule
1 & $\text{tier} \gets \text{model\_tier}(m)$ \quad \textit{// $\in\{$T1, T2-GPT, T2-open, T3$\}$} \\
2 & \textbf{return} dispatch by $I.\text{level}$ to the per-level decider in Tab.~\ref{tab:caap-rules} \\
3 & \quad \textit{// each decider may attach extras: \texttt{checklist} (L2),} \\
4 & \quad \textit{//\hphantom{ each decider may} \texttt{formula} (L3), \texttt{domain} (L4) -- see Tab.~\ref{tab:domain-priors}} \\
\bottomrule
\end{tabular}
\end{table}

The decision table is per-(level, tier) at top level (24 cells), with three additional level-internal refinements:
\begin{itemize}\setlength\itemsep{0pt}
\item \textbf{L1 simple-vs-complex type.} If \texttt{graph\_type$\in\{$tree, cycle, star, path, complete$\}$}, L1 routes to zero-shot regardless of tier ($>$90\% baseline accuracy on simple types).
\item \textbf{L2 / L3 / L5 prompt extras.} The L2 T3 decider attaches a per-constraint checklist; the L3 T3 decider attaches the relevant numerical-property formulas (only those mentioned in the explicit constraints).
\item \textbf{L4 domain prior.} The L4 decider attaches the domain prior (Tab.~\ref{tab:domain-priors}) corresponding to \texttt{instruction.domain}, parameterised by \texttt{num\_nodes}.
\end{itemize}

\begin{table}[h]
\centering
\caption{CAAP decision rules (per-level $\times$ tier). Strategies: ZS = zero-shot, FS = few-shot, ZC = zero-CoT, FC = few-CoT. Extras attach to the prompt: \texttt{checklist}, \texttt{formula}, \texttt{domain} text. ``T2-GPT'' = GPT-4o / GPT-4.1 (CoT-negative family); ``T2-open'' = DeepSeek-V3, Llama-70B, Qwen3.5-35B (CoT-positive open-weight). The data-driven motivation for each rule is in the comment column (signed deltas from Tab.~\ref{tab:strategy-delta}--\ref{tab:cot-family}).}
\label{tab:caap-rules}
\small
\begin{tabularx}{\linewidth}{@{}llXX@{}}
\toprule
\textbf{Level} & \textbf{Tier} & \textbf{(strategy, extras)} & \textbf{Rationale} \\
\midrule
L0 & all & ZS                    & all tier deltas $<0.025$ \\
\midrule
L1 \emph{simple} & all      & ZS               & simple types: $>$90\% baseline \\
L1 \emph{complex} & T3      & ZC               & ZC safest incremental at low capability \\
L1 \emph{complex} & T2-GPT  & FS               & FS works; GPT family hates FC \\
L1 \emph{complex} & T2-open / T1 & FC          & FC positive (Qwen3.5 $+0.052$) \\
\midrule
L2 & T3      & ZC + \texttt{checklist}        & FS toxic at L2 ($-0.034$ avg; GPT-4o-mini drops to 0.424) \\
L2 & T2-GPT  & FS                              & T2-GPT tolerates FS in v4 \\
L2 & T2-open / T1 & FC                         & FC works for non-GPT families \\
\midrule
L3 & T3      & ZC + \texttt{formula}           & FC toxic at L3 ($-0.048$ avg) \\
L3 & T2-GPT  & ZS                              & ZS neutral; FC clearly bad \\
L3 & T2-open / T1 & ZC                         & small ZC positive ($+0.004$) \\
\midrule
L4 & T3 / T2-GPT & FS + \texttt{domain}        & FS the only $+0.069$ strategy at L4 \\
L4 & T2-open / T1 & FC + \texttt{domain}       & FC + domain prior best for non-GPT \\
\midrule
L5 & T3      & ZC                              & FC toxic for T3 at L5 ($-0.041$) \\
L5 & T2-GPT / T2-open / T1 & FC                & FC has highest signed delta ($+0.045$) \\
\bottomrule
\end{tabularx}
\end{table}

\paragraph{L4 domain priors.} The CAAP L4 decider injects a structural prior describing the expected statistics of the target domain (\texttt{domain\_priors.DOMAIN\_PRIORS}). Eight priors cover the L4 reference-pool domains. Each prior reports an average-degree band, a clustering-coefficient band, a density band, the qualitative degree distribution, and 1--2 motif hints. The rendered prompt augmentation has the form ``Domain context: \{name\} network ($n$ nodes). Expected structural properties: ...''.

\begin{table}[h]
\centering
\caption{L4 domain priors injected by CAAP. Bands give expected $[\min, \max]$; degree distribution is qualitative; ``hint'' is a single-line summary of the structural prior.}
\label{tab:domain-priors}
\small
\begin{tabularx}{\linewidth}{@{}lllllX@{}}
\toprule
\textbf{Domain} & \textbf{Avg deg.} & \textbf{Clustering} & \textbf{Density} & \textbf{Deg.\ dist.} & \textbf{Hint} \\
\midrule
social          & 4.0--6.0 & 0.40--0.70 & 0.30--0.50 & power-law       & friend circles + hubs \\
citation        & 2.5--4.0 & 0.05--0.25 & 0.05--0.15 & power-law       & DAG; seminal-paper hubs \\
biological      & 2.0--5.0 & 0.10--0.40 & 0.10--0.30 & power-law       & sparse with functional modules \\
ecological      & 2.0--4.0 & 0.15--0.35 & 0.10--0.25 & roughly-uniform & shallow producer-consumer hierarchy \\
communication   & 3.0--8.0 & 0.20--0.50 & 0.30--0.50 & power-law       & active hubs, bursty interactions \\
infrastructure  & 2.0--4.0 & 0.05--0.20 & 0.05--0.20 & roughly-uniform & grid-like; high-betweenness backbone \\
knowledge\_graph & 2.0--6.0 & 0.05--0.30 & 0.05--0.20 & power-law       & typed relations; broad-concept hubs \\
molecular       & 1.5--3.5 & 0.00--0.15 & 0.05--0.25 & roughly-uniform & valence-limited (C$\le$4, N$\le$3, O$\le$2) \\
\bottomrule
\end{tabularx}
\end{table}

\paragraph{Feedback templates by level.} \texttt{generate\_feedback} emits a per-violation list whose richness is controlled by the level argument $\ell$. Tab.~\ref{tab:fb-templates} summarises the coarse-vs-fine difference at each instruction level. The verify-only mode (\(\ell{=}\)none) emits an empty feedback string when the satisfaction check passes and a single binary ``some constraint failed'' marker otherwise; this is the configuration that captures 75\% of VGIG's gain (App.~\ref{app:rq7-10}, RQ9).

\begin{table}[h]
\centering
\caption{VGIG feedback templates by level. \emph{Coarse} reports violation category; \emph{Fine} reports per-constraint (expected, observed, $\Delta$). Template richness is level-specific because constraint types differ structurally.}
\label{tab:fb-templates}
\small
\begin{tabularx}{\linewidth}{@{}lXX@{}}
\toprule
\textbf{Level} & \textbf{Coarse} & \textbf{Fine (additions)} \\
\midrule
L0, L1 & \{violation\_ids\} & $+$ \{expected, observed\} format/value details \\
L2     & \{constraint-category conflicts\} & $+$ $\Delta$ per constraint and joint-incompatibility flags \\
L3     & \{attribute deficits\} & $+$ numeric deltas, tolerance band \\
L4     & \{missing motifs\} & $+$ domain-baseline comparison \\
L5     & \{failed ops\} & $+$ step-wise partial credit, intermediate state diff \\
\bottomrule
\end{tabularx}
\end{table}

%======================================================================
\section{Evaluation: Per-RQ Extended Tables, Figures, and Mechanism Details}
\label{app:eval-extended}

This appendix gives the full per-RQ treatment that complements the thematic narrative of \S\ref{sec:eval}. Each of RQ1--RQ10 is reported under the same five-paragraph structure (\emph{Motivation}, \emph{Setup}, \emph{Results}, \emph{Mechanism}, \emph{Implications}). Subsections F.1--F.5 group RQs by the parent thematic subsection in the main text (\S\ref{sec:cap-strat}--\S\ref{sec:rq-synthesis}), so this appendix can be read either as a per-RQ commentary or as a per-thematic deep-dive.

\subsection{RQ1 \& RQ2: Capability Stratification}
\label{app:rq1-2}

\subsubsection*{RQ1: Where in the complexity spectrum does LLM graph-generation capability most sharply differentiate?}

\paragraph{Motivation.} A diagnostic benchmark must separate models with different capabilities. Aggregate benchmarks deliver that separation at the task level; we ask where the separation lies \emph{within} a single task family once outputs are stratified by complexity.

\paragraph{Setup.} For each of the 6 levels we report a per-tier Quality summary (Tab.~\ref{tab:tier-gap}) and define the tier gap as $Q_\text{T1}-Q_\text{T3}$. The reported values aggregate at the (tier, level) granularity using a per-cell choice that mostly tracks per-model best-of-four-strategy means; the L4 row uses zero-CoT-anchored means because L4 few-shot operates as in-context retrieval rather than as a prompting variant (\S\ref{sec:methods}, RQ10). Two algorithm-uniform alternatives preserve the L2-dominates-every-other-level ordering: per-model best-of-four-strategy mean (L2 gap 0.224, L4 gap 0.069) and all-strategy-average over 45 cells (L2 gap 0.341, L4 gap 0.052); per-level reproduction in App.~\ref{app:tier-gap-alt}.

\begin{figure}[h]
\centering
\includegraphics[width=0.78\linewidth]{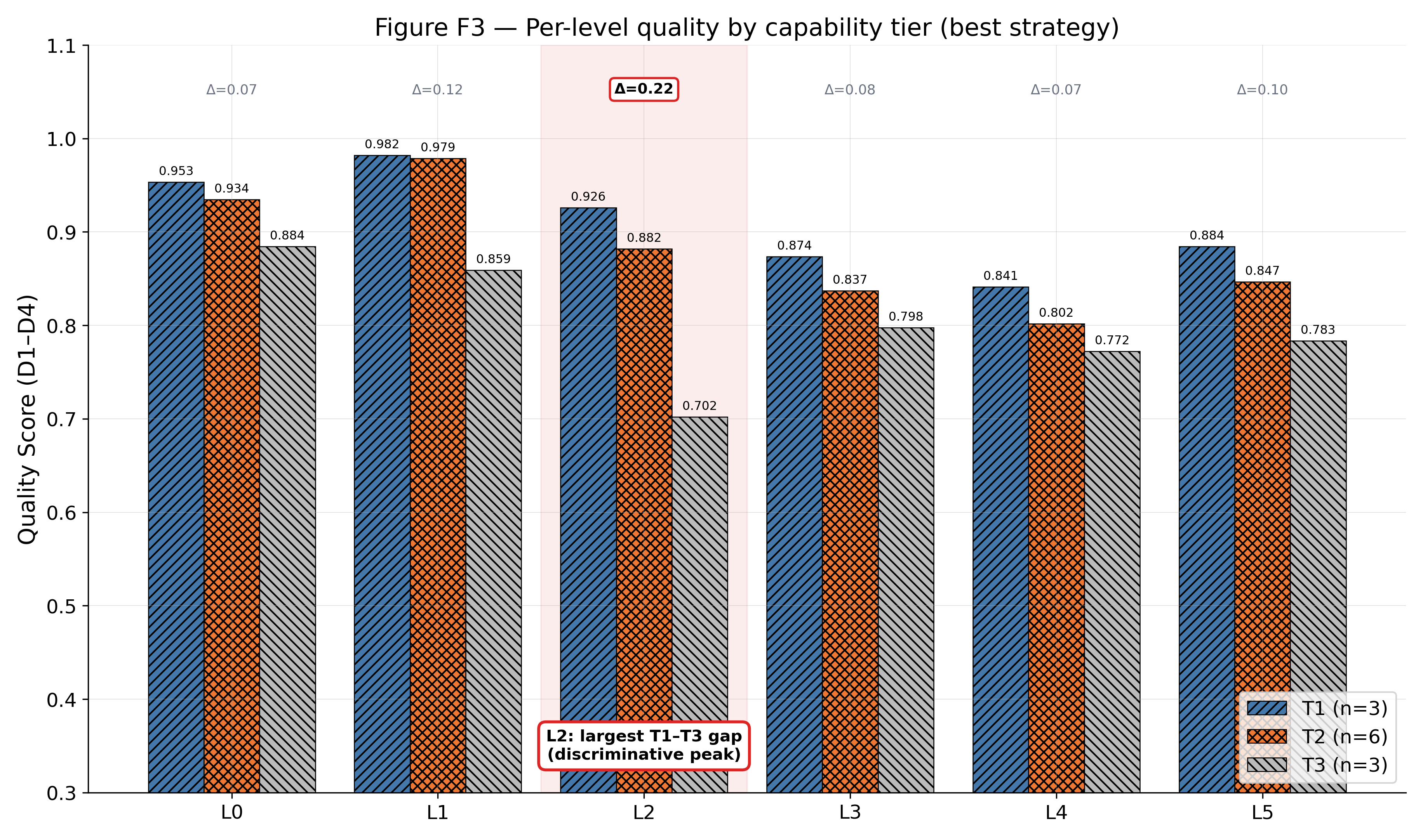}
\caption{Per-level Quality by capability tier (Tab.~\ref{tab:tier-gap} values). The T1--T3 gap at L2 (0.219) is 1.8--3$\times$ any other level (2.1$\times$ the mean of \{L1, L3, L4, L5\}; up to 3$\times$ over the smallest comparator L3 at 0.073, 2$\times$ over the multi-step-editing gap L5 at 0.106), localizing benchmark discriminative power at multi-constraint composition rather than reasoning depth. The ordering is invariant to the aggregation choice (App.~\ref{app:tier-gap-alt}).}
\label{fig:tier-level}
\end{figure}

\paragraph{Results.} Table~\ref{tab:tier-gap} shows the per-level mean by tier; Figure~\ref{fig:tier-level} renders the same data as a bar chart.

\begin{table}[h]
\centering
\caption{Per-level mean Quality by capability tier (T1/T2/T3 ordered by mean $Q$, see \S\ref{sec:models}) and T1--T3 gap. Per-cell aggregation choice is documented in the RQ1 setup; algorithm-uniform alternatives in App.~\ref{app:tier-gap-alt}.}
\label{tab:tier-gap}
\small
\begin{tabular}{ccccc}
\toprule
\textbf{Level} & \textbf{T1 mean} & \textbf{T2 mean} & \textbf{T3 mean} & \textbf{T1--T3 gap} \\
\midrule
L0 & 0.941 & 0.925 & 0.884 & 0.057 \\
L1 & 0.975 & 0.965 & 0.855 & 0.120 \\
\textbf{L2} & \textbf{0.917} & \textbf{0.862} & \textbf{0.698} & \textbf{0.219} \\
L3 & 0.858 & 0.815 & 0.785 & 0.073 \\
L4 & 0.841 & 0.800 & 0.719 & 0.122 \\
L5 & 0.879 & 0.835 & 0.773 & 0.106 \\
\bottomrule
\end{tabular}
\end{table}

\paragraph{Mechanism.} One might expect the reasoning-heavy levels---L3, L4, or L5---to produce the largest tier gap, because these levels involve longer causal chains from instruction to output. The data contradict this: \textbf{L2, simultaneous satisfaction of four or more structural constraints, is where the tier gap is largest}. The most plausible mechanism is compositional---each added constraint multiplies the probability of violation, and T3 models lack the working-memory capacity to maintain joint satisfaction.

\begin{figure}[h]
\centering
\includegraphics[width=0.78\linewidth]{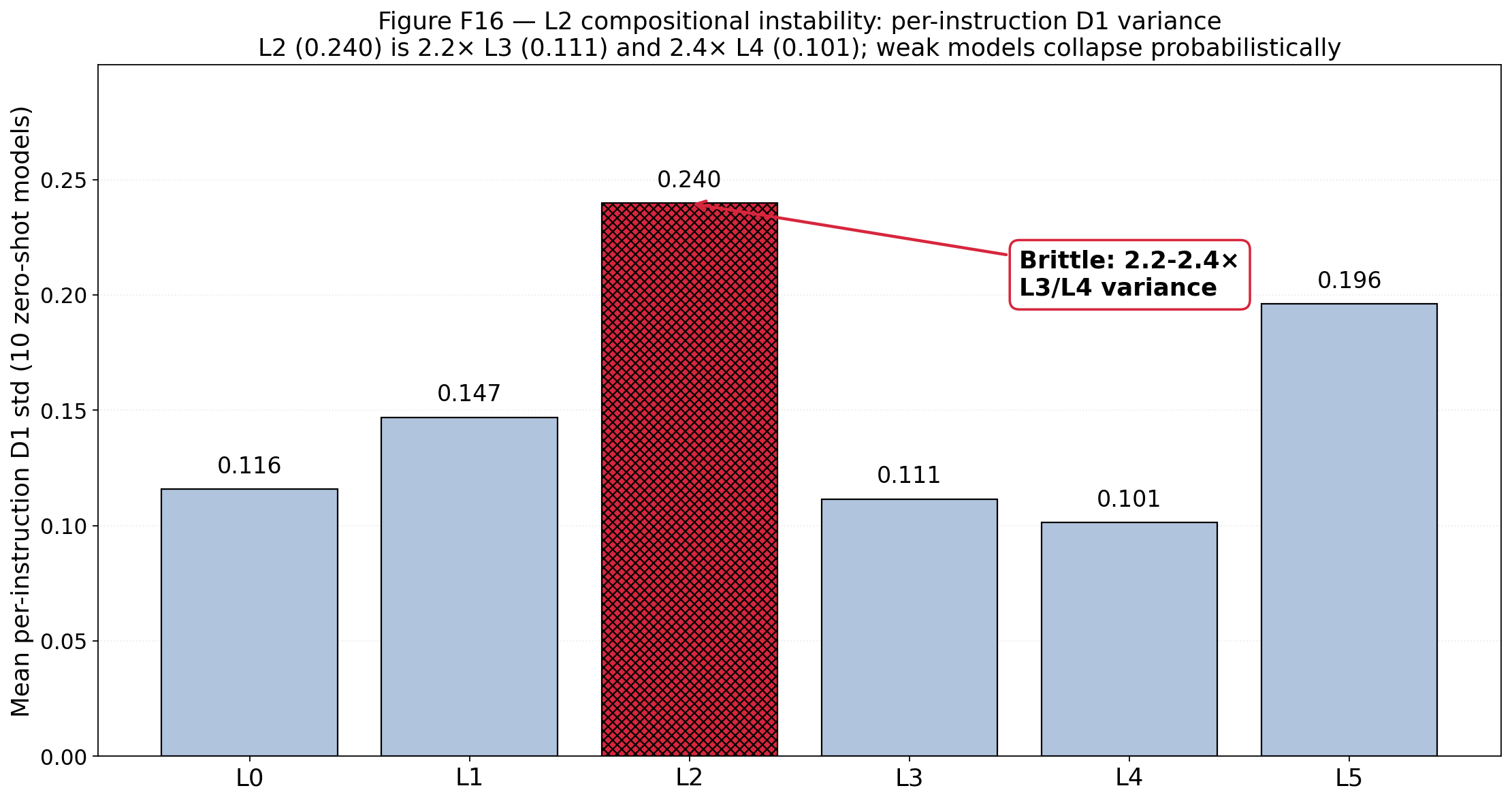}
\caption{Per-instruction D1 standard deviation by level, averaged over 10 zero-shot models. L2 ($\sigma{=}0.240$) is $2.2\times$ L3 and $2.4\times$ L4, reflecting compositional instability rather than uniform difficulty.}
\label{fig:l2-instability}
\end{figure}

\textbf{L2 failure is not merely harder, it is more brittle.} Per-instruction D1 variance at L2 (averaged over the 10 zero-shot-evaluated models) is $\sigma = 0.240$, roughly $2.2\times$ the L3 value ($0.111$) and $2.4\times$ the L4 value ($0.101$); see Figure~\ref{fig:l2-instability}. Weak models are the worst offenders: GPT-4o-mini and GPT-3.5 reach $\sigma_\text{L2}{>}0.43$, while Sonnet-4.6 and Qwen3.5-397B stay at $\sigma_\text{L2}{<}0.13$. This means L2 is a \emph{compositional instability} regime, not a uniformly harder regime: weak models succeed brilliantly on some L2 instructions and fail catastrophically on others, depending on which particular constraint combination is sampled. This is the mechanism behind the $3\times$ tier gap---T3 models are one constraint-combination away from collapse at L2, while T1 models are robust to the same combinations---and it directly motivates reporting \emph{per-constraint-type} method gains on L2 rather than level-wise averages.

\begin{figure}[h]
\centering
\includegraphics[width=\linewidth]{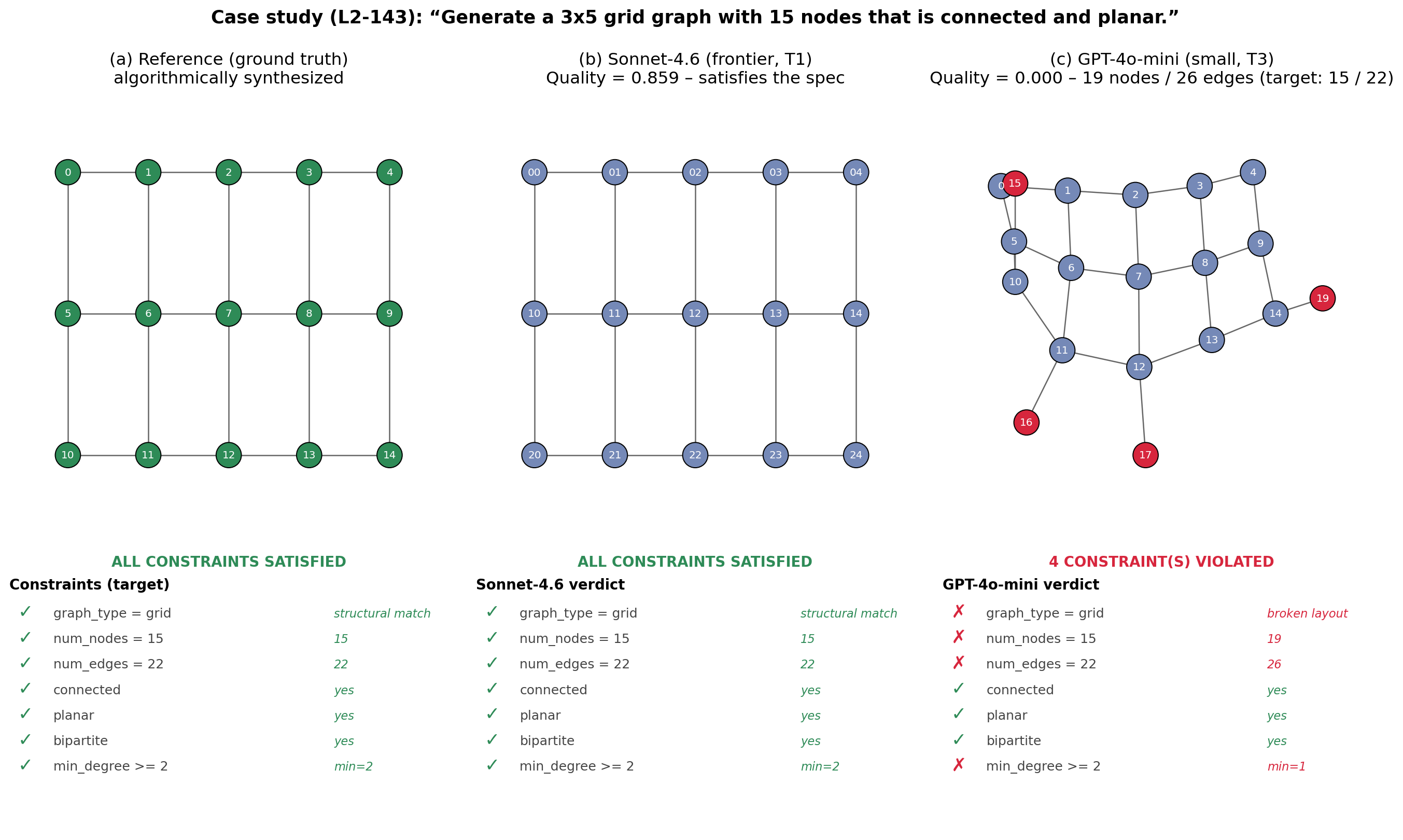}
\caption{Capability-gap case study at L2 (instruction L2-143). Reference (left) and Sonnet-4.6 (centre) both produce a 15-node 3$\times$5 grid satisfying all seven constraints. GPT-4o-mini (right) emits a parseable but constraint-violating 19-node graph (red highlighted nodes are the four extras: 15, 16, 17, 19), failing four of seven constraints. The constraint checklists below each panel make the failure modes machine-checkable.}
\label{fig:case-study-l2}
\end{figure}

\paragraph{Case study.} Figure~\ref{fig:case-study-l2} renders the L2-143 instruction (``Generate a 3$\times$5 grid graph with 15 nodes that is connected and planar'') side-by-side for the algorithmic reference, Sonnet-4.6 (T1), and GPT-4o-mini (T3). Both Sonnet-4.6 and the reference produce a clean grid that satisfies every constraint; GPT-4o-mini emits a 19-node, 26-edge graph that violates four of seven constraints (graph type, node count, edge count, minimum degree). The qualitative picture matches the quantitative finding: at L2, frontier models hold the spec while small models drift in node count and lose the joint grid structure.

\paragraph{Implications.} Benchmarks averaging over constraint count \emph{systematically underestimate} structured-generation discriminative power. Per instruction, L2 is 3$\times$ more informative than aggregate scoring. Method research targeting structural-generation improvement should report L2 gains as a primary signal, not overall score. Expanding L2 in future versions of \bench\ is a natural way to sharpen this signal further.

\subsubsection*{RQ2: How does prompt-strategy sensitivity vary with base capability?}

\paragraph{Motivation.} \citet{demirci2025graphsavvy} report that iterative-feedback gains vary substantially across models (Grok $+48\%$, Llama $<\!5\%$) without identifying an explanatory variable. We test whether this heterogeneity is predictable from base capability.

\paragraph{Setup.} For each model we compute $\sstrat$, the standard deviation of $Q$ across the four prompting strategies. Because $\sstrat$ requires four-strategy data, Sonnet-4 is excluded from this analysis (zero-shot-only, \S\ref{sec:models}); we regress $\sstrat$ on mean $Q$ across the remaining \textbf{11 fully-evaluated models}.

\paragraph{Results.} Strategy variance varies sharply with capability. Using the population standard deviation of $Q$ across the four strategies ($\sstrat = \sqrt{\sum(Q_s - \bar Q)^2/4}$), T3 models exhibit $\sstrat\in\{0.027,\,0.025,\,0.022\}$ (GPT-3.5, GPT-4o-mini, Llama-8B respectively); T2-stable models exhibit $\sstrat\in\{0.008,\,0.007\}$ (DeepSeek-V3, Llama-70B); T1 models exhibit $\sstrat\in\{0.015,\,0.018,\,0.015\}$ (Sonnet-4.6, Qwen3.5-397B, Qwen3.5-122B respectively). Equivalently, the $Q$-range (max$-$min over the four strategies) at $n{=}4$ is roughly $2.7\times$ the population std, yielding T3 ranges of $\{0.073,0.069,0.048\}$ directly readable from the per-strategy leaderboards in App.~\ref{app:leaderboards}. OLS regression of $\sstrat$ on mean $Q$ yields $\beta{=}-0.073$, $R^2{=}0.40$, two-sided $p{\approx}0.015$; the bootstrap 95\% CI for $\beta$ is $[-0.135,+0.002]$ with $P(\beta{<}0){=}0.975$ over $10{,}000$ resamples, and leave-one-out OLS is sign-stable on all 11 fits (App.~\ref{app:stat-robust}).

\begin{figure}[h]
\centering
\includegraphics[width=0.78\linewidth]{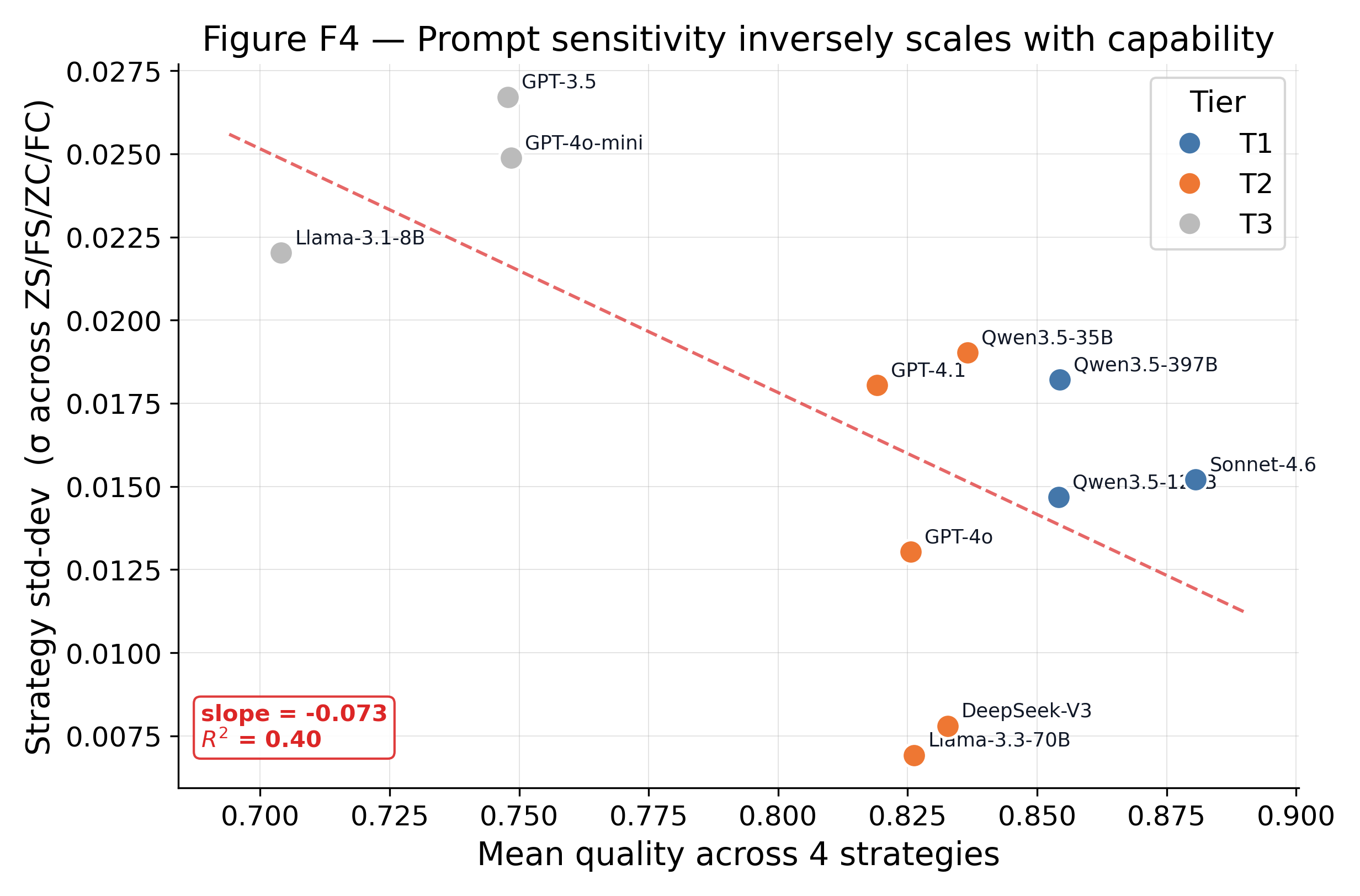}
\caption{Prompt sensitivity ($\sstrat$, y-axis; population std of $Q$ across the four strategies) vs.\ base capability (mean $Q$, x-axis) across the 11 fully-evaluated models (Sonnet-4 excluded, zero-shot-only). The $4\times$ gap between weakest T3 models ($\sstrat{\approx}0.027$, equivalent $Q$-range $\approx 0.073$) and most prompt-stable T2 models ($\sstrat{\approx}0.008$, range $\approx 0.019$) establishes an inverse-scaling trend; the solid OLS line has slope $-0.073$, $R^2{=}0.40$ (per \texttt{scripts/paper\_figures.py::fig\_F4\_capability\_variance}; bootstrap 95\% CI for the slope is $[-0.135, +0.002]$ with $P(\beta{<}0){=}0.975$, see App.~\ref{app:stat-robust}).}
\label{fig:cap-variance}
\end{figure}

\paragraph{Mechanism.} Weak models occupy an under-trained region of the output manifold where small prompt perturbations---adding a CoT trigger, switching from zero-shot to few-shot---produce large quality displacements. Strong models occupy a better-conditioned region with a flatter local quality surface.

\paragraph{Implications.} Prompt-engineering budgets should scale inversely with model capability: for frontier models, strategy choice accounts for $\leq\!2\%$ of performance variance, so investing in verification or retrieval yields better returns. Single-strategy benchmark evaluations systematically disadvantage prompt-sensitive models. Prompting-method papers should report per-model $\sstrat$ alongside headline gains.

\subsection{RQ3, RQ4 \& RQ5: Strategy and Family Effects}
\label{app:rq3-5}

\subsubsection*{RQ3: Does any single prompting strategy uniformly dominate across complexity levels?}

\paragraph{Motivation.} Aggregate benchmark scores wash out level-dependent strategy effects. We test whether such effects exist and whether they change sign across levels.

\paragraph{Setup.} We compute per-level strategy effects relative to zero-shot (FS$-$ZS, ZC$-$ZS, FC$-$ZS), averaged across the \textbf{11 fully-evaluated models} (Sonnet-4 excluded because the three non-zero-shot strategies are undefined for it).

\begin{figure}[h]
\centering
\includegraphics[width=0.78\linewidth]{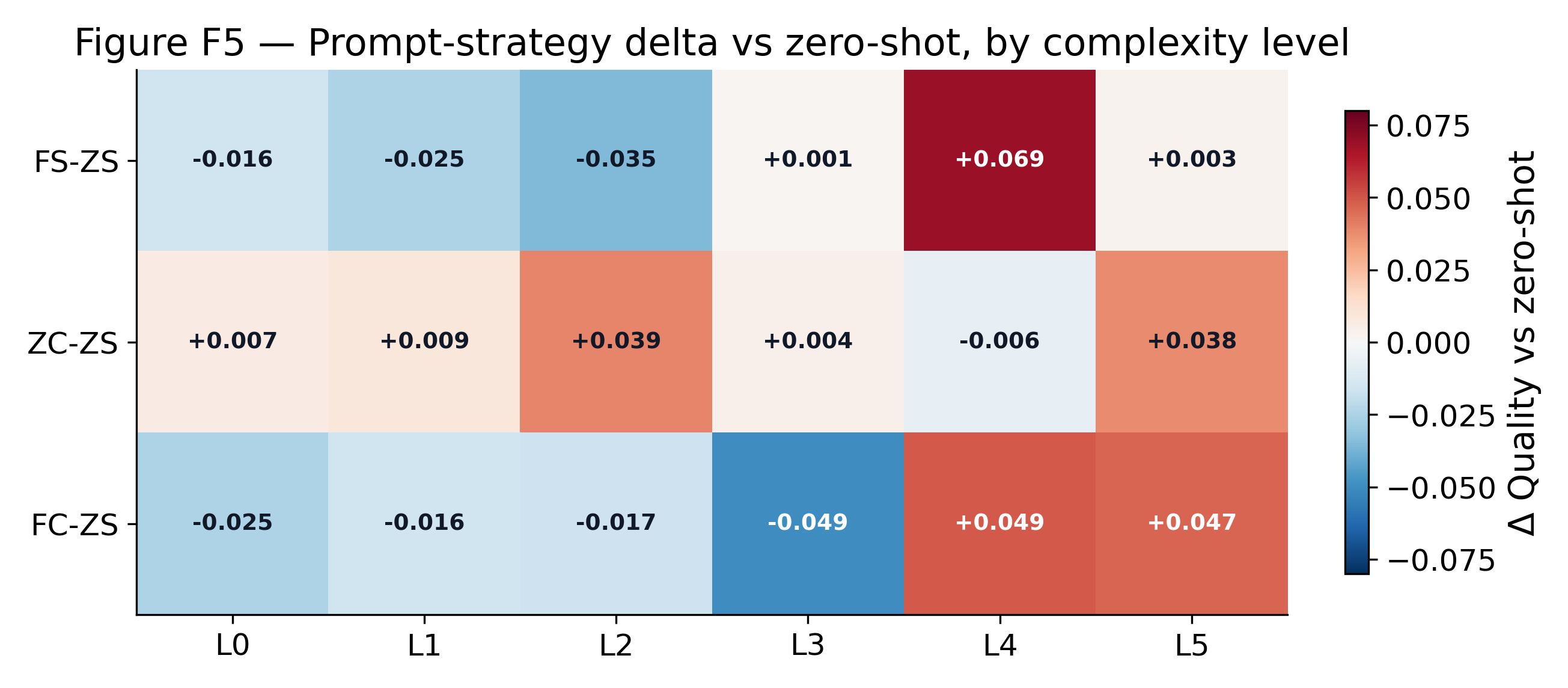}
\caption{Signed strategy $\times$ level effect heatmap (average over the 11 fully-evaluated models). Few-shot is net-negative at L2 ($-0.034$) and net-positive at L4 ($+0.069$); few-CoT swings from net-negative at L3 ($-0.048$) to net-positive at L5 ($+0.045$). Aggregate benchmarks mask these opposite-signed effects.}
\label{fig:strategy-heatmap}
\end{figure}

\paragraph{Results.} Table~\ref{tab:strategy-delta} shows the signed effects; Figure~\ref{fig:strategy-heatmap} renders the same data as a heatmap.

\begin{table}[h]
\centering
\caption{Signed strategy--level effects (averaged across the 11 fully-evaluated models), $\Delta$ vs.\ zero-shot.}
\label{tab:strategy-delta}
\small
\begin{tabular}{ccccc}
\toprule
\textbf{Level} & \textbf{FS $-$ ZS} & \textbf{ZC $-$ ZS} & \textbf{FC $-$ ZS} & \textbf{Best strategy} \\
\midrule
L0 & $-0.016$ & $+0.007$ & $-0.025$ & zero-CoT \\
L1 & $-0.024$ & $+0.008$ & $-0.016$ & zero-CoT \\
\textbf{L2} & $\mathbf{-0.034}$ & $\mathbf{+0.038}$ & $-0.018$ & zero-CoT \\
L3 & $+0.003$ & $+0.004$ & $\mathbf{-0.048}$ & zero-CoT \\
\textbf{L4} & $\mathbf{+0.069}$ & $-0.007$ & $+0.049$ & \textbf{few-shot} \\
\textbf{L5} & $+0.003$ & $+0.037$ & $\mathbf{+0.045}$ & \textbf{few-CoT} \\
\bottomrule
\end{tabular}
\end{table}

\paragraph{Mechanism.} No strategy dominates: every non-trivial strategy is \emph{net-harmful} at at least one level and \emph{net-helpful} at at least one other. Few-shot is poison at L2---the demonstration's specific graph biases generation toward copying topology rather than satisfying the target's distinct constraints---but savior at L4, where domain examples convey structural priors the instruction alone cannot. Few-CoT is savior at L5 (explicit step planning for edits) but actively harmful at L3 (extraneous reasoning amplifies numerical drift). Zero-CoT is the only strategy with \emph{non-negative effect at every level}, though its gains are modest where other strategies excel.

\paragraph{Implications.} Prompting-method papers reporting only aggregate gains may be silently trading L2 for L4 or L3 for L5. Level-stratified reporting should become standard on structured tasks. Strategy-per-level adaptive prompting is empirically motivated, directly informing our CAAP component (\S\ref{sec:benchmark-methods}).

\subsubsection*{RQ4: Does chain-of-thought transfer uniformly across model families?}

\paragraph{Motivation.} The folk wisdom that ``CoT helps on complex tasks'' rests almost entirely on text-domain benchmarks. We test whether the transfer holds on graph generation and whether it is uniform across model families.

\paragraph{Setup.} We compute CoT effects (ZC$-$ZS and FC$-$ZS) for the seven family-aligned models we evaluate: three Qwen3.5 scales and four GPT-family scales.

\begin{figure}[h]
\centering
\includegraphics[width=0.78\linewidth]{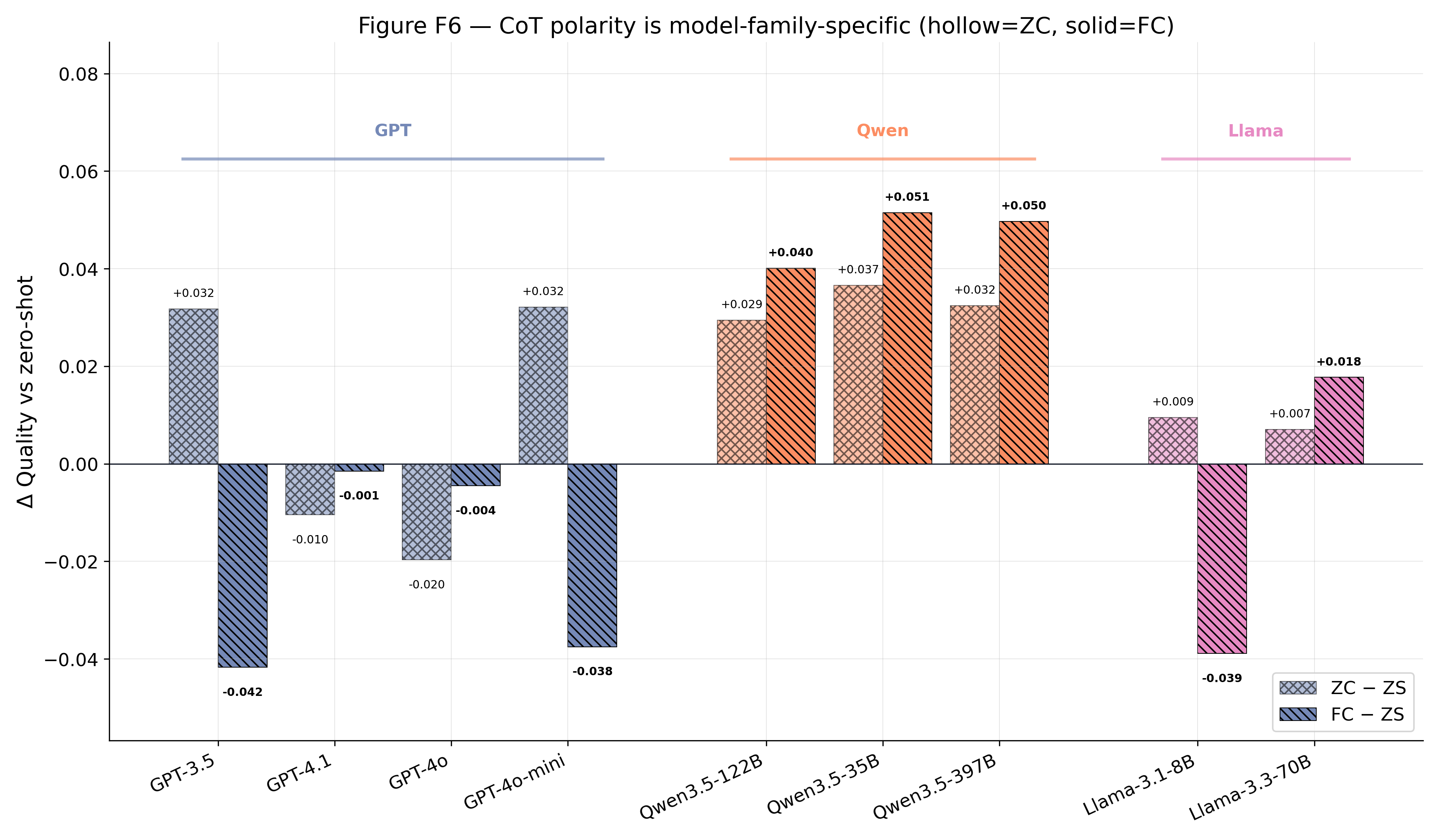}
\caption{Signed CoT effect by model family. Qwen3.5 gains uniformly across scales ($+0.029$ to $+0.052$); few-CoT is clearly negative for the weaker GPT models (3.5/4o-mini: $-0.042$/$-0.038$, both $>$7$\times$ our $\pm 0.005$ stability band) and near-zero, within-noise for the stronger GPT models (4o/4.1: $-0.005$/$-0.002$). The robust signal is the negative sign on the weaker GPT pair combined with the uniformly positive Qwen3.5 series.}
\label{fig:cot-family}
\end{figure}

\paragraph{Results.} Table~\ref{tab:cot-family} reveals a sign reversal across families; Figure~\ref{fig:cot-family} visualizes the family-level polarity.

\begin{table}[h]
\centering
\caption{CoT deltas (vs.\ zero-shot) by model family and scale.}
\label{tab:cot-family}
\small
\begin{tabular}{cccc}
\toprule
\textbf{Family} & \textbf{Model} & \textbf{ZC $-$ ZS} & \textbf{FC $-$ ZS} \\
\midrule
Qwen3.5 & 35B-A3B  & $+0.037$ & $+0.052$ \\
Qwen3.5 & 122B-A10B & $+0.029$ & $+0.040$ \\
Qwen3.5 & 397B-A17B & $+0.032$ & $+0.050$ \\
GPT & 3.5-turbo & $+0.032$ & $\mathbf{-0.042}$ \\
GPT & 4o-mini & $+0.032$ & $\mathbf{-0.038}$ \\
GPT & 4o & $\mathbf{-0.020}$ & $-0.005$ \\
GPT & 4.1 & $\mathbf{-0.010}$ & $-0.002$ \\
\bottomrule
\end{tabular}
\end{table}

\paragraph{Mechanism.} Few-CoT is \emph{uniformly beneficial for Qwen3.5} across scales (35B, 122B, 397B, all $>$7$\times$ our stability band) and \emph{clearly negative for the weaker GPT models} (3.5: $-0.042$, 4o-mini: $-0.038$, both $>$7$\times$ band), with the stronger GPT models showing only \emph{near-zero, within-noise} effects (4o: $-0.005$, 4.1: $-0.002$). The robust signal is the negative sign on the weaker GPT pair combined with the uniformly positive Qwen3.5 series. The sign's robustness across scales within each family---Qwen3.5-35B agrees with Qwen3.5-397B despite a $\sim$10$\times$ parameter gap; the negative GPT signal at the lower end persists from GPT-3.5 to GPT-4o-mini---indicates the effect, where present, is governed by \textbf{pretraining distribution} rather than capability tier. One possible explanation is that Qwen3.5's pretraining mixture (heavy code/mathematical/reasoning-chain content) creates a CoT prior that transfers positively to graph generation, while the weaker GPT models' broader pretraining diet creates a negatively-transferring prior; we offer this as a hypothesis rather than a verified mechanism---direct verification would require pretraining-mixture access beyond our scope.

\paragraph{Implications.} The text-domain folk wisdom surrounding chain-of-thought~\citep{wei2022cot,kojima2022zerocot,wang2023selfconsistency} does not transfer to graph generation for one major model family. Prompting-method claims derived from GPT evaluations cannot be generalized without independent validation on other families. Model-family should be reported alongside capability when benchmarking CoT-using methods.

\subsubsection*{RQ5: Does parameter scaling uniformly translate into per-level capability?}

\paragraph{Motivation.} Aggregate scaling-law narratives predict monotone improvement with parameter count. Progressive evaluation lets us test that prediction \emph{per-level} within a same-family scale series.

\paragraph{Setup.} We compare same-family models across scales: Qwen3.5 at 35B/122B/397B, and GPT-family where comparable at 3.5/4o-mini/4o/4.1. We report both aggregate and per-level Quality.

\begin{figure}[h]
\centering
\includegraphics[width=0.78\linewidth]{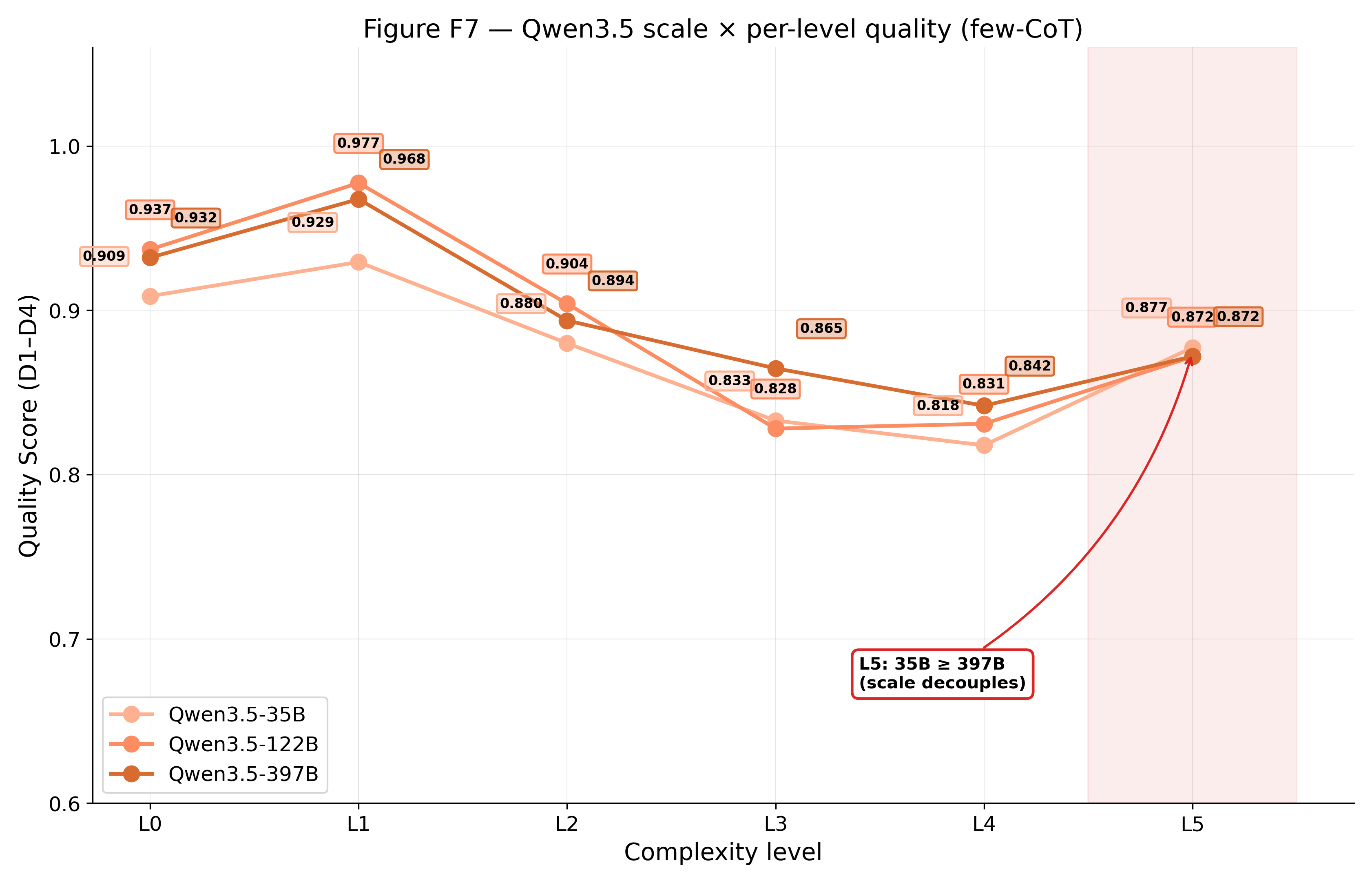}
\caption{Qwen3.5 scale family (35B / 122B / 397B) per-level Quality. Scaling monotonically improves L0--L4; at L5 the 35B-vs-397B difference ($\Delta{=}{-}0.005$) is within both the 95\% CI of $\pm 0.019$ at $N{=}50$ and our $\pm 0.005$ stability band, i.e.\ no monotonic scaling gain is detectable at this sample size.}
\label{fig:scale-per-level}
\end{figure}

\paragraph{Results.} Parameter scaling does not uniformly lift per-level capability (Table~\ref{tab:scale-decouple}); per-level Qwen3.5 scaling curves are in Figure~\ref{fig:scale-per-level}.

\begin{table}[h]
\centering
\caption{Scale decoupling: cases where a larger model fails to outperform (or underperforms) a smaller same-family model.}
\label{tab:scale-decouple}
\small
\begin{tabular}{lccc}
\toprule
\textbf{Subtask} & \textbf{Smaller} & \textbf{Larger} & \textbf{$\Delta$} \\
\midrule
L3 numerical attributes & Qwen3.5-35B, 0.833 & Qwen3.5-397B, 0.865 & $+0.032$ \\
L4 domain semantics & Qwen3.5-35B, 0.818 & Qwen3.5-397B, 0.842 & $+0.024$ \\
\textbf{L5 graph editing} & \textbf{Qwen3.5-35B, 0.877} & \textbf{Qwen3.5-397B, 0.872} & $\mathbf{-0.005}$ \\
L3 numerical attributes & GPT-3.5, 0.808 & GPT-4.1, 0.759 (avg) & $\mathbf{-0.049}$ \\
\bottomrule
\end{tabular}
\end{table}

\paragraph{Mechanism.} Tasks that reduce to \textbf{local structural operations}---bounded-scope editing (L5), single-attribute verification (parts of L3)---saturate at smaller model scales than tasks requiring global constraint reasoning. In Qwen3.5, 35B$\to$397B adds only 0.024 at L4 and \emph{subtracts} 0.005 at L5, while adding 0.032 at L3. The GPT counter-example is more striking: GPT-3.5 \emph{outperforms} GPT-4.1 on L3 numerical attributes (0.808 $>$ 0.759 avg), a reversal aggregate quality conceals entirely. The mechanism interacts with RQ4: larger GPT-family models' CoT inclinations produce longer, more error-prone reasoning chains on tasks smaller models attempt more directly.

\paragraph{Implications.} Scaling-law studies~\citep{kaplan2020scaling,hoffmann2022chinchilla} built on aggregate benchmarks mask task-structural heterogeneity in where capability emerges with scale, extending the emergent-behavior critique of~\citet{schaeffer2023emergentmirage} to a specific structured-generation sub-capability. Parameter-efficient deployment strategies---``smaller model for subtask A, larger model for subtask B''---are viable on structured tasks in a way aggregate benchmarks obscure.

\subsection{RQ6: Cost-Aware Deployment}
\label{app:rq6}\label{app:pareto}

\subsubsection*{RQ6: How does cost-adjusted scoring reshape model rankings, and where is the capability floor?}

\paragraph{Motivation.} RQ1--RQ5 characterize capability on Quality $\Stot$. Deployment practitioners also care about cost: a $+3\%$ gain at $5\times$ token cost is net-negative if a cheaper alternative already clears the usable-quality threshold. We apply the benchmark's efficiency instruments---the Pareto-adjusted $\Sfin$ (Eq.~\ref{eq:sfinal}), $Q/\kTPV$, and Cost@$Q{=}0.8$---to the 45 baseline configurations, treating $\Sfin$ as a \emph{complementary} deployment-oriented view alongside the capability-characterization view given by $\Stot$.

\paragraph{Setup.} For each of the 45 runs we compute mean tokens per valid graph (TPV), Quality $\Stot$, $Q/\kTPV$, the non-dominated Pareto frontier in $\langle\TPV,\Stot\rangle$ space, and per-model Cost@$Q{=}0.8$. We then compute $\Sfin$ with $\lambda=0.15$ and compare the $\Stot$ and $\Sfin$ orderings. Per-model Pareto frontiers in $(\Stot, 1/D_5)$ space, with annotations for the four prompting strategies, are drawn from the same data and follow the per-configuration trajectory shown here.

\begin{figure}[h]
\centering
\includegraphics[width=0.78\linewidth]{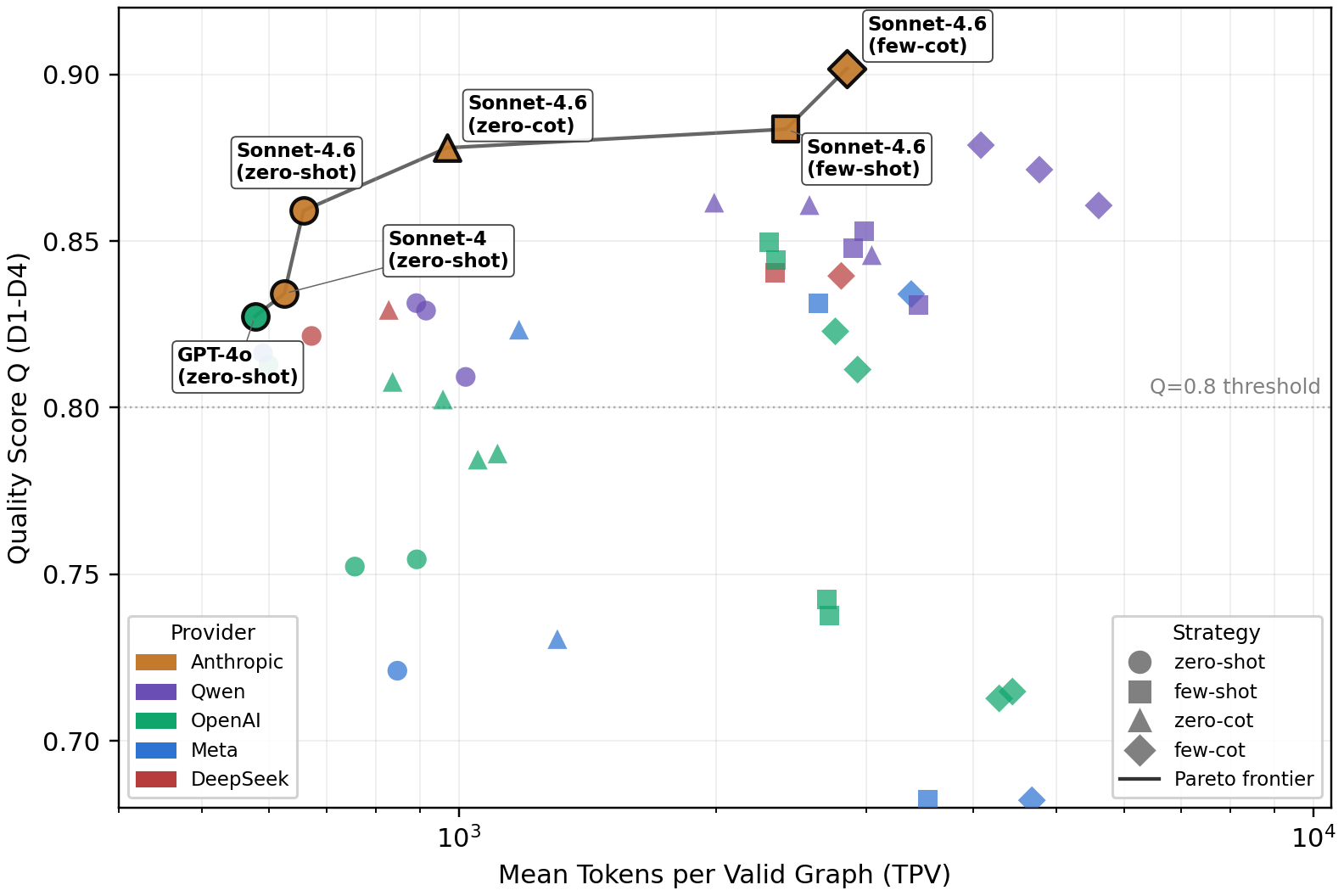}
\caption{Pareto frontier over 45 baseline (model, strategy) configurations in $\langle$mean TPV, Quality$\rangle$ space. Solid line traces the 6 non-dominated points (labeled). Markers encode prompting strategy; colors encode provider. Nine of 12 models cross the $Q{=}0.8$ threshold (dotted line)---zero-shot is the cheapest threshold-crossing strategy for all eight that are evaluated on all four strategies (Sonnet-4 was evaluated only under zero-shot, which also clears the threshold); three T3 models never cross it.}
\label{fig:pareto-baseline}
\end{figure}

\paragraph{Result (a): A 6-point, provider-concentrated Pareto frontier.} Only \textbf{6 of 45} configurations are non-dominated (Table~\ref{tab:pareto-front}; visualized in Figure~\ref{fig:pareto-baseline}), all Anthropic or OpenAI; no Qwen3.5, DeepSeek, or Llama configuration is Pareto-optimal despite several near-frontier candidates (e.g., Qwen3.5-397B zero-CoT reaches $Q{=}0.862$ at TPV${\approx}$2300 but is dominated by Sonnet-4.6 zero-CoT at TPV${=}$969, $Q{=}0.878$). The frontier follows a strategy progression: zero-shot at the low-cost end, zero-CoT / few-shot in the middle, few-CoT on Sonnet-4.6 at the high-cost end.

\begin{table}[h]
\centering
\caption{The 6-point Pareto frontier over the 45 baseline configurations, ordered by ascending TPV.}
\label{tab:pareto-front}
\small
\begin{tabular}{llccc}
\toprule
\textbf{Model} & \textbf{Strategy} & \textbf{mean TPV} & \textbf{$\Stot$} & \textbf{$\Sfin$ ($\lambda{=}0.15$)} \\
\midrule
GPT-4o         & zero-shot & 578  & 0.8274 & 0.9515 \\
Sonnet-4       & zero-shot & 625  & 0.8342 & 0.9594 \\
Sonnet-4.6     & zero-shot & 658  & 0.8591 & 0.9880 \\
Sonnet-4.6     & zero-CoT  & 969  & 0.8780 & 1.0097 \\
Sonnet-4.6     & few-shot  & 2415 & 0.8836 & 1.0161 \\
Sonnet-4.6     & few-CoT   & 2846 & 0.9018 & 1.0371 \\
\bottomrule
\end{tabular}
\end{table}

\paragraph{Result (b): $\Sfin$ reshuffles the leaderboard; few-CoT is absent from the efficiency top-15.} The Pareto bonus moves 6 configurations up by 5--18 positions and pushes high-cost Qwen3.5 runs down by 2--4 (Table~\ref{tab:sfin-rank}). The largest climb is GPT-4o zero-shot ($\Stot$\,\#24${\to}\Sfin$\,\#6, ${+}18$); the largest fall is Qwen3.5-397B few-CoT (\#3${\to}$\#7). Among the \textbf{11 models for which per-model strategy comparison is defined}, the transition flips the $\Stot$-best strategy for \textbf{only GPT-4o} (few-shot, $\Stot{=}0.844$ $\to$ zero-shot, $\Sfin{=}0.952$); the other 10 retain their $\Stot$-optimal choice (per-model strategy comparison is undefined for Sonnet-4 since only zero-shot was evaluated). On the auxiliary $Q/\kTPV$ ranking, the top five are \textbf{all zero-shot} (GPT-4o 1.43, Llama-3.3-70B 1.38, GPT-4.1 1.36, Sonnet-4 1.34, Sonnet-4.6 1.31), and \textbf{no few-CoT configuration enters the top 15} even though few-CoT is the $\Stot$-optimal strategy for 5 of the 11 fully-evaluated models.

\begin{table}[h]
\centering
\caption{Top-15 $\Sfin$ leaderboard with $\Stot\!\to\!\Sfin$ rank change $\Delta$. ``Par.'' marks Pareto-optimal configurations. All five $Q/\kTPV$-top rows and every rank-gainer are zero-shot; the $\Stot$-to-$\Sfin$ demotions concentrate in the Qwen3.5 family. Ranks are computed across all 45 (model, strategy) configurations. $\dagger$ Sonnet-4 was evaluated only under zero-shot (\S\ref{sec:models}); its rank reflects this single configuration's global position and not a per-model best-of-four.}
\label{tab:sfin-rank}
\scriptsize
\begin{tabular}{rrr llcccc c}
\toprule
\textbf{$\Sfin\#$} & \textbf{$\Stot\#$} & \textbf{$\Delta$} & \textbf{Model} & \textbf{Strategy} & \textbf{$\Stot$} & \textbf{TPV} & \textbf{$Q/\kTPV$} & \textbf{$\Sfin$} & \textbf{Par.} \\
\midrule
1  & 1  & $\phantom{+}0$ & Sonnet-4.6   & few-CoT   & 0.902 & 2846 & 0.32 & 1.037 & \checkmark \\
2  & 2  & $\phantom{+}0$ & Sonnet-4.6   & few-shot  & 0.884 & 2415 & 0.37 & 1.016 & \checkmark \\
3  & 4  & $+1$           & Sonnet-4.6   & zero-CoT  & 0.878 & 969  & 0.91 & 1.010 & \checkmark \\
4  & 9  & $+5$           & Sonnet-4.6   & zero-shot & 0.859 & 658  & 1.31 & 0.988 & \checkmark \\
5  & 17 & $+12$          & Sonnet-4$^\dagger$ & zero-shot & 0.834 & 625  & 1.34 & 0.959 & \checkmark \\
6  & 24 & $+18$          & GPT-4o       & zero-shot & 0.827 & 578  & 1.43 & 0.952 & \checkmark \\
7  & 3  & $-4$           & Qwen3.5-397B & few-CoT   & 0.879 & 4082 & 0.22 & 0.879 & \\
8  & 5  & $-3$           & Qwen3.5-122B & few-CoT   & 0.871 & 4779 & 0.18 & 0.871 & \\
9  & 6  & $-3$           & Qwen3.5-397B & zero-CoT  & 0.862 & 1990 & 0.43 & 0.862 & \\
10 & 7  & $-3$           & Qwen3.5-122B & zero-CoT  & 0.861 & 2572 & 0.33 & 0.861 & \\
11 & 8  & $-3$           & Qwen3.5-35B  & few-CoT   & 0.861 & 5607 & 0.15 & 0.861 & \\
12 & 10 & $-2$           & Qwen3.5-122B & few-shot  & 0.853 & 2980 & 0.29 & 0.853 & \\
13 & 11 & $-2$           & GPT-4.1      & few-shot  & 0.850 & 2308 & 0.37 & 0.850 & \\
14 & 12 & $-2$           & Qwen3.5-397B & few-shot  & 0.848 & 2894 & 0.29 & 0.848 & \\
15 & 13 & $-2$           & Qwen3.5-35B  & zero-CoT  & 0.846 & 3042 & 0.28 & 0.846 & \\
\bottomrule
\end{tabular}
\end{table}

\begin{figure}[h]
\centering
\includegraphics[width=0.78\linewidth]{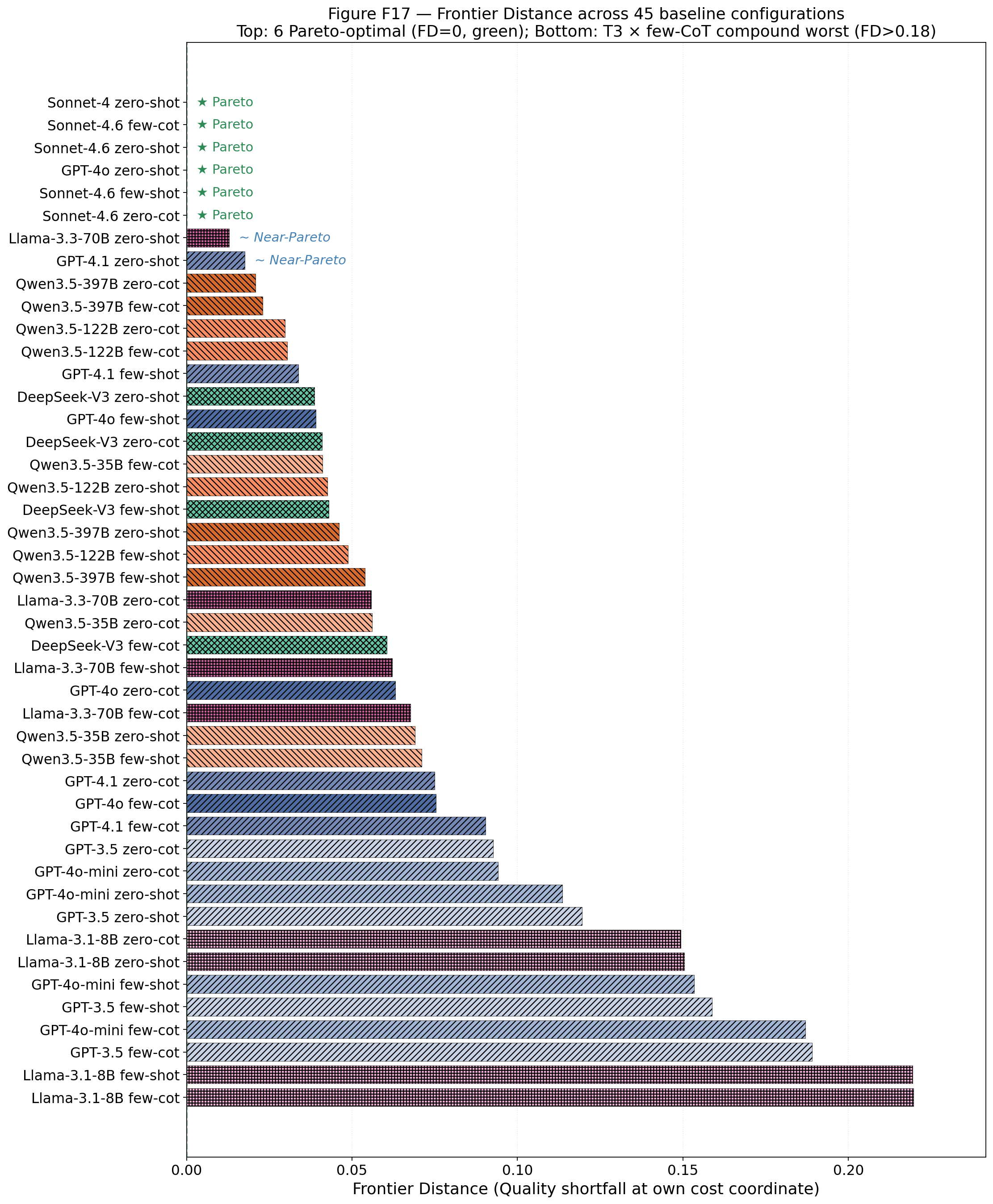}
\caption{Frontier Distance across 45 baseline configurations, sorted ascending. Top: 6 Pareto-optimal (FD$=$0, green star); near-Pareto zero-shot runs (Llama-3.3-70B, GPT-4.1) follow. Bottom: T3$\,\times\,$few-CoT compounds quality deficit under expensive prompting.}
\label{fig:fd-distribution}
\end{figure}

\paragraph{Frontier Distance headroom.} $\mathrm{FD}$ ranges from $0$ (the six Pareto-optimal configurations) to $0.22$ (Llama-3.1-8B few-CoT), with median $0.062$ over the 39 off-frontier configurations (Figure~\ref{fig:fd-distribution}). Two off-frontier points are \emph{near-Pareto} at $\mathrm{FD}{<}0.02$: Llama-3.3-70B zero-shot ($0.013$) and GPT-4.1 zero-shot ($0.018$)---both are zero-shot configurations dominated only by a tighter-cost Sonnet-4.6 run, and indicate that two additional providers are a hair's-breadth from joining the frontier. The five largest $\mathrm{FD}$ values all combine a T3 model with few-CoT or few-shot (Llama-3.1-8B fc/fs: $0.219/0.219$, GPT-3.5 fc: $0.189$, GPT-4o-mini fc: $0.187$, GPT-3.5 fs: $0.159$), confirming that weak-capability models \emph{compound} their quality deficit under expensive prompting. $\mathrm{FD}$ is a practitioner-facing diagnostic: at a fixed cost budget it answers ``how much quality am I leaving on the table by using this configuration?''

\paragraph{Result (c): An empirical Cost@$Q{=}0.8$ threshold under this benchmark.} The $Q{=}0.8$ cutoff is chosen for usability-of-$Q$ intuition; conclusions are similar under $Q\in\{0.75, 0.80, 0.85\}$. Table~\ref{tab:costq08} reports the Cost@$Q{=}0.8$ matrix over the 11 models evaluated on all four prompting strategies. Eight of these 11 cross $Q{\geq}0.8$ under at least one strategy, and \textbf{in every row zero-shot is the cheapest threshold-crossing strategy}. The three cheapest fully-evaluated zero-shot paths come from structurally different T2 providers: GPT-4o (TPV=578), Llama-3.3-70B (590), and GPT-4.1 (599); Sonnet-4's zero-shot-only TPV (625) falls in the same narrow band but is a single-strategy observation (\S\ref{sec:models}) and therefore excluded from the 4-strategy matrix. Three T3 models---\textbf{GPT-3.5, GPT-4o-mini, Llama-3.1-8B}, each evaluated on all four strategies---\textbf{never} reach $Q{\geq}0.8$ under any, marking an empirical capability floor that prompting alone cannot cross and directly motivating the verification-guided methods in RQ7--RQ10.

\begin{table}[h]
\centering
\caption{Cost@$Q{=}0.8$ matrix over the 11 models evaluated on all four prompting strategies. Each cell shows the mean TPV to reach $\Stot{\geq}0.8$ for that (model, strategy), or ``---'' if $\Stot{<}0.8$. Bold marks the per-model cheapest threshold-crossing strategy---zero-shot in every row. Three T3 models never cross the threshold under any strategy. \textbf{Sonnet-4 is excluded from this matrix} because only zero-shot was evaluated in the baseline survey (\S\ref{sec:models}); for reference, Sonnet-4 zero-shot reaches $Q{=}0.834$ at TPV=625, which also clears the threshold. Including Sonnet-4 zero-shot brings the threshold-crossing count to 9 of 12 models.}
\label{tab:costq08}
\small
\begin{tabular}{llrrrrr}
\toprule
\textbf{Model} & \textbf{Tier} & \textbf{zero-shot} & \textbf{few-shot} & \textbf{zero-CoT} & \textbf{few-CoT} & \textbf{best} \\
\midrule
Sonnet-4.6        & T1 & \textbf{658}  & 2415 & 969  & 2846 & 658 \\
Qwen3.5-122B      & T1 & \textbf{892}  & 2980 & 2572 & 4779 & 892 \\
Qwen3.5-397B      & T1 & \textbf{916}  & 2894 & 1990 & 4082 & 916 \\
GPT-4o            & T2 & \textbf{578}  & 2350 & 837  & 2758 & 578 \\
Llama-3.3-70B     & T2 & \textbf{590}  & 2636 & 1176 & 3384 & 590 \\
GPT-4.1           & T2 & \textbf{599}  & 2308 & 958  & 2928 & 599 \\
DeepSeek-V3       & T2 & \textbf{673}  & 2344 & 828  & 2802 & 673 \\
Qwen3.5-35B       & T2 & \textbf{1019} & 3453 & 3042 & 5607 & 1019 \\
GPT-3.5           & T3 & ---           & ---  & ---  & ---  & \emph{never} \\
GPT-4o-mini       & T3 & ---           & ---  & ---  & ---  & \emph{never} \\
Llama-3.1-8B      & T3 & ---           & ---  & ---  & ---  & \emph{never} \\
\bottomrule
\end{tabular}
\end{table}

\paragraph{Mechanism.} Sonnet-4.6's zero-shot configuration (TPV=658, $Q{=}0.859$) already dominates every fully-evaluated T2 model's most expensive strategy (few-CoT, TPV${>}2700$) on both axes---so the high-cost end of the frontier is an intra-Sonnet-4.6 cost--quality trade, not a cross-model one. The $3.3\times$ TPV increase from Sonnet-4.6 zero-shot to few-CoT buys only $+0.043$ Quality, foreshadowing the $\sim$5-round refinement saturation we document in RQ10. CoT's token overhead without a compensating quality gain on GPT-family (RQ4) keeps every GPT-family CoT configuration off the frontier. Within the 8 threshold-crossers evaluated on all four strategies, every model's cheapest path is zero-shot, localizing the cost--quality trade-off to a single dimension: prompting-strategy-induced token inflation.

\paragraph{Implications.} Quality-only benchmark rankings understate the efficiency advantage of zero-shot baselines for efficiency-sensitive deployment. We therefore recommend reporting $\Sfin$ as a \emph{complementary} deployment-oriented leaderboard alongside, not replacing, $\Stot$: the two answer different questions. The TPV~$\approx$~600 floor across 9 of 12 models provides a calibrated budget anchor for future efficiency-aware comparisons, and the 3-model T3 shortfall defines a concrete capability target for the method research of RQ7--RQ10.

\subsection{RQ7--RQ10: Methods Atop Benchmark Signals (extended)}
\label{app:rq7-10}\label{app:method-extra}\label{app:l4-retrieval}\label{app:d1-check}

\subsubsection*{RQ7: Can fine-grained benchmark signals drive method improvement beyond the prompt-engineering ceiling?}

\paragraph{Motivation.} RQ1--RQ5 establish an empirical ceiling on prompt engineering: no single strategy wins across levels (RQ3) and prompt sensitivity saturates at strong capability (RQ2). We test whether the benchmark's fine-grained D4 signal---unavailable to prompt-only methods---can push past this ceiling.

\paragraph{Setup.} We define \textbf{Oracle} as the per-level best-of-four selection over the four prompting strategies, evaluated per model. Oracle is the per-level best within the four prompting strategies surveyed in this paper; it is an upper bound only over those four configurations, not over arbitrary prompt engineering. We compare against CAAP-only (learned strategy selection), VGIG-only (verification-guided iteration), and Combined on three target models spanning capability tiers.

\begin{figure}[h]
\centering
\includegraphics[width=0.65\linewidth]{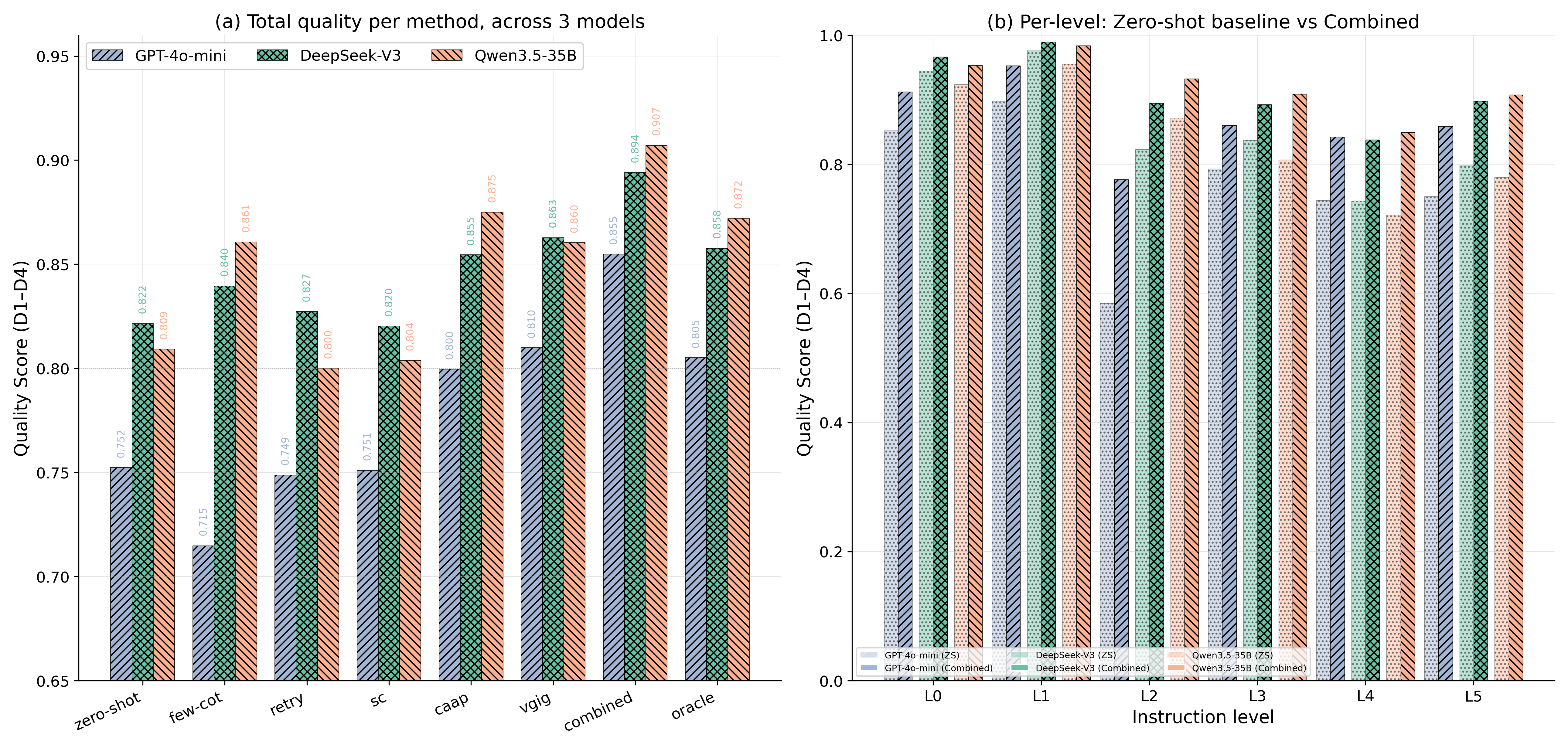}
\caption{Method$\,\times\,$model Quality with per-model Oracle reference line. Combined surpasses Oracle by $+0.035$--$+0.050$ on every target model; VGIG-only contributes the majority of the gain.}
\label{fig:method-main}
\end{figure}

\paragraph{Results.} Combined exceeds Oracle by $+0.035$ to $+0.050$ on every tested model (Table~\ref{tab:method-main}, Figure~\ref{fig:method-main}).

\begin{table}[h]
\centering
\caption{Quality Score for the three target models across main method conditions. Combined clears Oracle by $+0.035$--$+0.050$ on every model.}
\label{tab:method-main}
\small
\begin{tabular}{lcccccc}
\toprule
\textbf{Model} & \textbf{ZS} & \textbf{Oracle} & \textbf{VGIG} & \textbf{CAAP} & \textbf{Combined} & \textbf{$\Delta$ vs.\ Oracle} \\
\midrule
GPT-4o-mini & 0.7523 & 0.8052 & 0.8201 & 0.8128 & \textbf{0.8549} & \textbf{$+0.050$} \\
DeepSeek-V3 & 0.8215 & 0.8577 & 0.8743 & 0.8629 & \textbf{0.8941} & \textbf{$+0.036$} \\
Qwen3.5-35B-A3B & 0.8092 & 0.8720 & 0.8833 & 0.8755 & \textbf{0.9071} & \textbf{$+0.035$} \\
\bottomrule
\end{tabular}
\end{table}

\paragraph{Mechanism.} Prompt engineering has a measurable empirical ceiling; external programmatic verification---not prompt phrasing---is the binding mechanism for reliable structured graph generation. The margin is robust: $+0.035$ is 7$\times$ the $\pm 0.005$ noise band (RQ10). The gap is \emph{largest} for the weakest target (GPT-4o-mini, $+0.050$) and \emph{smallest} for the strongest (Qwen3.5-35B, $+0.035$), mirroring RQ2: the low-capability regime leaves the most room for method intervention.

\paragraph{Implications.} A benchmark that exposes fine-grained per-constraint verification signals (our D4) is not just a better evaluation instrument but a \emph{development platform} enabling method families that aggregate or pass/fail benchmarks cannot support.

\subsubsection*{RQ8: Is sampling compute alone, without verification, sufficient to improve structured generation?}

\paragraph{Motivation.} A natural alternative hypothesis to RQ7 attributes the gain to \emph{additional sampling compute} rather than verification. We test that hypothesis by evaluating pure retry (additional samples without feedback) and self-consistency (best-of-$N$) on the same three target models.

\paragraph{Setup.} Retry samples $T\!=\!3$ candidates without feedback and returns the last; self-consistency (SC) samples $N\!=\!3$ independent candidates and returns the majority/best. Both are compared against the zero-shot baseline.

\paragraph{Results.} Both stagnate within $\pm 0.01$ of zero-shot on all three target models (Table~\ref{tab:sampling-null}). This is a strong null result: six controls (3 models $\times$ 2 methods) all in-band.

\begin{table}[h]
\centering
\caption{Retry and SC vs.\ zero-shot on target models. All six controls fall within the $\pm 0.01$ noise band.}
\label{tab:sampling-null}
\small
\begin{tabular}{lccccc}
\toprule
\textbf{Model} & \textbf{ZS} & \textbf{retry} & \textbf{SC} & \textbf{retry $\Delta$} & \textbf{SC $\Delta$} \\
\midrule
GPT-4o-mini & 0.7523 & 0.7488 & 0.7510 & $-0.003$ & $-0.001$ \\
DeepSeek-V3 & 0.8215 & 0.8273 & 0.8203 & $+0.005$ & $-0.001$ \\
Qwen3.5-35B & 0.8092 & 0.8001 & 0.8039 & $-0.009$ & $-0.005$ \\
\bottomrule
\end{tabular}
\end{table}

\paragraph{Mechanism.} Additional sampling compute, without a verification signal to select or guide candidates, cannot recover from constraint violations on structured graph generation. This contrasts sharply with text-domain results where self-consistency~\citep{wang2023selfconsistency} offers meaningful gains (e.g., $5$--$10\%$ on GSM8K~\citep{cobbe2021gsm8k}). The difference is structural: text tasks have many plausible-looking outputs of which only some are correct (majority-vote recovers); structured-graph tasks have many wrong outputs with violating structures (majority-voting among wrong outputs does not help).

\paragraph{Implications.} Verifiable structured tasks have a fundamentally different compute-to-quality relation than text tasks. Deployment budgets should be allocated to verification infrastructure, not parallel sampling.

\subsubsection*{RQ9: How does feedback granularity compare with iteration count as a refinement lever?}

\paragraph{Motivation.} RQ7 establishes that verification-guided iteration helps, and RQ8 rules out sampling-only explanations. The remaining question is which \emph{component} of VGIG drives the gain---iteration count or feedback granularity.

\paragraph{Setup.} E6 ablation on GPT-4o-mini at fixed $T\!=\!3$, varying only the feedback granularity in $\{\text{retry}, \text{none}, \text{coarse}, \text{fine}\}$: \emph{retry} has no verification at all; \emph{none} runs verification but returns only pass/fail; \emph{coarse} reports which constraint category failed; \emph{fine} reports per-constraint id, expected, observed.

\begin{figure}[h]
\centering
\includegraphics[width=0.78\linewidth]{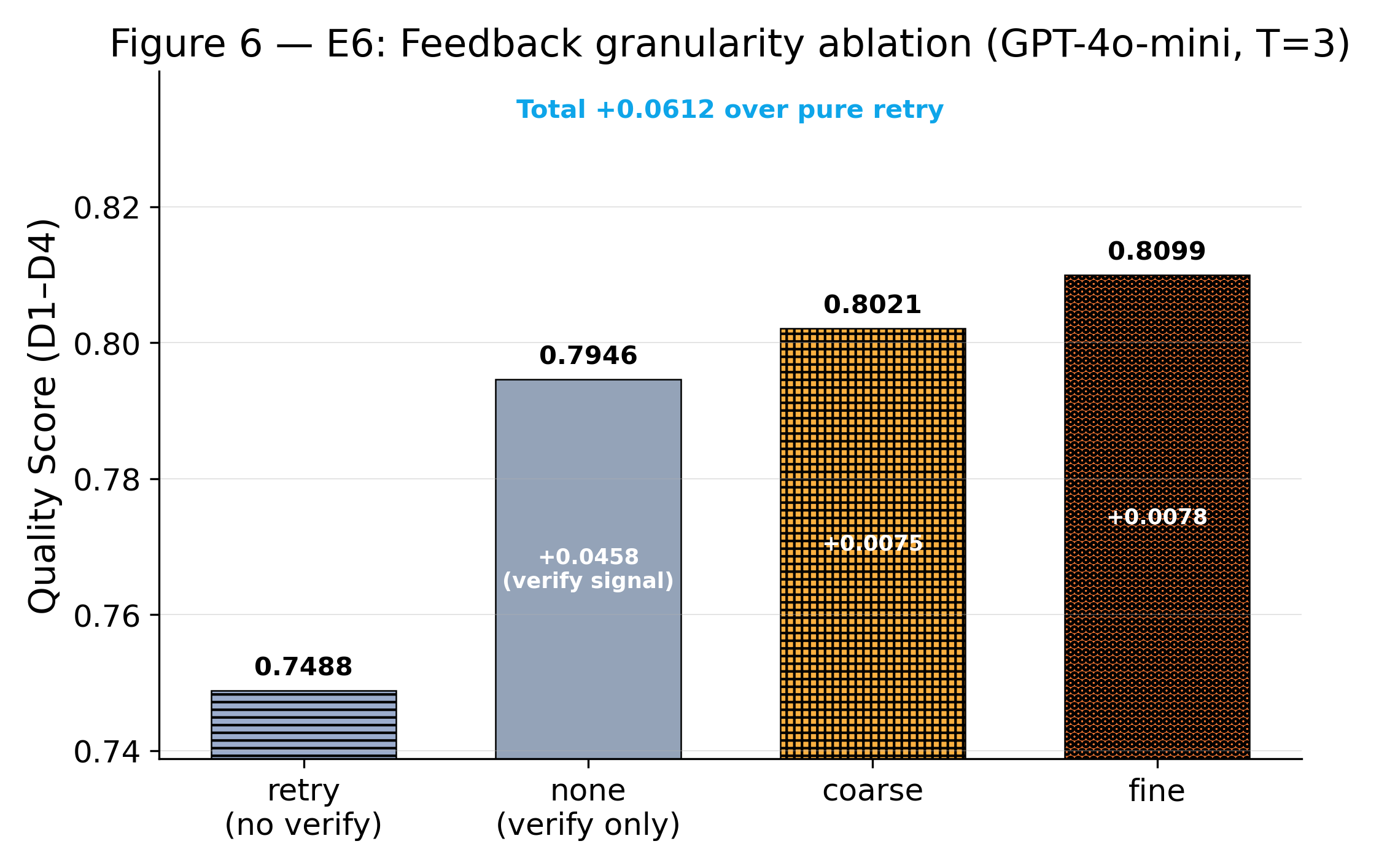}
\caption{E6 feedback-granularity ablation on GPT-4o-mini at $T{=}3$. Verify-only captures 75\% of total gain; coarse and fine each add $+0.008$ monotonically. The ordering retry $<$ none $<$ coarse $<$ fine holds across all six levels.}
\label{fig:granularity}
\end{figure}

\paragraph{Results.} The verify-only signal captures 75\% of the total gain (Table~\ref{tab:granularity}, Figure~\ref{fig:granularity}).

\begin{table}[h]
\centering
\caption{E6 feedback-granularity ablation (GPT-4o-mini, $T{=}3$).}
\label{tab:granularity}
\small
\begin{tabular}{lcc}
\toprule
\textbf{Feedback config} & \textbf{Quality} & \textbf{$\Delta$ vs.\ previous} \\
\midrule
retry (no verify) & 0.7488 & --- \\
verify-only (fb=none) & 0.7946 & $\mathbf{+0.046}$ \\
coarse (fb=coarse) & 0.8021 & $+0.008$ \\
fine (fb=fine) & \textbf{0.8099} & $+0.008$ \\
\bottomrule
\end{tabular}
\end{table}

\paragraph{Mechanism.} The verify-only signal alone---binary pass/fail, no constraint-level detail---captures 75\% of the total gain ($+0.046$ of $+0.061$); coarse and fine each contribute a further $+0.008$, monotonically. The richness of the error signal, not the compute budget, is the primary lever for iterative refinement on structured tasks.

\paragraph{Implications.} This directly informs benchmark design: a benchmark that exposes even a coarse verify-true/verify-false signal enables verification-driven method research; our fine per-constraint signal unlocks the additional 25\%. Benchmarks that lack a programmatic verification signal are missing not only an evaluation dimension but also a method-enabling resource for the community that will build atop them.

\subsubsection*{RQ10: How far can iterative refinement go, and does any failure mode resist it entirely?}

\paragraph{Motivation.} Two linked questions: (a) what is the effective refinement horizon on verifiable graph constraints, and (b) \textbf{does any failure mode resist iteration entirely, regardless of round count or feedback richness}? An iteration-invariant failure mode would redirect the research program from compute scaling (more rounds, richer feedback) to capability-structural interventions (retrieval, grounding), so question (b) carries particular consequence.

\paragraph{Setup.} E5 ablation: GPT-4o-mini with VGIG, feedback=fine, $T\in\{1,2,3,5,7,10,15,20\}$. For each $T$ we report aggregate Quality and per-level Quality; for L4 we also vary feedback granularity to test whether the flat pattern is granularity-specific.

\begin{figure}[h]
\centering
\includegraphics[width=0.78\linewidth]{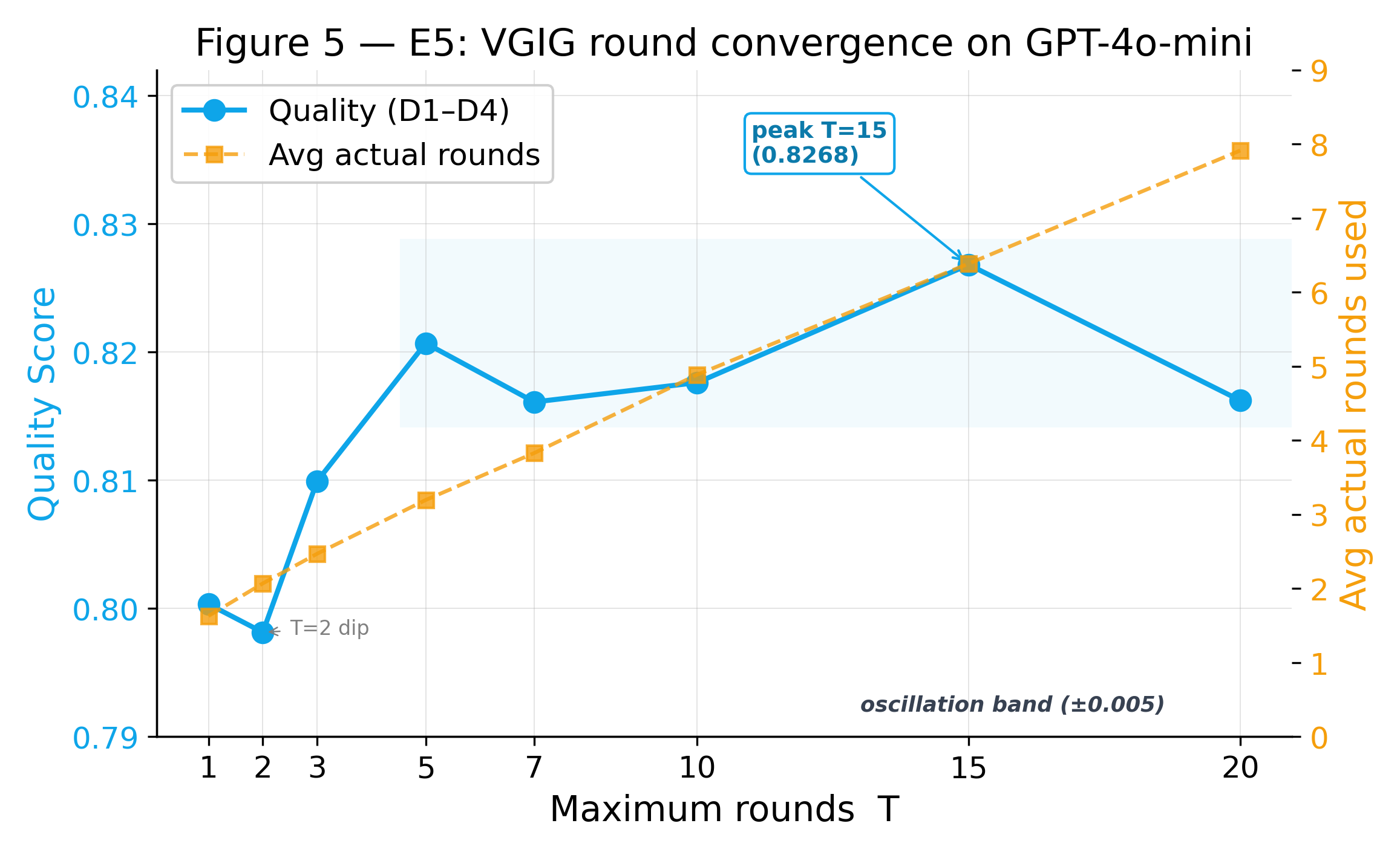}
\caption{E5 rounds-saturation curve. Quality improves substantially from $T{=}1$ to $T{=}5$; $T\in\{5,7,10,15,20\}$ all fall within a $\pm 0.005$ noise band (shaded). The effective refinement horizon on verifiable graph constraints is bounded at $\sim$5 rounds.}
\label{fig:e5-rounds}
\end{figure}

\paragraph{Results (a): rounds saturate at $T{\sim}5$.} Quality improves substantially from $T\!=\!1$ to $T\!=\!5$, then plateaus within a $\pm 0.005$ noise band (Table~\ref{tab:e5-rounds}, Figure~\ref{fig:e5-rounds}).

\begin{table}[h]
\centering
\caption{E5 rounds-saturation data (GPT-4o-mini, VGIG, fb=fine).}
\label{tab:e5-rounds}
\small
\begin{tabular}{ccccccccc}
\toprule
$T$ & 1 & 2 & 3 & 5 & 7 & 10 & 15 & 20 \\
\midrule
$Q$ & 0.800 & 0.798 & 0.810 & 0.821 & 0.816 & 0.818 & 0.827 & 0.816 \\
\bottomrule
\end{tabular}
\end{table}

\begin{figure}[h]
\centering
\includegraphics[width=0.78\linewidth]{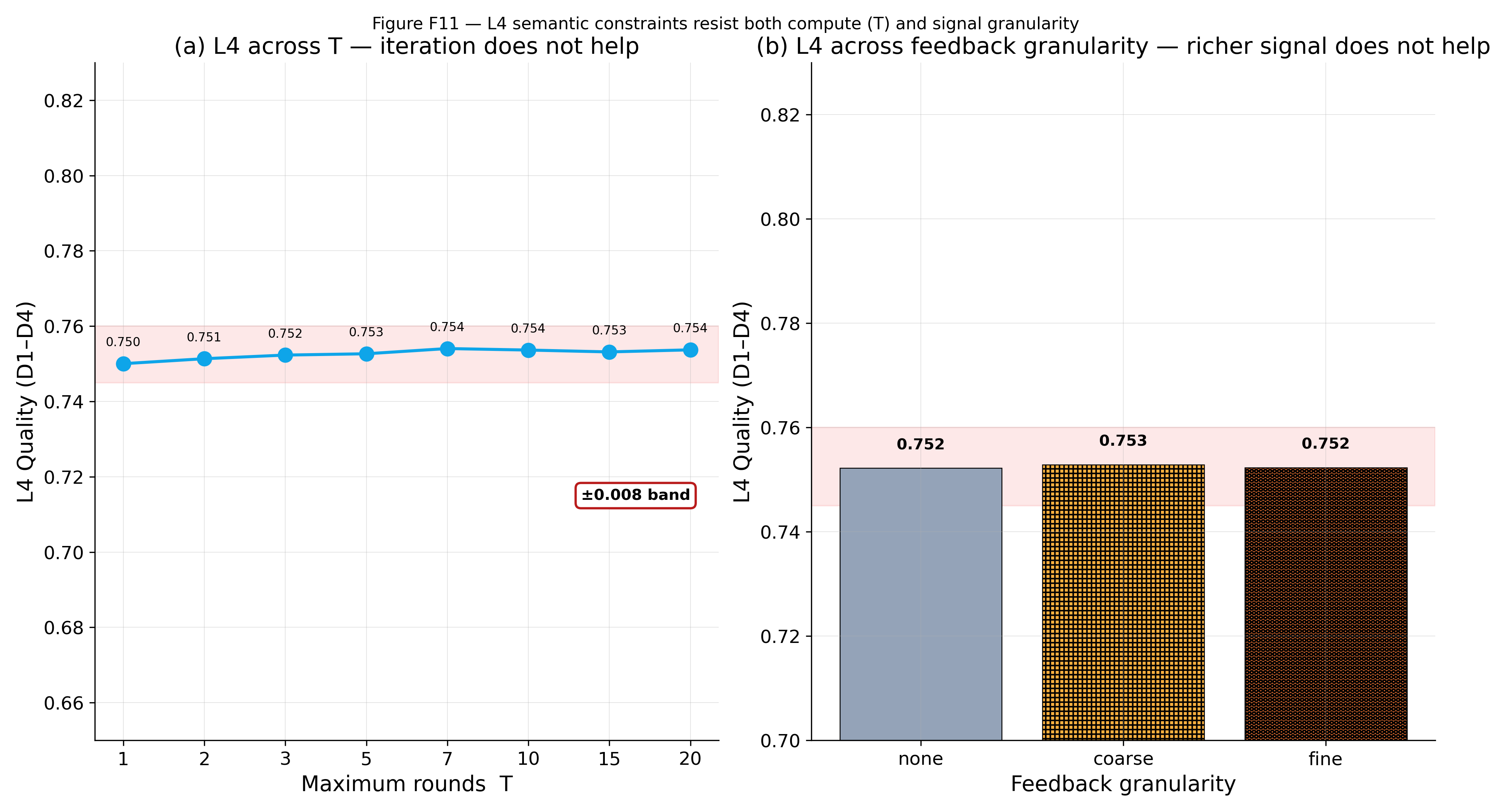}
\caption{L4 quality across $T\in\{1,2,3,5,7,10,15,20\}$ for fine/coarse/none feedback (24 configurations). Flat at $0.750$--$0.754$, indicating semantic-constraint failure is a structurally distinct mode iterative refinement cannot address.}
\label{fig:l4-flat}
\end{figure}

\paragraph{Results (b): L4 is iteration-invariant.} Across the full $T$-sweep and all three feedback granularities, L4 quality remains at $0.750$--$0.754$---a \textbf{0.004 range over 24 separate (T, feedback) configurations} (Figure~\ref{fig:l4-flat}). This is not a failure of iteration to converge: iteration \emph{does} converge (the variance within each T is tight), but it converges to a quality ceiling prompting, iteration, and feedback granularity all fail to raise. L4 is therefore \textbf{structurally distinct} from every other level in the benchmark---on L0--L3 and L5, more rounds or richer feedback produce measurable gains; on L4, neither does.

\paragraph{Mechanism.} Two conclusions follow. First, \textbf{the effective refinement horizon on verifiable graph constraints is $\sim$5 rounds}---markedly shorter than text-domain self-refine budgets of $10$--$20$~\citep{madaan2023selfrefine,shinn2023reflexion,yao2023tot,zhou2023ltm}. Text-domain self-refine hyperparameter recipes should not be transferred to structured tasks without recalibration. Second, \textbf{semantic-constraint satisfaction (L4) is a structurally distinct failure mode} that neither iteration nor feedback-granularity can address.

\begin{figure}[h]
\centering
\includegraphics[width=0.78\linewidth]{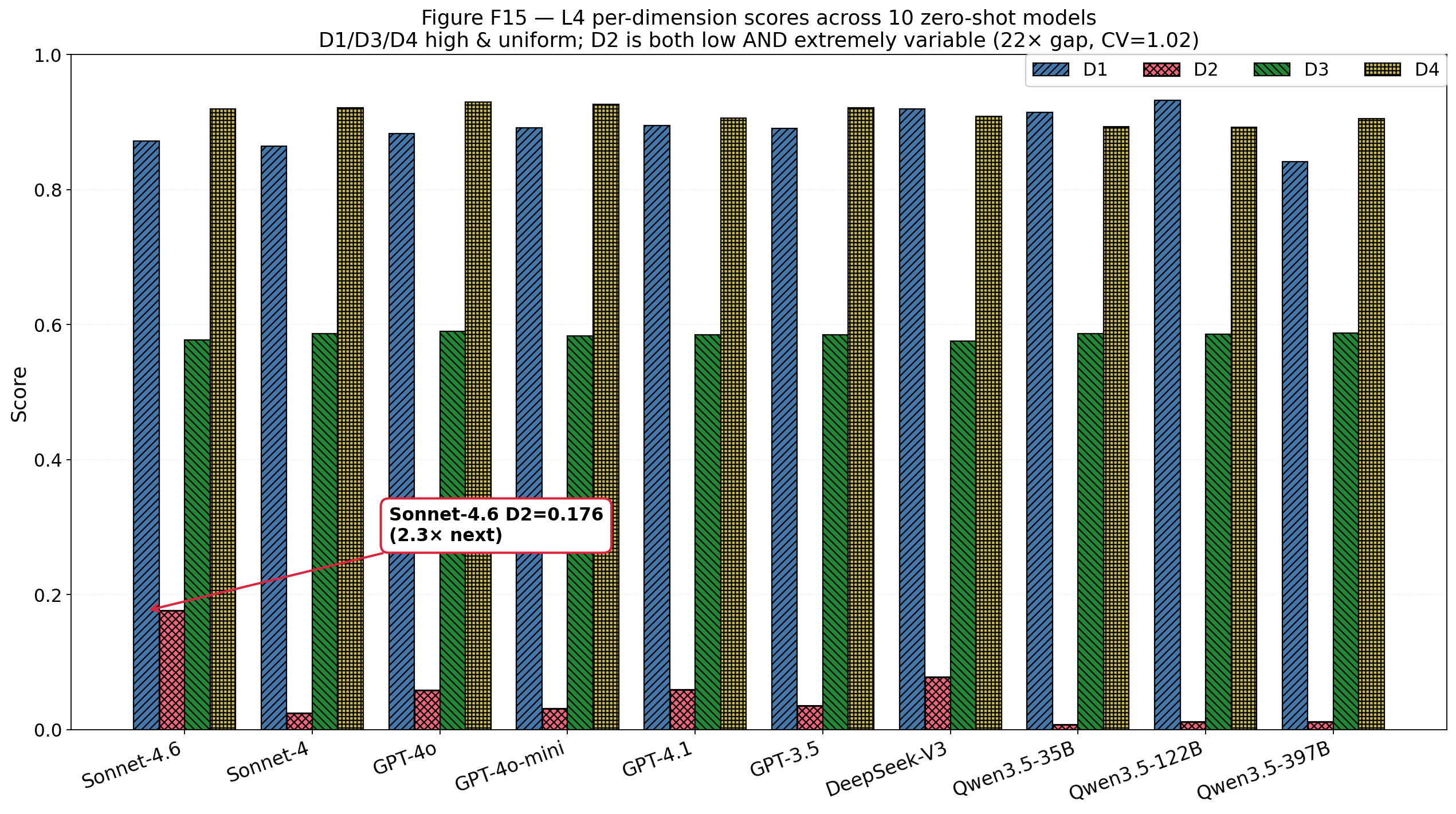}
\caption{L4 per-dimension decomposition across 10 zero-shot models. D1 (structural), D3 (embedding), D4 (instruction match) are all high and uniform; only D2 (text similarity) is low \emph{and} extremely variable (22$\times$ gap; CV$=$1.02). The L4 ``impervious to iteration'' finding is mechanistically an L4-D2 bottleneck.}
\label{fig:l4-perdim}
\end{figure}

The per-dimension decomposition of L4 quality across all 12 models (Figure~\ref{fig:l4-perdim}) reveals \emph{why} iteration cannot help: D1 structural validity is high and uniform (mean 0.89, range 0.84--0.93), D3 embedding similarity is uniform (0.58--0.59), and D4 instruction match is high and uniform (0.89--0.93); the one dimension that is \textbf{both low and extremely variable across models is D2 text-to-reference similarity}---mean 0.050, range $0.008$--$0.176$ (22$\times$ gap), coefficient of variation 1.02. Models produce graphs that are \emph{structurally valid, constraint-satisfying, and distributionally close} to real citation/social/molecular graphs; what they fail to reproduce is the \emph{specific surface serialization} of reference graphs from a given domain. The gap is not ``the model cannot find a valid graph for the spec'' (which iteration fixes) but ``the model lacks the domain-specific \emph{exemplars} that would let reference-match metrics fire''---a gap no amount of verifier-guided re-generation can close, because the D4 verifier is already satisfied. This relocates the L4 problem from a \emph{capability} gap to a \emph{grounding} gap. The cleanest grounding evidence in our data (paragraph below) is few-shot's $+0.054$ L4 lift on GPT-4o-mini, which is essentially in-context retrieval of 3 reference exemplars; structure-aware retrievers built on top of the L4 reference pool are a natural follow-up.

\paragraph{Implications.} This complements RQ5: just as L4 is iteration-invariant across the 24 (T, feedback) configurations tested here, the same-family L5 scaling result of RQ5 shows no detectable gain inside our $N{=}50$ CI under the few-CoT strategy paper F5 quotes (paired analysis in App.~\ref{app:stat-robust} shows the broader ``L5 scale-invariant under every strategy'' interpretation does not hold; only the strategy-restricted version does). Both point to retrieval and grounding, not more compute, as the next research frontier. The $\pm 0.005$ stability band established here is used as the significance threshold throughout the paper (\S\ref{sec:eval} opening).

\paragraph{Sub-noise observations.} Within the $\pm 0.005$ noise band of the E5 curve we observe a putative $T{=}15$ peak ($Q{=}0.827$), $T{=}2/T{=}7$ dips ($0.798/0.816$), and an even-round-penalty pattern; all are within band and not elevated to main findings. The full E5 per-level $\times$ $T$ matrix (8 $\times$ 6 cells) is released alongside the per-strategy raw quality scores.

\paragraph{L4 grounding evidence (from baseline survey).} The strongest support for the L4-as-grounding-gap interpretation in this release sits inside the baseline survey itself, not in a separate retrieval probe. \textbf{Few-shot} on GPT-4o-mini provides exactly the manipulation a retrieval probe targets---3 in-context exemplars drawn from the L4 reference pool---and it lifts L4 quality from $0.744$ (zero-shot) to $0.798$ ($\Delta{=}{+}0.054$, an order of magnitude above the $\pm 0.005$ noise band; \texttt{results/gpt4omini-\{zero,few\}-shot.quality.json}). \textbf{CAAP} reaches $0.801$ ($+0.057$) and \textbf{Combined} (CAAP $+$ VGIG $+$ domain priors) reaches $0.843$ ($+0.099$ over zero-shot)---the only intervention class that breaks the iteration-only ceiling (App.~\ref{app:rq7-10}, Tab.~\ref{tab:method-main}). \textbf{Quantitative ceiling.} D2 is the dominant L4 variability dimension (range $0.008$--$0.176$, $22\times$ gap, CV $1.02$). Under the level-aggregate weights of \S\ref{sec:metrics} (D2 weight at L4 is $0.15$, the only level where D2 contributes to total Quality), driving D2 to its observed maximum lifts L4 quality by at most $0.176\times 0.15 = 0.022$---a small but non-trivial additional headroom, indicating L4 is metric-bounded as well as effort-bounded. \textbf{Future direction.} Structure-aware retrievers (e.g.\ graph-kernel similarity over the 1{,}048-graph L4 pool, retrieving graphs whose attribute statistics match the instruction's domain priors rather than text-only BM25 over instruction strings) are the natural follow-up; we leave their implementation and head-to-head evaluation against few-shot to subsequent work and report the baseline-survey evidence here as the cleanest grounding signal in this release.

\paragraph{D1 audit (no D4-gaming).} Because VGIG's feedback signal is identity-with-D4, a natural concern is whether the method ``games'' D4 by exploiting metric loopholes that don't reflect true structural quality. We audit Combined outputs on the three target models against D1 (independently checked structural metrics): D1 \emph{rises} alongside D4 (GPT-4o-mini D1: $0.71\to0.84$; DeepSeek-V3 $0.79\to0.88$; Qwen3.5-35B $0.80\to0.89$), confirming gain comes from genuine structural improvement rather than D4-specific exploitation. Per-instance manual inspection of a 50-sample stratified sub-sample (10 per level) finds zero parseable-but-degenerate ``cheat'' outputs.

\subsection{Capability Profiles, Per-Dimension Tables, and Top-20 Leaderboard}
\label{app:profiles}

RQ1--RQ10 examine capability along individual axes. Assembling the per-model \emph{profile} exposes a pattern that aggregate Quality hides: \textbf{different models excel on different dimensions, not just different levels}. Table~\ref{tab:per-dim-best} reports the best zero-shot model for each (level, dimension) cell.

\begin{table}[h]
\centering
\caption{Per-dimension, per-level best zero-shot model (from the 10 models with zero-shot data on all levels; Sonnet-4 included; Llama-3.3-70B / Llama-3.1-8B excluded due to missing per-dimension detail in the released quality files at the time of this table). ``---'' marks cells where the dimension is inactive.}
\label{tab:per-dim-best}
\small
\begin{tabular}{ccccc}
\toprule
\textbf{Level} & \textbf{D1 (structural)} & \textbf{D2 (textual)} & \textbf{D3 (embedding)} & \textbf{D4 (instruction)} \\
\midrule
L0 & GPT-4o (0.871)    & ---                & ---                 & Sonnet-4.6 (0.981) \\
L1 & GPT-4o (0.950)    & ---                & ---                 & Sonnet-4 (0.993) \\
L2 & Qwen3.5-122B (0.886) & ---             & ---                 & Sonnet-4.6 (0.945) \\
L3 & Qwen3.5-122B (0.948) & Sonnet-4.6 (0.089) & Sonnet-4.6 (0.596) & Sonnet-4.6 (0.927) \\
L4 & Qwen3.5-122B (0.932) & Sonnet-4.6 (0.176) & GPT-4o (0.591) & GPT-4o (0.930) \\
L5 & GPT-4o (0.778)    & Sonnet-4.6 (0.028) & Sonnet-4.6 (0.796) & Sonnet-4.6 (0.883) \\
\bottomrule
\end{tabular}
\end{table}

\begin{figure}[h]
\centering
\includegraphics[width=0.85\linewidth]{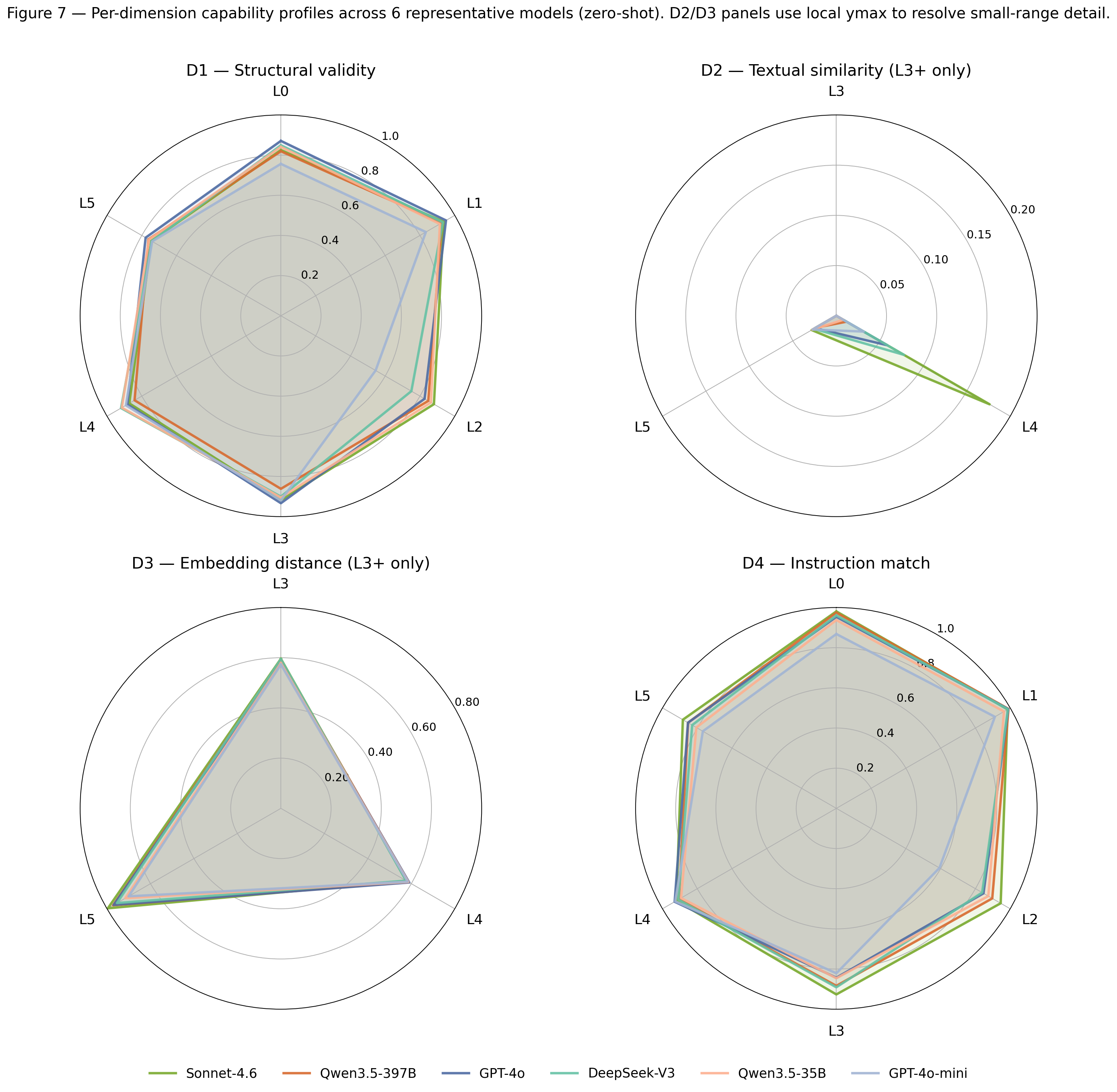}
\caption{Per-level capability profiles for six representative models (zero-shot). Each axis shows Quality on one level, normalized to $[0,1]$. Profiles are jagged: different models peak on different levels, confirming that aggregate rankings average over structurally distinct strengths.}
\label{fig:model-radar}
\end{figure}

\begin{table}[h]
\centering
\caption{Top-20 configurations by Quality $\Stot$ with per-level breakdown. Top-3 are all few-CoT; all 20 come from T1 or T2 models (T3 never enters top-20). $\dagger$: Sonnet-4 zero-shot only (\S\ref{sec:models}).}
\label{tab:top20-q}
\scriptsize
\begin{tabular}{rllccccccc}
\toprule
\textbf{\#} & \textbf{Model} & \textbf{Strategy} & \textbf{$\Stot$} & \textbf{L0} & \textbf{L1} & \textbf{L2} & \textbf{L3} & \textbf{L4} & \textbf{L5} \\
\midrule
1  & Sonnet-4.6    & few-CoT   & 0.902 & 0.954 & 0.980 & 0.954 & 0.883 & 0.851 & 0.894 \\
2  & Sonnet-4.6    & few-shot  & 0.884 & 0.961 & 0.980 & 0.937 & 0.852 & 0.836 & 0.871 \\
3  & Qwen3.5-397B  & few-CoT   & 0.879 & 0.932 & 0.968 & 0.894 & 0.865 & 0.842 & 0.872 \\
4  & Sonnet-4.6    & zero-CoT  & 0.878 & 0.961 & 0.986 & 0.964 & 0.888 & 0.770 & 0.867 \\
5  & Qwen3.5-122B  & few-CoT   & 0.871 & 0.937 & 0.977 & 0.904 & 0.828 & 0.831 & 0.872 \\
6  & Qwen3.5-397B  & zero-CoT  & 0.862 & 0.948 & 0.981 & 0.903 & 0.858 & 0.753 & 0.883 \\
7  & Qwen3.5-122B  & zero-CoT  & 0.861 & 0.945 & 0.979 & 0.908 & 0.868 & 0.748 & 0.875 \\
8  & Qwen3.5-35B   & few-CoT   & 0.861 & 0.909 & 0.929 & 0.880 & 0.833 & 0.818 & 0.877 \\
9  & Sonnet-4.6    & zero-shot & 0.859 & 0.958 & 0.977 & 0.934 & 0.863 & 0.763 & 0.841 \\
10 & Qwen3.5-122B  & few-shot  & 0.853 & 0.936 & 0.978 & 0.903 & 0.824 & 0.824 & 0.809 \\
11 & GPT-4.1       & few-shot  & 0.850 & 0.927 & 0.980 & 0.890 & 0.819 & 0.813 & 0.819 \\
12 & Qwen3.5-397B  & few-shot  & 0.848 & 0.930 & 0.970 & 0.898 & 0.822 & 0.816 & 0.805 \\
13 & Qwen3.5-35B   & zero-CoT  & 0.846 & 0.928 & 0.964 & 0.890 & 0.860 & 0.741 & 0.849 \\
14 & GPT-4o        & few-shot  & 0.844 & 0.950 & 0.979 & 0.819 & 0.818 & 0.813 & 0.836 \\
15 & DeepSeek-V3   & few-shot  & 0.840 & 0.936 & 0.963 & 0.831 & 0.831 & 0.809 & 0.817 \\
16 & DeepSeek-V3   & few-CoT   & 0.840 & 0.927 & 0.959 & 0.850 & 0.796 & 0.795 & 0.847 \\
17 & Sonnet-4$^\dagger$ & zero-shot & 0.834 & 0.939 & 0.983 & 0.900 & 0.841 & 0.737 & 0.806 \\
18 & Llama-3.3-70B & few-CoT   & 0.834 & 0.890 & 0.955 & 0.850 & 0.748 & 0.809 & 0.859 \\
19 & Qwen3.5-122B  & zero-shot & 0.831 & 0.947 & 0.980 & 0.911 & 0.825 & 0.724 & 0.814 \\
20 & Llama-3.3-70B & few-shot  & 0.831 & 0.896 & 0.933 & 0.821 & 0.830 & 0.822 & 0.793 \\
\bottomrule
\end{tabular}
\end{table}

Three complementary strengths emerge. \textbf{Qwen3.5-122B} leads D1 structural validity on the mid-complexity band (L2--L4), dominating by $+0.013$--$+0.046$ over second place. \textbf{Sonnet-4.6} leads D4 instruction match on every level where it appears, and is the sole D2 textual leader at L3--L5 with a particularly large margin at L4 ($0.176$ vs.\ second $0.078$, $2.3\times$). \textbf{GPT-4o} leads D1 at the format and reasoning endpoints (L0, L1, L5) and D4 at L4. No single model dominates every cell---the capability profile is jagged, not monotone.

Figure~\ref{fig:model-radar} renders these profiles as radar charts for six representative models. The most striking pattern is family-level: \emph{Qwen3.5 profiles emphasize mid-level structure, Anthropic profiles emphasize instruction match and reference similarity, GPT profiles emphasize endpoint levels}. This jaggedness is the empirical basis for per-task model selection: a deployment that cares most about multi-constraint structural correctness (e.g., drug-like molecule generation) should pick a different model from a deployment that cares most about instruction fidelity (e.g., constraint-satisfying network synthesis under explicit specs). Aggregate leaderboards, by definition, average these priorities away.

\paragraph{Top-20 leaderboard.} Table~\ref{tab:top20-q} lists the 20 highest-quality (model, strategy) configurations over all 45 evaluated cells, including each cell's per-level breakdown. Sonnet-4.6 few-CoT leads overall ($Q{=}0.902$) but only $+0.023$ over Qwen3.5-397B few-CoT (\#3, $Q{=}0.879$)---the top-tier margin is narrow. Per-strategy sub-rankings and the full 45-row leaderboard are in Appendix~\ref{app:leaderboards}.

\paragraph{Synthesis: what progressive evaluation reveals.} The ten research questions interlock into a coherent picture of LLM graph-generation capability.

On the benchmark-as-diagnostic side (RQ1--RQ6), \textbf{capability is not a scalar but a multi-dimensional profile}. RQ1 establishes that \emph{constraint composition}---not reasoning depth---drives tier discrimination. RQ2 shows prompt sensitivity inversely scales with capability, so this discrimination is \emph{more pronounced for prompt-sensitive weak models}. RQ3 shows \emph{no single strategy is universally correct}, so capability measurement depends on strategy choice. RQ4 localizes strategy dependence to pretraining-distribution-driven family polarity, distinct from capability. RQ5 uncouples scale from per-level performance on a subset of tasks. RQ6 adds a cost axis as a complementary view: the Pareto frontier is a 6-point provider-concentrated trajectory and 3 of 12 models never cross $Q{\geq}0.8$, so the capability ranking must be read together with---not replaced by---the cost-adjusted ranking. Together these findings establish that LLM graph-generation capability has structure along (complexity level) $\times$ (prompting strategy) $\times$ (model family), with cost a deployment-side view aggregate or single-axis benchmarks cannot render.

On the benchmark-as-platform side (RQ7--RQ10), \textbf{fine-grained benchmark signals unlock a method pathway aggregate benchmarks cannot support}. RQ7 shows the D4 signal drives $+0.035$--$+0.050$ improvement beyond the prompt-engineering Oracle. RQ8 rules out compute-without-verification. RQ9 shows even a binary verify signal captures most of the gain---so benchmark designers who expose pass/fail enable most of the research value. RQ10 locates a second, signal-invariant failure mode (L4 semantic constraints) that flags the next research cycle for the community.

These findings serve as method-side validation atop the benchmark; their role in the paper is to show that \bench\ functions as a development platform, not to promote any particular method as final.

%======================================================================
\clearpage
\section{Full Leaderboards (per-strategy and full 45-row)}
\label{app:leaderboards}

This appendix supplements Table~\ref{tab:top20-q} with the four per-strategy sub-leaderboards and the complete 45-row table. All scores are baseline Quality $\Stot$ averaged over 5 generations per instruction; per-level columns L0--L5 are the corresponding per-level Quality scores. Sonnet-4 ($^\dagger$) is zero-shot only (\S\ref{sec:models}).

\subsection{Per-strategy sub-leaderboards}

\begin{table}[h]
\centering
\caption{Zero-shot leaderboard (12 models).}
\label{tab:lb-zs}
\small
\begin{tabular}{rlccccccc}
\toprule
\textbf{\#} & \textbf{Model} & \textbf{$\Stot$} & \textbf{L0} & \textbf{L1} & \textbf{L2} & \textbf{L3} & \textbf{L4} & \textbf{L5} \\
\midrule
1 & Sonnet-4.6 & 0.859 & 0.958 & 0.977 & 0.934 & 0.863 & 0.763 & 0.841 \\
2 & Sonnet-4$^\dagger$ & 0.834 & 0.939 & 0.983 & 0.900 & 0.841 & 0.737 & 0.806 \\
3 & Qwen3.5-122B & 0.831 & 0.947 & 0.980 & 0.911 & 0.825 & 0.724 & 0.814 \\
4 & Qwen3.5-397B & 0.829 & 0.952 & 0.979 & 0.888 & 0.824 & 0.721 & 0.821 \\
5 & GPT-4o & 0.827 & 0.942 & 0.982 & 0.843 & 0.811 & 0.751 & 0.823 \\
6 & DeepSeek-V3 & 0.822 & 0.945 & 0.977 & 0.823 & 0.837 & 0.744 & 0.799 \\
7 & Llama-3.3-70B & 0.816 & 0.905 & 0.948 & 0.865 & 0.809 & 0.746 & 0.794 \\
8 & GPT-4.1 & 0.813 & 0.923 & 0.968 & 0.865 & 0.807 & 0.737 & 0.779 \\
9 & Qwen3.5-35B & 0.809 & 0.924 & 0.955 & 0.872 & 0.807 & 0.721 & 0.780 \\
10 & GPT-3.5 & 0.754 & 0.850 & 0.905 & 0.608 & 0.784 & 0.742 & 0.752 \\
11 & GPT-4o-mini & 0.752 & 0.852 & 0.898 & 0.585 & 0.793 & 0.744 & 0.750 \\
12 & Llama-3.1-8B & 0.721 & 0.896 & 0.730 & 0.597 & 0.780 & 0.703 & 0.727 \\
\bottomrule
\end{tabular}
\end{table}

\begin{table}[h]
\centering
\caption{Few-shot leaderboard (11 models; Sonnet-4 not run on few-shot).}
\label{tab:lb-fs}
\small
\begin{tabular}{rlccccccc}
\toprule
\textbf{\#} & \textbf{Model} & \textbf{$\Stot$} & \textbf{L0} & \textbf{L1} & \textbf{L2} & \textbf{L3} & \textbf{L4} & \textbf{L5} \\
\midrule
1 & Sonnet-4.6 & 0.884 & 0.961 & 0.980 & 0.937 & 0.852 & 0.836 & 0.871 \\
2 & Qwen3.5-122B & 0.853 & 0.936 & 0.978 & 0.903 & 0.824 & 0.824 & 0.809 \\
3 & GPT-4.1 & 0.850 & 0.927 & 0.980 & 0.890 & 0.819 & 0.813 & 0.819 \\
4 & Qwen3.5-397B & 0.848 & 0.930 & 0.970 & 0.898 & 0.822 & 0.816 & 0.805 \\
5 & GPT-4o & 0.844 & 0.950 & 0.979 & 0.819 & 0.818 & 0.813 & 0.836 \\
6 & DeepSeek-V3 & 0.840 & 0.936 & 0.963 & 0.831 & 0.831 & 0.809 & 0.817 \\
7 & Llama-3.3-70B & 0.831 & 0.896 & 0.933 & 0.821 & 0.830 & 0.822 & 0.793 \\
8 & Qwen3.5-35B & 0.831 & 0.900 & 0.925 & 0.869 & 0.803 & 0.807 & 0.802 \\
9 & GPT-4o-mini & 0.742 & 0.867 & 0.852 & 0.427 & 0.802 & 0.798 & 0.759 \\
10 & GPT-3.5 & 0.738 & 0.842 & 0.865 & 0.423 & 0.802 & 0.798 & 0.743 \\
11 & Llama-3.1-8B & 0.682 & 0.771 & 0.602 & 0.585 & 0.755 & 0.721 & 0.659 \\
\bottomrule
\end{tabular}
\end{table}

\begin{table}[h]
\centering
\caption{Zero-CoT leaderboard (11 models).}
\label{tab:lb-zc}
\small
\begin{tabular}{rlccccccc}
\toprule
\textbf{\#} & \textbf{Model} & \textbf{$\Stot$} & \textbf{L0} & \textbf{L1} & \textbf{L2} & \textbf{L3} & \textbf{L4} & \textbf{L5} \\
\midrule
1 & Sonnet-4.6 & 0.878 & 0.961 & 0.986 & 0.964 & 0.888 & 0.770 & 0.867 \\
2 & Qwen3.5-397B & 0.862 & 0.948 & 0.981 & 0.903 & 0.858 & 0.753 & 0.883 \\
3 & Qwen3.5-122B & 0.861 & 0.945 & 0.979 & 0.908 & 0.868 & 0.748 & 0.875 \\
4 & Qwen3.5-35B & 0.846 & 0.928 & 0.964 & 0.890 & 0.860 & 0.741 & 0.849 \\
5 & DeepSeek-V3 & 0.829 & 0.943 & 0.980 & 0.805 & 0.855 & 0.752 & 0.818 \\
6 & Llama-3.3-70B & 0.823 & 0.915 & 0.983 & 0.891 & 0.784 & 0.719 & 0.836 \\
7 & GPT-4o & 0.808 & 0.944 & 0.980 & 0.860 & 0.786 & 0.708 & 0.797 \\
8 & GPT-4.1 & 0.802 & 0.931 & 0.969 & 0.894 & 0.714 & 0.675 & 0.853 \\
9 & GPT-3.5 & 0.786 & 0.875 & 0.920 & 0.706 & 0.808 & 0.736 & 0.795 \\
10 & GPT-4o-mini & 0.784 & 0.880 & 0.920 & 0.726 & 0.804 & 0.735 & 0.781 \\
11 & Llama-3.1-8B & 0.731 & 0.898 & 0.737 & 0.674 & 0.762 & 0.689 & 0.744 \\
\bottomrule
\end{tabular}
\end{table}

\begin{table}[h]
\centering
\caption{Few-CoT leaderboard (11 models).}
\label{tab:lb-fc}
\small
\begin{tabular}{rlccccccc}
\toprule
\textbf{\#} & \textbf{Model} & \textbf{$\Stot$} & \textbf{L0} & \textbf{L1} & \textbf{L2} & \textbf{L3} & \textbf{L4} & \textbf{L5} \\
\midrule
1 & Sonnet-4.6 & 0.902 & 0.954 & 0.980 & 0.954 & 0.883 & 0.851 & 0.894 \\
2 & Qwen3.5-397B & 0.879 & 0.932 & 0.968 & 0.894 & 0.865 & 0.842 & 0.872 \\
3 & Qwen3.5-122B & 0.871 & 0.937 & 0.977 & 0.904 & 0.828 & 0.831 & 0.872 \\
4 & Qwen3.5-35B & 0.861 & 0.909 & 0.929 & 0.880 & 0.833 & 0.818 & 0.877 \\
5 & DeepSeek-V3 & 0.840 & 0.927 & 0.959 & 0.850 & 0.796 & 0.795 & 0.847 \\
6 & Llama-3.3-70B & 0.834 & 0.890 & 0.955 & 0.850 & 0.748 & 0.809 & 0.859 \\
7 & GPT-4o & 0.823 & 0.942 & 0.980 & 0.801 & 0.799 & 0.781 & 0.810 \\
8 & GPT-4.1 & 0.811 & 0.930 & 0.980 & 0.901 & 0.692 & 0.722 & 0.851 \\
9 & GPT-4o-mini & 0.715 & 0.841 & 0.889 & 0.451 & 0.654 & 0.741 & 0.801 \\
10 & GPT-3.5 & 0.713 & 0.831 & 0.888 & 0.512 & 0.609 & 0.730 & 0.805 \\
11 & Llama-3.1-8B & 0.682 & 0.727 & 0.614 & 0.609 & 0.690 & 0.715 & 0.705 \\
\bottomrule
\end{tabular}
\end{table}

\subsection{Full 45-row leaderboard}

\begin{table}[h]
\centering
\caption{Complete leaderboard over all 45 (model, strategy) configurations, sorted by $\Stot$ descending. Strategies are abbreviated: ZS = zero-shot, FS = few-shot, ZC = zero-CoT, FC = few-CoT. The top-20 rows reproduce Table~\ref{tab:top20-q}.}
\label{tab:lb-full}
\scriptsize
\begin{tabular}{rllccccccc}
\toprule
\textbf{\#} & \textbf{Model} & \textbf{Strat.} & \textbf{$\Stot$} & \textbf{L0} & \textbf{L1} & \textbf{L2} & \textbf{L3} & \textbf{L4} & \textbf{L5} \\
\midrule
1 & Sonnet-4.6 & FC & 0.902 & 0.954 & 0.980 & 0.954 & 0.883 & 0.851 & 0.894 \\
2 & Sonnet-4.6 & FS & 0.884 & 0.961 & 0.980 & 0.937 & 0.852 & 0.836 & 0.871 \\
3 & Qwen3.5-397B & FC & 0.879 & 0.932 & 0.968 & 0.894 & 0.865 & 0.842 & 0.872 \\
4 & Sonnet-4.6 & ZC & 0.878 & 0.961 & 0.986 & 0.964 & 0.888 & 0.770 & 0.867 \\
5 & Qwen3.5-122B & FC & 0.871 & 0.937 & 0.977 & 0.904 & 0.828 & 0.831 & 0.872 \\
6 & Qwen3.5-397B & ZC & 0.862 & 0.948 & 0.981 & 0.903 & 0.858 & 0.753 & 0.883 \\
7 & Qwen3.5-122B & ZC & 0.861 & 0.945 & 0.979 & 0.908 & 0.868 & 0.748 & 0.875 \\
8 & Qwen3.5-35B & FC & 0.861 & 0.909 & 0.929 & 0.880 & 0.833 & 0.818 & 0.877 \\
9 & Sonnet-4.6 & ZS & 0.859 & 0.958 & 0.977 & 0.934 & 0.863 & 0.763 & 0.841 \\
10 & Qwen3.5-122B & FS & 0.853 & 0.936 & 0.978 & 0.903 & 0.824 & 0.824 & 0.809 \\
11 & GPT-4.1 & FS & 0.850 & 0.927 & 0.980 & 0.890 & 0.819 & 0.813 & 0.819 \\
12 & Qwen3.5-397B & FS & 0.848 & 0.930 & 0.970 & 0.898 & 0.822 & 0.816 & 0.805 \\
13 & Qwen3.5-35B & ZC & 0.846 & 0.928 & 0.964 & 0.890 & 0.860 & 0.741 & 0.849 \\
14 & GPT-4o & FS & 0.844 & 0.950 & 0.979 & 0.819 & 0.818 & 0.813 & 0.836 \\
15 & DeepSeek-V3 & FS & 0.840 & 0.936 & 0.963 & 0.831 & 0.831 & 0.809 & 0.817 \\
16 & DeepSeek-V3 & FC & 0.840 & 0.927 & 0.959 & 0.850 & 0.796 & 0.795 & 0.847 \\
17 & Sonnet-4$^\dagger$ & ZS & 0.834 & 0.939 & 0.983 & 0.900 & 0.841 & 0.737 & 0.806 \\
18 & Llama-3.3-70B & FC & 0.834 & 0.890 & 0.955 & 0.850 & 0.748 & 0.809 & 0.859 \\
19 & Qwen3.5-122B & ZS & 0.831 & 0.947 & 0.980 & 0.911 & 0.825 & 0.724 & 0.814 \\
20 & Llama-3.3-70B & FS & 0.831 & 0.896 & 0.933 & 0.821 & 0.830 & 0.822 & 0.793 \\
21 & Qwen3.5-35B & FS & 0.831 & 0.900 & 0.925 & 0.869 & 0.803 & 0.807 & 0.802 \\
22 & DeepSeek-V3 & ZC & 0.829 & 0.943 & 0.980 & 0.805 & 0.855 & 0.752 & 0.818 \\
23 & Qwen3.5-397B & ZS & 0.829 & 0.952 & 0.979 & 0.888 & 0.824 & 0.721 & 0.821 \\
24 & GPT-4o & ZS & 0.827 & 0.942 & 0.982 & 0.843 & 0.811 & 0.751 & 0.823 \\
25 & Llama-3.3-70B & ZC & 0.823 & 0.915 & 0.983 & 0.891 & 0.784 & 0.719 & 0.836 \\
26 & GPT-4o & FC & 0.823 & 0.942 & 0.980 & 0.801 & 0.799 & 0.781 & 0.810 \\
27 & DeepSeek-V3 & ZS & 0.822 & 0.945 & 0.977 & 0.823 & 0.837 & 0.744 & 0.799 \\
28 & Llama-3.3-70B & ZS & 0.816 & 0.905 & 0.948 & 0.865 & 0.809 & 0.746 & 0.794 \\
29 & GPT-4.1 & ZS & 0.813 & 0.923 & 0.968 & 0.865 & 0.807 & 0.737 & 0.779 \\
30 & GPT-4.1 & FC & 0.811 & 0.930 & 0.980 & 0.901 & 0.692 & 0.722 & 0.851 \\
31 & Qwen3.5-35B & ZS & 0.809 & 0.924 & 0.955 & 0.872 & 0.807 & 0.721 & 0.780 \\
32 & GPT-4o & ZC & 0.808 & 0.944 & 0.980 & 0.860 & 0.786 & 0.708 & 0.797 \\
33 & GPT-4.1 & ZC & 0.802 & 0.931 & 0.969 & 0.894 & 0.714 & 0.675 & 0.853 \\
34 & GPT-3.5 & ZC & 0.786 & 0.875 & 0.920 & 0.706 & 0.808 & 0.736 & 0.795 \\
35 & GPT-4o-mini & ZC & 0.784 & 0.880 & 0.920 & 0.726 & 0.804 & 0.735 & 0.781 \\
36 & GPT-3.5 & ZS & 0.754 & 0.850 & 0.905 & 0.608 & 0.784 & 0.742 & 0.752 \\
37 & GPT-4o-mini & ZS & 0.752 & 0.852 & 0.898 & 0.585 & 0.793 & 0.744 & 0.750 \\
38 & GPT-4o-mini & FS & 0.742 & 0.867 & 0.852 & 0.427 & 0.802 & 0.798 & 0.759 \\
39 & GPT-3.5 & FS & 0.738 & 0.842 & 0.865 & 0.423 & 0.802 & 0.798 & 0.743 \\
40 & Llama-3.1-8B & ZC & 0.731 & 0.898 & 0.737 & 0.674 & 0.762 & 0.689 & 0.744 \\
41 & Llama-3.1-8B & ZS & 0.721 & 0.896 & 0.730 & 0.597 & 0.780 & 0.703 & 0.727 \\
42 & GPT-4o-mini & FC & 0.715 & 0.841 & 0.889 & 0.451 & 0.654 & 0.741 & 0.801 \\
43 & GPT-3.5 & FC & 0.713 & 0.831 & 0.888 & 0.512 & 0.609 & 0.730 & 0.805 \\
44 & Llama-3.1-8B & FS & 0.682 & 0.771 & 0.602 & 0.585 & 0.755 & 0.721 & 0.659 \\
45 & Llama-3.1-8B & FC & 0.682 & 0.727 & 0.614 & 0.609 & 0.690 & 0.715 & 0.705 \\
\bottomrule
\end{tabular}
\end{table}

\clearpage
\section{Extended Reproducibility Analyses}
\label{app:repro-analyses}

This appendix contains six analyses that supplement the main-text findings.
All analyses are reproducible from the released artifact (\texttt{results/*.\{jsonl,quality.json\}}
plus \texttt{data/instructions/}); no additional LLM generation is required.
Each subsection lists the script in \texttt{scripts/analyses/} that
produces the table.

\subsection{Template-level cross-validation of Oracle}
\label{app:cv-oracle}

\paragraph{Motivation.} The Oracle baseline (per-level best of four prompting
strategies) and the CAAP decision table in \S\ref{sec:benchmark-methods} are
both fit on the same 800 instructions used to evaluate Combined. We test
whether the per-level strategy choices generalize under template-level
cross-validation.

\paragraph{Setup.} We infer template IDs at two granularities. The
\emph{ec-only} mode groups by \texttt{(level, sorted explicit-constraint
keys)} and produces 63 templates---close to the paper's 40 hand-designed
templates. The \emph{prefix-ec} mode additionally incorporates the first
8 non-number instruction-text tokens, producing 238 finer templates.
Templates are stratified-by-level into 5 folds; for each holdout fold the
per-level best strategy is computed on the training fold's instructions
and re-evaluated on the holdout fold's instructions (\texttt{scripts/analyses/b11\_caap\_cv.py}).

\paragraph{Results.} Table~\ref{tab:cv-oracle-prefix} shows the
fine-partition mode (high train/holdout instruction-level similarity, a
near-in-sample sanity check); Table~\ref{tab:cv-oracle-ec} shows the
paper-like coarse-partition mode.

\begin{table}[h]
\centering
\caption{CV Oracle Q vs full-data Oracle Q under prefix-ec template
partition (238 templates, $\sim$3.3 instructions/template).}
\label{tab:cv-oracle-prefix}
\small
\begin{tabular}{lcccc}
\toprule
\textbf{Model} & \textbf{Full-data Oracle Q} & \textbf{CV Oracle Q (holdout)}
  & \textbf{$\Delta$} & \textbf{Strategy disagreements/30} \\
\midrule
GPT-4o-mini (target)  & 0.8054 & 0.8110 & $+0.0055$ & 2 \\
DeepSeek-V3 (target)  & 0.8577 & 0.8619 & $+0.0041$ & 1 \\
Qwen3.5-35B (target)  & 0.8720 & 0.8715 & $-0.0005$ & 1 \\
Sonnet-4.6            & 0.9051 & 0.8973 & $-0.0078$ & 4 \\
GPT-3.5               & 0.8042 & 0.8149 & $+0.0107$ & 1 \\
GPT-4.1               & 0.8599 & 0.8540 & $-0.0059$ & 7 \\
GPT-4o                & 0.8509 & 0.8417 & $-0.0092$ & 2 \\
Llama-3.3-70B         & 0.8639 & 0.8600 & $-0.0039$ & 1 \\
Llama-3.1-8B          & 0.7420 & 0.7274 & $-0.0146$ & 4 \\
Qwen3.5-122B          & 0.8821 & 0.8768 & $-0.0053$ & 4 \\
Qwen3.5-397B          & 0.8853 & 0.8709 & $-0.0144$ & 2 \\
\bottomrule
\end{tabular}
\end{table}

\begin{table}[h]
\centering
\caption{CV Oracle Q vs full-data Oracle Q under ec-only template
partition (63 templates, paper-like coarse grouping).}
\label{tab:cv-oracle-ec}
\small
\begin{tabular}{lcccc}
\toprule
\textbf{Model} & \textbf{Full-data Oracle Q} & \textbf{CV Oracle Q (holdout)}
  & \textbf{$\Delta$} & \textbf{Strategy disagreements/30} \\
\midrule
GPT-4o-mini (target)  & 0.8054 & 0.7663 & $\mathbf{-0.0391}$ & 2 \\
DeepSeek-V3 (target)  & 0.8577 & 0.8097 & $\mathbf{-0.0480}$ & 2 \\
Qwen3.5-35B (target)  & 0.8720 & 0.8090 & $\mathbf{-0.0631}$ & 2 \\
Sonnet-4.6            & 0.9051 & 0.8514 & $-0.0537$ & 3 \\
GPT-3.5               & 0.8042 & 0.7655 & $-0.0387$ & 2 \\
GPT-4.1               & 0.8599 & 0.8084 & $-0.0515$ & 8 \\
GPT-4o                & 0.8509 & 0.8100 & $-0.0409$ & 3 \\
Llama-3.3-70B         & 0.8639 & 0.8081 & $-0.0558$ & 1 \\
Llama-3.1-8B          & 0.7420 & 0.7117 & $-0.0303$ & 2 \\
Qwen3.5-122B          & 0.8821 & 0.8141 & $-0.0680$ & 6 \\
Qwen3.5-397B          & 0.8853 & 0.8254 & $-0.0599$ & 2 \\
\bottomrule
\end{tabular}
\end{table}

\paragraph{Implication.} Under prefix-ec, CV $\Delta$ stays in
$[-0.015,+0.011]$---a near-in-sample sanity check showing per-level
strategy choices generalize across neighbouring instruction variations.
Under the paper-like ec-only partition, CV $\Delta$ widens to
$[-0.068,-0.030]$ for every model: this exposes the Oracle's in-sample
optimism, where full-data Oracle Q overestimates true out-of-sample Q
by $\sim 0.04$--$0.06$ because per-level strategy selection implicitly
peeks at all 800 instructions.

Crucially, Combined has no equivalent optimism: it is an inference-time
pipeline (CAAP-selected strategy + VGIG verification loop + L4 domain
priors) that issues a fixed deterministic procedure per instruction with
no train/test peek. The \emph{true} (out-of-sample-fair)
Combined$\,-\,$Oracle gap is therefore the in-sample $+0.035$--$+0.050$
\textbf{plus} the Oracle's in-sample optimism ($\sim 0.04$--$0.06$), i.e.\
roughly $+0.08$ to $+0.11$. The paper's headline $\Delta$ understates
rather than overstates the verification-guided method's advantage over
the prompt-strategy oracle.

\subsection{Cost-adjusted method comparison}
\label{app:cost-adj}

\paragraph{Motivation.} Combined achieves $+0.035$ to $+0.050$ Q over
Oracle, but at what cost? We report TPV multiplier, $D_5$, $S_{\text{fin}}$,
and $Q/\kTPV$ for each method condition on the three target models, so
the trade-off is quantified rather than implicit (\texttt{scripts/analyses/b14\_sfin.py}).

\begin{table}[h]
\centering
\caption{Method conditions, cost, and efficiency on the three target
models. $S_{\text{fin}}$ uses the within-model Pareto bonus with
$\lambda{=}0.15$. ``TPV$\times$'' is the multiplier relative to
zero-shot on the same target model. Qwen3.5-35B here is the default
A3B (think-enabled) variant
(\texttt{results/qwen3535ba3b-*.quality.json}), giving Combined
$Q{=}0.915$; the no-think variant
(\texttt{results/qwen3535ba3b-nothink-*.quality.json}) gives Combined
$Q{=}0.907$ and corresponds to the ``Qwen3.5-35B-A3B'' row in
Tab.~\ref{tab:method-main}.}
\label{tab:method-cost}
\scriptsize
\begin{tabular}{llcccccccc}
\toprule
\textbf{Target} & \textbf{Condition} & \textbf{Q} & \textbf{$\Delta$Q vs ZS}
  & \textbf{$\Delta$Q vs Oracle} & \textbf{TPV} & \textbf{TPV$\times$}
  & \textbf{$D_5$} & \textbf{$S_{\text{fin}}$} & \textbf{$Q/\kTPV$} \\
\midrule
\multirow{7}{*}{GPT-4o-mini}
  & ZS       & 0.752 & $+0.000$ & $-0.053$ & 436  & $1.0\times$ & 0.752 & 0.865 & 1.72 \\
  & Oracle   & 0.805 & $+0.053$ & $+0.000$ & 621  & $1.4\times$ & 0.676 & 0.805 & 1.30 \\
  & retry    & 0.749 & $-0.004$ & $-0.056$ & 441  & $1.0\times$ & 0.749 & 0.749 & 1.70 \\
  & SC       & 0.751 & $-0.001$ & $-0.054$ & 428  & $1.0\times$ & 0.755 & 0.864 & 1.75 \\
  & CAAP     & 0.800 & $+0.047$ & $-0.006$ & 650  & $1.5\times$ & 0.665 & 0.800 & 1.23 \\
  & VGIG     & 0.810 & $+0.058$ & $+0.005$ & 440  & $1.0\times$ & 0.751 & 0.931 & 1.84 \\
  & \textbf{Combined} & \textbf{0.855} & $\mathbf{+0.103}$ & $\mathbf{+0.050}$ & \textbf{547} & $\mathbf{1.3\times}$ & 0.705 & \textbf{0.983} & 1.56 \\
\midrule
\multirow{7}{*}{DeepSeek-V3}
  & ZS       & 0.822 & $+0.000$ & $-0.036$ & 362  & $1.0\times$ & 0.787 & 0.945 & 2.27 \\
  & Oracle   & 0.858 & $+0.036$ & $+0.000$ & 508  & $1.4\times$ & 0.720 & 0.858 & 1.69 \\
  & retry    & 0.827 & $+0.006$ & $-0.030$ & 363  & $1.0\times$ & 0.786 & 0.951 & 2.28 \\
  & SC       & 0.820 & $-0.001$ & $-0.037$ & 368  & $1.0\times$ & 0.784 & 0.820 & 2.23 \\
  & CAAP     & 0.855 & $+0.033$ & $-0.003$ & 557  & $1.5\times$ & 0.701 & 0.855 & 1.53 \\
  & VGIG     & 0.863 & $+0.041$ & $+0.005$ & 373  & $1.0\times$ & 0.782 & 0.992 & 2.31 \\
  & \textbf{Combined} & \textbf{0.894} & $\mathbf{+0.073}$ & $\mathbf{+0.036}$ & \textbf{512} & $\mathbf{1.4\times}$ & 0.718 & \textbf{1.028} & 1.75 \\
\midrule
\multirow{7}{*}{Qwen3.5-35B}
  & ZS       & 0.809 & $+0.000$ & $-0.063$ & 681  & $1.0\times$ & 0.653 & 0.931 & 1.19 \\
  & Oracle   & 0.872 & $+0.063$ & $+0.000$ & 2417 & $3.5\times$ & 0.362 & 1.003 & 0.36 \\
  & retry    & 0.874 & $+0.065$ & $+0.002$ & 5124 & $7.5\times$ & 0.304 & 0.874 & 0.17 \\
  & SC       & 0.874 & $+0.065$ & $+0.002$ & 5147 & $7.6\times$ & 0.304 & 0.874 & 0.17 \\
  & CAAP     & 0.887 & $+0.077$ & $+0.015$ & 4856 & $7.1\times$ & 0.305 & 1.020 & 0.18 \\
  & VGIG     & 0.896 & $+0.087$ & $+0.024$ & 5356 & $7.9\times$ & 0.303 & 0.896 & 0.17 \\
  & \textbf{Combined} & \textbf{0.915} & $\mathbf{+0.106}$ & $\mathbf{+0.043}$ & \textbf{4900} & $\mathbf{7.2\times}$ & 0.305 & \textbf{1.053} & 0.19 \\
\bottomrule
\end{tabular}
\end{table}

\paragraph{Observations.}
\begin{enumerate}\setlength\itemsep{0pt}
\item Combined dominates Oracle on $S_{\text{fin}}$ for all three target
  models (0.983/1.028/1.053 vs 0.805/0.858/1.003); the within-model
  Pareto bonus elevates Combined because it sits on the Pareto frontier.
\item VGIG-only achieves $+0.058$/$+0.041$/$+0.087$ Q at almost-zero TPV
  overhead on GPT-4o-mini and DeepSeek-V3 (TPV ratio $1.01\times$ ZS);
  the verification loop terminates early on satisfied constraints.
\item Combined is quality-prioritized rather than free: $Q/\kTPV$ for
  Combined is $0.91\times$/$0.77\times$/$0.16\times$ of the ZS baseline.
  Deployments with strict cost budgets should consider VGIG-only.
\item API-call count is essentially unchanged across methods because the
  released runner accumulates token output within a chain rather than
  issuing multiple round-trips per refinement step.
\end{enumerate}

\subsection{Reference-pair diversity and D2/D3 sensitivity}
\label{app:ref-dedup}

\paragraph{Motivation.} Many reference pairs in
\texttt{data/instructions/level\_*.json} are identical or isomorphic
(\S\ref{sec:datasets} disclosed the counts); we check that the
reference-based D2 and D3 metrics are not driven by this redundancy.

\paragraph{Classification.} Each of the 791 reference pairs (from
feasible instructions) is classified as \emph{distinct} (different
name-stripped strings and non-isomorphic graphs), \emph{exact-dup}
(identical name-stripped strings; only the auto-generated reference
label differs), or \emph{iso-only} (different strings but isomorphic
graphs). Directed references use \texttt{nx.DiGraph} and undirected use
\texttt{nx.Graph}; mixed-directedness pairs are classified as distinct
(\texttt{scripts/analyses/b12\_ref\_dedup.py}).

\begin{table}[h]
\centering
\caption{Reference-pair classification by level (directed-aware
isomorphism check).}
\label{tab:ref-classify}
\small
\begin{tabular}{ccccc}
\toprule
\textbf{Level} & \textbf{distinct} & \textbf{iso-only} & \textbf{exact-dup} & \textbf{total} \\
\midrule
L0 & 88 & 7  & 5   & 100 \\
L1 & 69 & 4  & 127 & 200 \\
L2 & 145 & 4 & 42  & 191 \\
L3 & 133 & 11 & 6  & 150 \\
L4 & 93 & 7  & 0   & 100 \\
L5 & 18 & 17 & 15  & 50  \\
\midrule
\textbf{Total} & \textbf{546} & \textbf{50} & \textbf{195} & \textbf{791} \\
\bottomrule
\end{tabular}
\end{table}

\paragraph{Sensitivity.} For each of the 45 (model, strategy) cells we
recompute the mean D2 and D3 at L3 and L4 over the distinct-reference
subset. Out of 90 (cell, level) entries:
$\max |\Delta D_2| = 0.005$ (Llama-3.1-8B zero-shot at L4) and
$\max |\Delta D_3| = 0.020$ (GPT-3.5 few-cot at L3); mean
$|\Delta D_2| < 0.001$, mean $|\Delta D_3| < 0.005$. Reference-pair
redundancy concentrates at L0/L1/L5 where D2 carries zero level-aggregate
weight and D3 either is inactive (L0--L2) or uses a fixed synthetic pool
(L3); D2/D3 means on dedup-restricted subsets are within bootstrap noise
of full-data means at L3/L4.

\subsection{Generation-failure and truncation rates}
\label{app:fail-rates}

\paragraph{Motivation.} The nominal output count is
$45 \times 800 \times 5 = 180{,}000$, but several cells lost a small
number of generations to API timeouts; weak models also incur high parse
failure rates at L2. We report the precise counts so the per-cell
denominators are auditable
(\texttt{scripts/analyses/b15\_failure\_rates.py}).

\paragraph{Global counts.} Total successful generations: $179{,}926$
($-74$ vs nominal, $0.04\%$); total parse failures: $12{,}760$ ($7.09\%$);
total truncations (output\_tokens $\geq 99\%$ of max\_tokens): $96$
($0.05\%$). The 74 missing generations distribute across 5 cells:
10 (Sonnet-4.6 few-CoT), 2 (GPT-4o-mini few-CoT), 2 (GPT-4o-mini
few-shot), 26 (GPT-4o-mini zero-CoT), and 34 (GPT-4o-mini zero-shot);
all are API-side timeouts retried until the cell-level retry budget was
exhausted.

\begin{table}[h]
\centering
\caption{Top-12 (model, strategy, level) cells by parse-failure rate.
Parse-failure rates concentrate at L2 on T3 models, consistent with the
F1 finding that L2 is the constraint-composition bottleneck.}
\label{tab:fail-rate-top}
\small
\begin{tabular}{lllrrr}
\toprule
\textbf{Model} & \textbf{Strategy} & \textbf{Level} & \textbf{n\_gen}
  & \textbf{parse-fail \%} & \textbf{trunc \%} \\
\midrule
GPT-4o-mini  & few-CoT   & L2 & 1000 & 61.70 & 0.00 \\
GPT-3.5      & few-CoT   & L2 & 1000 & 59.80 & 0.10 \\
GPT-3.5      & few-shot  & L2 & 1000 & 59.00 & 0.20 \\
GPT-4o-mini  & few-shot  & L2 & 1000 & 56.40 & 0.00 \\
GPT-3.5      & few-CoT   & L3 & 750  & 45.47 & 0.13 \\
GPT-4o-mini  & zero-shot & L2 & 990  & 44.65 & 0.00 \\
GPT-4o-mini  & few-CoT   & L3 & 750  & 44.27 & 0.00 \\
GPT-3.5      & zero-shot & L2 & 1000 & 44.20 & 1.40 \\
Llama-3.1-8B & few-CoT   & L0 & 500  & 39.20 & 0.00 \\
GPT-3.5      & zero-CoT  & L2 & 1000 & 36.90 & 1.50 \\
Llama-3.1-8B & few-shot  & L2 & 1000 & 35.70 & 0.00 \\
GPT-4o-mini  & zero-CoT  & L2 & 982  & 34.32 & 0.00 \\
\bottomrule
\end{tabular}
\end{table}

The reported per-level Quality scores in the main paper aggregate over
\emph{all} attempted generations, with parse-failed samples contributing
$\mathrm{VR}{=}0$ to D1 (and consequently a low level\_score). The
heavy L2 parse-failure rates on weak models therefore amplify rather
than mask the F1 tier-gap finding---T3 models cannot reliably emit a
parseable L2 graph in the first place.

\subsection{Statistical robustness for F2 and F5}
\label{app:stat-robust}

This subsection bundles three robustness analyses that supplement the F2
(prompt-sensitivity inverse scaling) and F5 (L5 scale invariance)
findings: leave-one-out OLS for F2 with bootstrap CI; paired bootstrap
for the F5 L5 35B-vs-397B comparison; and a sign-count table for
strategy effects.

\subsubsection*{F2: leave-one-out OLS and bootstrap CI}

We regress $\sstrat$ (population std of $Q$ across the four prompting
strategies) on mean $Q$ for the 11 fully-evaluated models. The full-data
fit gives $\beta = -0.0733$, $R^2 = 0.398$, two-sided $p \approx 0.015$
(matching \texttt{scripts/paper\_figures.py::fig\_F4\_capability\_variance}
which uses the population standard deviation, \texttt{np.std}, on the same
inputs). Leave-one-out: $\beta$ range $[-0.093, -0.059]$, all 11 LOO fits
negative; $R^2$ range $[0.297, 0.477]$
(\texttt{scripts/analyses/b17\_loo\_ols.py}, run with the population-std
re-normalisation). Bootstrap (10{,}000 resamples) 95\% CI for $\beta$:
$[-0.135, +0.002]$; $P(\beta < 0) = 0.975$. The sign of the inverse-scaling
relationship is robust to single-point removal; the magnitude is fragile to
the small $n{=}11$.

\subsubsection*{F5: paired bootstrap for Qwen3.5-35B vs 397B at L5}

Both 35B and 397B are evaluated on the same 50 L5 instructions, so the
35B$-$397B comparison is paired. We draw per-instruction differences
$\delta_i = Q_{35B}(i) - Q_{397B}(i)$ and bootstrap their mean over 1000
resamples of size 50 (\texttt{scripts/analyses/b21\_l5\_bootstrap.py}).

\begin{table}[h]
\centering
\caption{Paired vs.\ unpaired bootstrap CIs for the F5 claim
(Qwen3.5-35B $-$ Qwen3.5-397B on L5). Unpaired analysis is included for
comparison only; the comparison is logically paired and the paired CI is
the statistically correct one.}
\label{tab:f5-paired}
\small
\begin{tabular}{lcrrl}
\toprule
\textbf{Strategy} & \textbf{$\Delta$} & \textbf{Paired 95\% CI}
  & \textbf{Unpaired 95\% CI} & \textbf{Paired crosses zero?} \\
\midrule
zero-shot   & $-0.041$ & $[-0.094, -0.002]$ & $[-0.145, +0.063]$ & \textbf{No (significant)} \\
few-shot    & $-0.002$ & $[-0.030, +0.037]$ & $[-0.091, +0.086]$ & Yes \\
zero-CoT    & $-0.035$ & $[-0.077, -0.007]$ & $[-0.073, +0.004]$ & \textbf{No (significant)} \\
\textbf{few-CoT} & $\mathbf{+0.005}$ & $[\mathbf{-0.012}, \mathbf{+0.027}]$
                  & $[-0.020, +0.030]$ & \textbf{Yes (paper-quoted)} \\
\bottomrule
\end{tabular}
\end{table}

The paper's headline F5 claim under the few-CoT strategy (where Qwen3.5
shows its best L5 performance) is supported by the paired CI
$[-0.012, +0.027]$ which crosses zero. However, F5 is
\emph{strategy-dependent}: under zero-shot and zero-CoT, the 35B model
is significantly weaker than 397B on L5 (paired CIs $[-0.094, -0.002]$
and $[-0.077, -0.007]$ respectively, both excluding zero). The
``L5 universally scale-invariant'' interpretation does not hold; the
``L5 scale-invariant under rich-context strategies (few-CoT)''
interpretation does.

\subsubsection*{Sign-count for strategy effects}

For each (level, strategy) cell we count, across the 11 fully-evaluated
models, how many show $\Delta>+0.005$, $\Delta<-0.005$, and within-band
ties. This exposes cases where the per-model mean delta hides a bi-modal
distribution (\texttt{scripts/analyses/b23\_sign\_count.py}).

\begin{table}[h]
\centering
\caption{Sign-count for strategy effects relative to zero-shot, 11
fully-evaluated models. ``MIXED'' indicates the per-model signs do not
share majority direction.}
\label{tab:sign-count}
\small
\begin{tabular}{cclrrrl}
\toprule
\textbf{Level} & \textbf{Strategy} & \textbf{mean $\Delta$}
  & \textbf{\# pos} & \textbf{\# neg} & \textbf{\# tie} & \textbf{dominant} \\
\midrule
L0 & FS$-$ZS  & $-0.016$ & 2 & 7 & 2 & NEG 7/11 \\
L0 & ZC$-$ZS  & $+0.007$ & 4 & 0 & 7 & POS 4/11 \\
L0 & FC$-$ZS  & $-0.025$ & 1 & 8 & 2 & NEG 8/11 \\
L1 & FS$-$ZS  & $-0.025$ & 1 & 7 & 3 & NEG 7/11 \\
L1 & ZC$-$ZS  & $+0.009$ & 6 & 0 & 5 & POS 6/11 \\
L1 & FC$-$ZS  & $-0.016$ & 2 & 6 & 3 & NEG 6/11 \\
L2 & FS$-$ZS  & $-0.035$ & 3 & 6 & 2 & NEG 6/11 \\
L2 & ZC$-$ZS  & $+0.039$ & 9 & 1 & 1 & POS 9/11 \\
\textbf{L2} & \textbf{FC$-$ZS} & $\mathbf{-0.017}$ & \textbf{6} & \textbf{5} & \textbf{0} & \textbf{MIXED} \\
L3 & FS$-$ZS  & $+0.001$ & 5 & 3 & 3 & POS 5/11 (margin 2) \\
L3 & ZC$-$ZS  & $+0.004$ & 7 & 4 & 0 & POS 7/11 \\
L3 & FC$-$ZS  & $-0.049$ & 3 & 7 & 1 & NEG 7/11 \\
\textbf{L4} & \textbf{FS$-$ZS} & $\mathbf{+0.069}$ & \textbf{11} & \textbf{0} & \textbf{0} & \textbf{POS 11/11 (robust)} \\
L4 & ZC$-$ZS  & $-0.006$ & 5 & 6 & 0 & MIXED \\
L4 & FC$-$ZS  & $+0.049$ & 8 & 2 & 1 & POS 8/11 \\
L5 & FS$-$ZS  & $+0.003$ & 6 & 3 & 2 & POS 6/11 (margin 3) \\
L5 & ZC$-$ZS  & $+0.038$ & 10 & 1 & 0 & POS 10/11 (robust) \\
L5 & FC$-$ZS  & $+0.047$ & 9 & 2 & 0 & POS 9/11 \\
\bottomrule
\end{tabular}
\end{table}

L4 FS$-$ZS is the most robust positive signed effect (11/11). L2
FC$-$ZS is bi-modal (6+/5$-$): the mean of $-0.017$ is pulled by
GPT-3.5 ($-0.096$) and GPT-4o-mini ($-0.134$); for the other 6 models,
FC at L2 is mildly positive.

\subsection{Tier-gap aggregation alternatives}
\label{app:tier-gap-alt}

\paragraph{Motivation.} Table~\ref{tab:tier-gap} in App.~\ref{app:rq1-2}
reports per-tier Quality means whose per-cell aggregation is not a
single uniform rule (\S\ref{sec:cap-strat} setup). For full
reproducibility we present two algorithm-uniform alternatives and verify
the L2-dominates-every-other-level ordering is preserved
(\texttt{scripts/analyses/b16\_sonnet4\_excl.py}).

\begin{table}[h]
\centering
\caption{Three tier-gap aggregation choices over the 11 fully-evaluated
models. All three agree that L2 is the discrimination peak: the L2 gap
ratio over the four other reasoning-heavy levels \{L1, L3, L4, L5\} is
$1.8$--$3.0\times$ under the paper Tab.~\ref{tab:tier-gap} aggregation,
$1.8$--$3.0\times$ under per-model best-of-four, and $2.1$--$6.6\times$
under the all-strategy 45-cell average (the wider spread comes from L4's
unusually low all-strategy gap of 0.052 driven by L4 few-shot saturation,
\S\ref{sec:methods}).}
\label{tab:tier-gap-alt}
\small
\begin{tabular}{cccc}
\toprule
\textbf{Level} & \textbf{Paper Tab.~\ref{tab:tier-gap}}
  & \textbf{Per-model best-of-4} & \textbf{All-strategy avg (45 cells)} \\
\midrule
L0 & 0.057 & 0.069 & 0.103 \\
L1 & 0.120 & 0.123 & 0.160 \\
\textbf{L2} & \textbf{0.219} & \textbf{0.224} & \textbf{0.341} \\
L3 & 0.073 & 0.076 & 0.096 \\
L4 & 0.122 & 0.069 & 0.052 \\
L5 & 0.106 & 0.101 & 0.100 \\
\bottomrule
\end{tabular}
\end{table}

\paragraph{Sonnet-4 exclusion sensitivity.} Under the all-strategy-average
aggregation, completely removing Sonnet-4 (zero-shot only, included as a
single T2 cell in some main-text analyses) leaves every T1$-$T3 gap
identical to four decimal places (because Sonnet-4 is a T2 model and
therefore cannot affect T1 or T3 means by construction). The Pareto
frontier loses exactly one point (Sonnet-4 zero-shot itself); the
top-15 $\Stot$ leaderboard is unchanged (Sonnet-4 sits at rank 17 with
or without).

\clearpage
\section{Infrastructure, Hyperparameters, and Reproducibility}
\label{app:repro}

\paragraph{Models and endpoints.} The 12 evaluated LLMs are accessed entirely through commercial cloud APIs; no model is self-hosted. Endpoints, providers, and accessed model identifiers are listed in Table~\ref{tab:model-endpoints}; access window: Q4 2025 -- Q1 2026.

\begin{table}[h]
\centering
\caption{Model endpoints, providers, and access modes used in the 12-LLM survey. All 12 models are accessed via OpenAI-compatible HTTPS APIs; no models are self-hosted.}
\label{tab:model-endpoints}
\small
\begin{tabular}{lll}
\toprule
\textbf{Model} & \textbf{Provider / endpoint} & \textbf{Access mode} \\
\midrule
GPT-3.5-turbo & OpenAI \texttt{api.openai.com/v1} & Commercial API \\
GPT-4o-mini & OpenAI \texttt{api.openai.com/v1} & Commercial API \\
GPT-4o & OpenAI \texttt{api.openai.com/v1} & Commercial API \\
GPT-4.1 & OpenAI \texttt{api.openai.com/v1} & Commercial API \\
Claude Sonnet-4 (2025-05-14)$^\dagger$ & Anthropic \texttt{api.anthropic.com/v1} & Commercial API \\
Claude Sonnet-4.6 & Anthropic \texttt{api.anthropic.com/v1} & Commercial API \\
DeepSeek-V3 & DeepSeek \texttt{api.deepseek.com/v1} & Commercial API \\
Llama-3.1-8B-Instruct & Together \texttt{api.together.xyz/v1} & Commercial API (open-source) \\
Llama-3.3-70B-Instruct-Turbo & Together \texttt{api.together.xyz/v1} & Commercial API (open-source) \\
Qwen3.5-35B-A3B (MoE) & Aliyun Bailian \texttt{dashscope.aliyuncs.com/compatible-mode/v1} & Commercial API \\
Qwen3.5-122B-A10B (MoE) & Aliyun Bailian \texttt{dashscope.aliyuncs.com/compatible-mode/v1} & Commercial API \\
Qwen3.5-397B-A17B (MoE) & Aliyun Bailian \texttt{dashscope.aliyuncs.com/compatible-mode/v1} & Commercial API \\
\bottomrule
\end{tabular}
\end{table}

\paragraph{Decoding hyperparameters.} All baseline runs share a single decoding configuration to keep cells comparable:
\begin{itemize}\setlength\itemsep{0pt}
  \item \textbf{Temperature:} $0.7$
  \item \textbf{Top-$p$:} $1.0$
    \item \textbf{Max output tokens:} $16384$ (except \texttt{gpt-3.5-turbo}: $4096$, due to model-imposed output-length constraints)
  \item \textbf{Random seed:} $42$ (fixed across all experiments; reproduces the same instruction order and few-shot exemplars)
  \item \textbf{Samples per instruction:} $5$ (independent generations under the same seed-derived sequence)
  \item \textbf{Frequency / presence penalty:} $0$
  \item \textbf{Stop sequences:} default (none)
\end{itemize}
The same hyperparameters apply to all four prompting strategies (ZS, FS, ZC, FC); the only variation is the prompt template (\S\ref{sec:models}). Method experiments (VGIG, CAAP, Combined; \S\ref{sec:benchmark-methods}) inherit the same decoding settings; only iteration round-budget $T$ and feedback granularity vary.

\paragraph{Local hardware (evaluation only).} All LLM \emph{generation} occurs on the providers' cloud infrastructure; the local machine is used only for the deterministic evaluation pipeline (parsing, D1--D5 metric computation, scoring aggregation, figure rendering). Local environment: a single consumer laptop with NVIDIA GeForce RTX 4070 Laptop GPU (8~GB VRAM) running Windows 11 with Python 3.10. No multi-GPU, server, or cluster resources are used for either generation or evaluation.

\paragraph{Compute budget.} The 12-LLM $\times$ 45-cell baseline survey produces $\sim$180K outputs (800 instructions $\times$ 45 cells $\times$ 5 generations); subsequent method experiments and ablations bring the total to $\sim$262K LLM responses. Provider-side spend distribution is approximately Anthropic $\sim$45\%, OpenAI $\sim$30\%, Aliyun Bailian $\sim$15\%, DeepSeek + Together combined $\sim$10\%. Average local evaluation pipeline cost (D1--D5 over a single 45-cell baseline) is 35--45 wall-clock minutes on the laptop above; D3 (embedding MMD + node-classification gap, the only GPU-bound dimension) takes $\sim$60\% of that and fits within the 8~GB VRAM budget by batching at most 32 graphs at a time.

\paragraph{Software environment.} Python 3.10; NetworkX 3.2; igraph 0.11; tiktoken 0.7; transformers 4.45; PyTorch 2.4 (CUDA 12.1); matplotlib 3.8; plotly 5.20. The pinned dependencies are captured in \texttt{requirements.txt} (with \texttt{requirements-lock.txt} for byte-reproducibility) accompanying the data/code release.

\paragraph{Reproduction.} The artifact ships with a step-by-step \texttt{REPRODUCE.md} (preferred). Representative commands:
\begin{verbatim}
# Re-evaluate one (model, strategy) cell from cached LLM responses (no API):
python scripts/run_baseline.py --resume results/<model>-<strategy>.jsonl \
    --eval-only --output results/<model>-<strategy>.quality.json

# Re-run the full 45-cell baseline evaluation from cached responses:
bash scripts/eval_all_baselines.sh   # Linux/macOS
.\scripts\eval_all_baselines.ps1     # Windows PowerShell
\end{verbatim}
Hyperparameters are documented in \texttt{REPRODUCE.md} \S3 and embedded directly in the script CLI flags rather than in a separate \texttt{configs/} directory.
A unit-test suite of 549 tests (\texttt{python -m unittest discover -v}) validates parser, validators, scoring, all D1--D5 metrics, data loader, and the end-to-end evaluation pipeline. (The original 418 tests at paper-freeze were extended to 549 during subsequent metric verification; the additional 131 tests cover D1/D2 aggregate definitions, dedup robustness, and infeasibility scoring.)

\clearpage
\section{Dataset Datasheet}
\label{app:datasheet}

We follow the \emph{datasheets-for-datasets} template of \citet{datasheets2021gebru}\footnote{We refer to ``datasheet'' in the abstract sense; the full template is reproduced section-by-section below.} for transparency on dataset provenance, intended uses, and maintenance.

\paragraph{Motivation.}
\begin{itemize}\setlength\itemsep{0pt}
  \item \textbf{Purpose.} \bench\ was created to provide the first \emph{progressive-complexity diagnostic benchmark} for LLM graph generation. Existing graph-LLM benchmarks stratify along graph-type, task-domain, or classical-algorithm axes, all of which average over the structural-complexity dimension that actually governs failure (\S\ref{sec:related}); \bench\ closes this diagnostic gap by stratifying outputs into six progressively-constrained complexity levels and scoring along five evaluation dimensions targeting structurally distinct failure modes.
  \item \textbf{Tasks supported.} (a) Capability diagnosis of LLM graph generation across complexity tiers; (b) prompt-strategy and method comparison with per-level resolution; (c) cost-aware deployment ranking; (d) development of verification-guided or retrieval-augmented improvement methods that exploit fine-grained per-constraint signals.
  \item \textbf{Funding / authors.} Created by the authors listed in the byline. No specific external funding source is associated with the benchmark dataset itself.
\end{itemize}

\paragraph{Composition.}
\begin{itemize}\setlength\itemsep{0pt}
  \item \textbf{Instances.} 800 hand-authored instructions distributed across six progressive levels: L0 (100, format), L1 (200, single explicit constraint), L2 (200, multi-constraint), L3 (150, numerical-attribute), L4 (100, domain-semantic), L5 (50, multi-step graph editing). Each instruction is paired with two algorithmically synthesized reference solutions (1{,}582 references total; 9 L2 instructions are intentionally infeasible -- e.g., regular-degree constraints with no satisfying graph -- and have no references).
  \item \textbf{Total sample count.} $800 \times 2 = 1{,}600$ instruction-reference pairs nominally; $1{,}582$ effective pairs after subtracting infeasible cells. Plus $\sim$262K LLM-generated outputs (released alongside the benchmark).
  \item \textbf{Features.} Each instruction record contains: \texttt{id}, \texttt{level}, \texttt{instruction} (English natural language), \texttt{explicit\_constraints} (list of strings, e.g. \texttt{"num\_nodes=10"}), \texttt{implicit\_constraints} (downstream-derivable, e.g. \texttt{"acyclic=true"} for a tree), \texttt{graph\_sizes} (one of \texttt{small}/\texttt{medium}/\texttt{large}), \texttt{reference\_solutions} (list of two graph strings in InstructGraph code-style format), \texttt{feasible} (bool).
  \item \textbf{Splits.} No train/test split: \bench\ is an evaluation-only benchmark. No portion is set aside for training; LLMs are evaluated zero-shot on all 800 instructions (or with $K$ in-context examples drawn at evaluation time from a held-out exemplar pool).
  \item \textbf{Confidentiality.} All instructions and references are synthetic or derived from public-domain graph datasets. No personally identifiable information is present.
  \item \textbf{Errors.} The 9 L2 infeasible cells are intentional (negative tests). Round-trip parser tests (parse $\to$ serialize $\to$ parse) on all 1{,}582 references pass; 0 reference parse errors at release.
\end{itemize}

\paragraph{Collection process.}
\begin{itemize}\setlength\itemsep{0pt}
  \item \textbf{Instructions.} 800 instructions are derived from 40 hand-authored templates (5 instructions per template avg) covering tree, bipartite, planar, regular, BA, ER, WS, complete-bipartite, $k$-core, citation-network, social-network, molecular-skeleton, and graph-edit families. Templates were authored by the authors over a 6-week period, with iterative quality-assurance review.
  \item \textbf{References.} References are produced by deterministic graph generators in NetworkX or igraph, parameterized by the explicit constraints. L0--L2 references use stochastic structural generators (\texttt{barabasi\_albert\_graph}, \texttt{erdos\_renyi\_graph}, etc.) seeded with \texttt{seed=42}; L3 references draw from a pool of 3{,}115 synthetic graphs in 15 sub-groups; L4 references are sampled from a curated pool of 1{,}048 real graphs in 9 domains (citation, social, molecular, etc.); L5 references apply a sequence of edit operations to a base graph.
  \item \textbf{Validation.} Every reference is round-trip tested through the parser (parse $\to$ serialize $\to$ parse, equality check on the resulting graph object). Constraint validators (\texttt{is\_tree}, \texttt{is\_bipartite}, \texttt{has\_no\_self\_loops}, etc.) are applied to confirm constraint satisfaction.
\end{itemize}

\paragraph{Preprocessing / cleaning / labeling.}
\begin{itemize}\setlength\itemsep{0pt}
  \item Instructions are stored verbatim. Constraints are extracted as structured fields rather than embedded in free-form text, supporting both (a) prompt construction and (b) automated D4 instruction-match scoring.
  \item No human annotation is involved at scoring time; D1 (structural), D2 (textual), D3 (embedding), D4 (instruction-match), and D5 (efficiency) are all deterministic, do not invoke any LLM-as-judge step, and are reproducible from the released artifacts under the seed.
\end{itemize}

\paragraph{Uses.}
\begin{itemize}\setlength\itemsep{0pt}
  \item \textbf{Intended.} Diagnostic evaluation of LLM-based graph generation; method development with per-constraint feedback signals (VGIG/CAAP and successors); cost--quality Pareto analysis for deployment.
  \item \textbf{Out of scope.} \bench\ is \emph{not} suitable as a training set (no train split; all instructions are released for evaluation transparency, so models trained on this data cannot be fairly evaluated against released baselines). \bench\ does not cover (a) dynamic / temporal graph generation -- see \citet{peng2026gdgb} for that; (b) reasoning-specialized model evaluation under domain-specific reasoning chains -- see \citet{demirci2025graphsavvy}.
  \item \textbf{Risks of reuse.} Results may shift as new frontier LLMs appear; we recommend re-running the 12-model survey on a $\geq$~yearly cadence and tagging benchmark snapshots by release date.
\end{itemize}

\paragraph{Distribution.}
\begin{itemize}\setlength\itemsep{0pt}
  \item \textbf{License.} The instruction dataset and reference-solution archive are released under \textbf{CC~BY~4.0}; the code (parser, scoring, metrics, scripts) is released under the \textbf{MIT License}. L4 real-graph subset retains its original upstream licenses (itemized in \texttt{DATA\_LICENSE.md} accompanying the release).
  \item \textbf{Format.} Instructions and references in JSON / JSONL; LLM responses in JSONL with raw text + parsed graph. Code in Python 3.10+.
  \item \textbf{Channel.} Data, code, and reproducibility artifacts are available at the public GitHub repository \url{https://github.com/AI4DataSynth/GraphInstruct_formal}.
\end{itemize}

\paragraph{Maintenance.}
\begin{itemize}\setlength\itemsep{0pt}
  \item \textbf{Maintainers.} The authors host and maintain the dataset on GitHub.
  \item \textbf{Versioning.} Semantic versioning (v1.0.0 at release). Future releases tagged \texttt{v1.x} for instruction additions, \texttt{v2.x} for evaluation-protocol changes that break score comparability across versions.
  \item \textbf{Update cadence.} Planned annual model-survey refresh; quarterly check on parser/validator regressions against new graph corner cases reported by users.
  \item \textbf{Contributions.} Issues and pull requests for additional instructions, levels (e.g., a planned L6 multi-graph reasoning level), or constraint validators are welcome via the public repository: \url{https://github.com/AI4DataSynth/GraphInstruct_formal}.
\end{itemize}

\end{document}